\documentclass{mn2e}
\usepackage{epsfig,verbatim}
\usepackage{amssymb,amsmath,bm}
\usepackage{natbib209}
\newcommand{\vv}[1]{\textbf{#1}}	

\newcommand{\Msun}{M_{\odot}}
\newcommand{\Ma}{M_{\rm a}}
\newcommand{\Mc}{M_{\rm c}}
\def\bxi{\mbox{\boldmath$\xi$}}

\def\bb{{\hat{\bf b}}}

\def\tbB{{\tilde{\bf B}}}
\def\tbJ{{\tilde{\bf J}}}

\def\bB{{\bf B}}

\def\bxi{{\bf\xi}}
\def\bJ{{\bf J}}

\def\bkappa{\mbox{\boldmath$\kappa$}}
\def\bxip{{{\mbox{\boldmath$\xi$}}_{\perp}}}

\def\cs2{{c_{s}^{2}}}

\begin{document}
\title{
Burial of the polar magnetic field of an accreting neutron star.
II.
Hydromagnetic stability of axisymmetric equilibria
}
\author[D. J. B. Payne \& A. Melatos]
{D. J. B. Payne$^1$  \& A. Melatos$^1$ \\
$^1$
School of Physics, University of Melbourne,
Parkville, VIC 3010, Australia. \\
}

\maketitle
\begin{abstract}
The theory of polar magnetic burial in accreting neutron stars
predicts that a mountain of accreted material accumulates
at the magnetic poles of the star,
and that, as the mountain spreads equatorward, it is confined by,
and compresses, the equatorial magnetic field.
Here, we extend previous, axisymmetric, Grad-Shafranov
calculations of the hydromagnetic structure of a magnetic mountain
up to accreted masses as high as $\Ma = 6\times 10^{-4}\Msun$,
by importing the output from previous calculations
(which were limited by numerical problems and the formation
of closed bubbles to $\Ma < 10^{-4}\Msun$)
into the time-dependent, ideal-magnetohydrodynamic code ZEUS-3D and
loading additional mass onto the star dynamically.
The rise of buoyant magnetic bubbles through the accreted layer
is observed in these experiments.
We also investigate the stability of the resulting hydromagnetic equilibria
by perturbing them in ZEUS-3D.
Surprisingly, it is observed that the equilibria 
are marginally stable for all $\Ma\leq 6\times 10^{-4}\Msun$;
the mountain oscillates persistently when perturbed,
in a combination of Alfv\'en and acoustic modes,
without appreciable damping or growth,
and is therefore not disrupted
(apart from a transient Parker instability initially, which expels
$< 1 \%$ of the mass and magnetic flux).
\end{abstract}

\begin{keywords}
{accretion, accretion discs ---
 pulsars ---
 stars: magnetic fields ---
 stars: neutron}
\end{keywords}
%
\section{Introduction \label{sec:acc1}}
The magnetic dipole moments $\mu$ of neutron stars are observed to
decrease with accreted mass, $\Ma$.
Evidence of this trend is found in a variety of systems,
e.g. low- and high-mass X-ray binaries, and binary radio pulsars
with white-dwarf and neutron-star companions
\citep{taa86,shi89,van95},
although there is some debate over whether the trend is monotonic
\citep{wij97}.
Several theoretical mechanisms have been proposed to explain the
$\mu(\Ma)$ data, including accelerated Ohmic decay
\citep{urp95,urp97},
fluxoid-vortex interactions \citep{mus85,sri90}, and
magnetic screening or burial \citep{bis74,rom90}.
With regard to the latter mechanism,
\citet{pay04} (hereafter PM04) calculated a sequence of
two-dimensional, hydromagnetic (Grad-Shafranov)
equilibria describing the structure of the magnetically
confined mountain of material accreted at the magnetic
poles of the neutron star.
The mountain is confined by magnetic stresses near
the equator, where the field is compressed
\citep{mel01}.
These solutions are the first of their kind
to explicitly disallow
cross-field transport of material as the mountain evolves
from its initial to its final state
\citep[cf.][]{mou74},
as required in the ideal-magnetohydrodynamic (ideal-MHD)
limit.
PM04 found that $\mu$ is screened substantially
above a critical accreted mass $\Mc \sim 10^{-5}\Msun$,
well above previous estimates
of $\Mc\lesssim 10^{-10}\Msun$
\citep[]{ham83,bro98,lit01}.

PM04 calculated equilibria up to $\Ma\lesssim 10^{-4}\Msun$,
falling short of the mass required ($\sim 0.1\Msun$)
to spin up a neutron
star to millisecond periods \citep{bur99}. This is
supplied by a large class of mass donors like LMXBs
\citep{str03},
even given nonconservative mass transfer \citep{tau00}.
Grad-Shafranov calculations are stymied above
$\Ma \sim 10^{-4}\Msun$ by physical effects
(e.g. magnetic bubbles form above the stellar surface)
and numerical effects
(e.g. steep magnetic gradients hinder iterative convergence).
In this paper, we extend the $\mu(\Ma)$ relationship
to $\Ma \sim 10^{-3}\Msun$ by 
loading equilibria with $\Ma \sim 10^{-4}\Msun$ into ZEUS-3D,
a multi-purpose time-dependent, ideal-MHD code for astrophysical
fluid dynamics, and adding extra mass quasistatically through
the outer boundary of the simulation volume.

PM04 also left open the important question of the
stability of the hydromagnetic equilibria.
Distorted magnetic configurations, in which the polar flux
is buried beneath the accreted overburden and
compressed into a narrow belt at the magnetic equator,
are expected  prima facie to be unstable.
Indeed, the Grad-Shafranov analysis in PM04 hints at
the existence of an instability
by predicting (informally)
the existence of magnetic bubbles
as steady-state solutions.
In this paper, we systematically explore the stability of the equilibria
by evolving them in ZEUS-3D, subject to
linear and nonlinear perturbations.

The structure of the paper is as follows.
In section \ref{sec:stability2}, the necessary
theory is summarised and the solution method described.
The formalism of PM04 is again used here.
The numerical Grad-Shafranov solver is described in
appendix B of PM04 and
appendix C of \citet{mou74}.
In section \ref{sec:burialsetup},
we verify ZEUS-3D against a set of test cases relevant
to the problem of magnetic burial;
the implementation is described thoroughly in Appendix A.
In section \ref{sec:lateburial},
we explore the late stages of magnetic burial
($10^{-5} \lesssim\Ma/\Msun\lesssim 10^{-3}$)
by adding mass quasistatically to equilibria from PM04 in ZEUS-3D.
In section \ref{sec:stability},
we discuss the linear and nonlinear stability of the
equilibria in the regime $10^{-5}\lesssim\Ma/\Msun\lesssim 10^{-3}$.
The paper concludes, in section \ref{sec:stabilityconclusion},
with a discussion of the limitations of our analysis and suggestions
for future numerical work.

\section{Physics of magnetic burial}
\label{sec:stability2}
\subsection{Axisymmetric equilibria}
\begin{center}
\begin{figure*}
\centering
\begin{tabular}{cc}
 \begin{tabular}{c}
 (a) \\
\includegraphics[height=60mm]{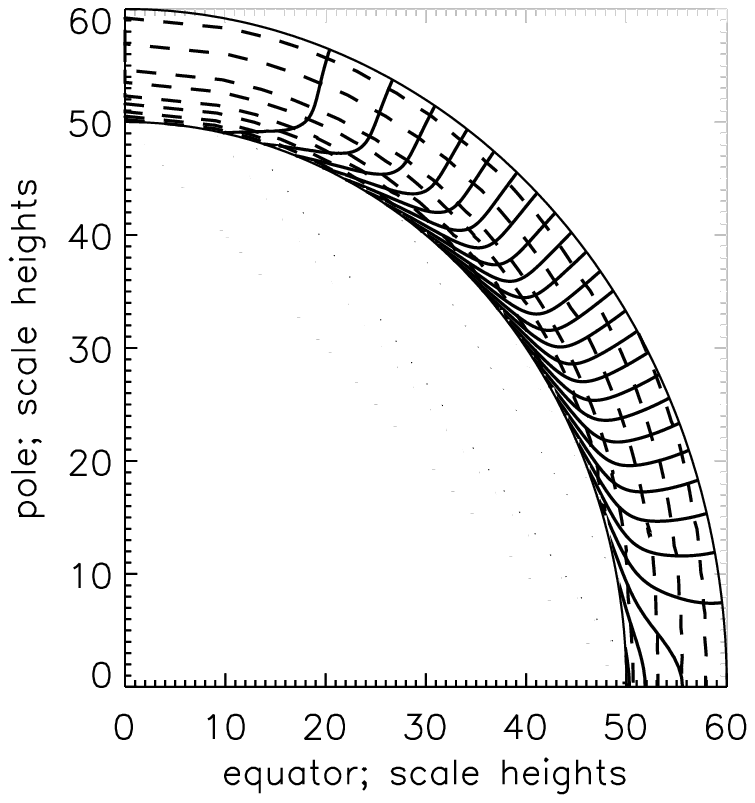} 
 \end{tabular}
&
 \begin{tabular}{c}
 (b) \\
\includegraphics[height=60mm]{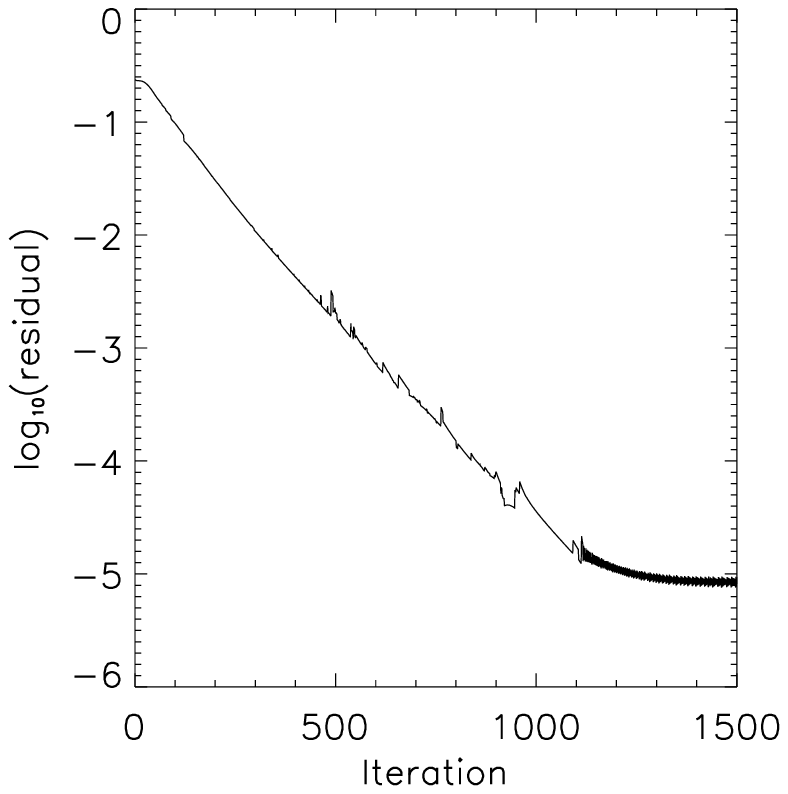}
 \end{tabular}
\\
 \begin{tabular}{c}
 (c) \\
\includegraphics[height=60mm]{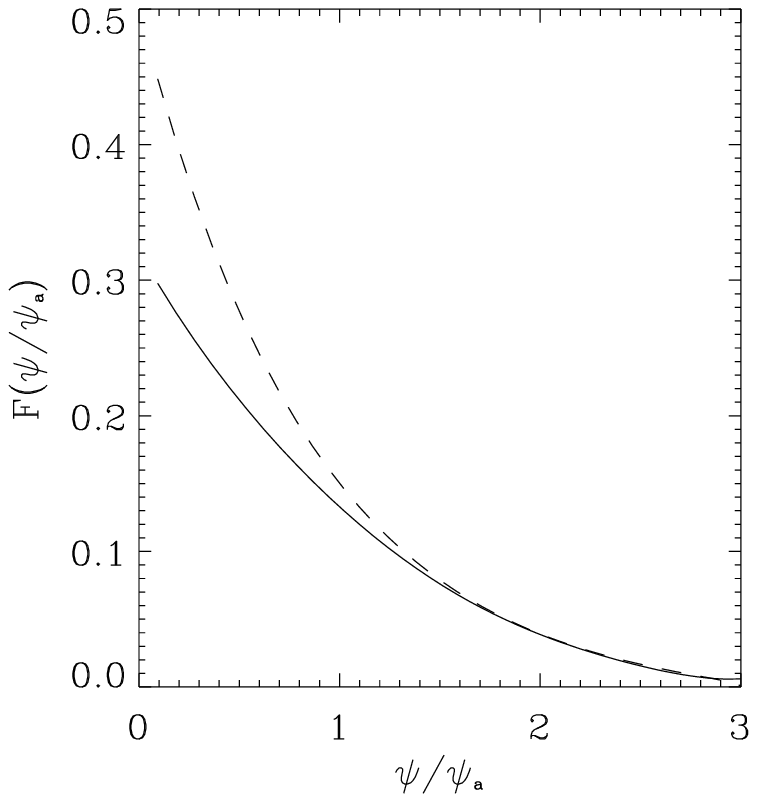}
 \end{tabular}
&
 \begin{tabular}{c}
 (d) \\
\includegraphics[height=60mm]{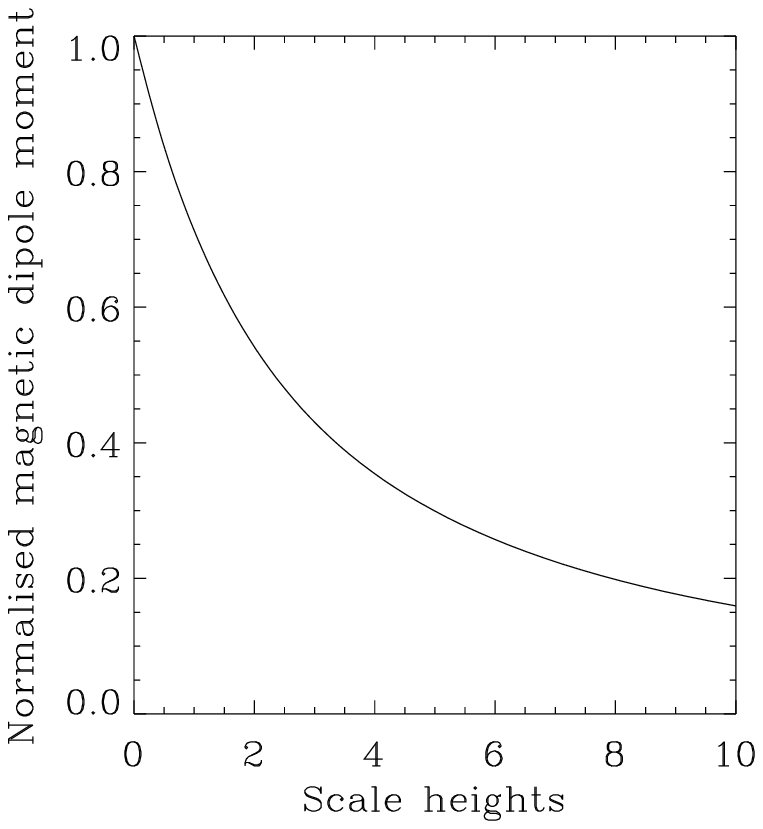}
 \end{tabular}
\\
\end{tabular}
\caption{\small
Hydromagnetic equilibrium for $a = 50$
(scaled from real stellar dimensions as discussed in section \ref{sec:scalea})
and $\Ma/\Mc = 0.16$,
with $G_x = G_y = 64$,
showing (a) magnetic field lines (\emph{solid})
and density contours (\emph{dashed})
and (b)
the corresponding mean residual versus iteration number for
an underrelaxation parameter
$\Theta = 0.99$.
(c) Final $F(\psi)$ (\emph{solid}) and initial $F(\psi)$ (\emph{dashed})
versus $\psi$.
(d) Magnetic dipole moment (normalized by its surface value)
as a function of altitude above the stellar surface (in units of $h_{0}$).
}
\label{fig:smalla}
\end{figure*}
\end{center}

During accretion onto a neutron star from a binary companion,
matter piles up on the polar cap, funnelled
by the magnetic tension of the polar magnetic flux tube.
Once $\Ma$ exceeds $\sim 10^{-5}\Msun$, the hydrostatic pressure at
the base of the accretion column overcomes the magnetic tension
and matter spreads over the stellar surface towards the equator,
dragging along frozen-in polar field lines (PM04).
The distorted magnetic field leads to screening currents
which act to decrease the magnetic dipole moment.
Figure \ref{fig:smalla}(a) illustrates the magnetic `tutu'
formed for $\Ma = 10^{-5}\Msun$,
cut off at ten density scale heights.
The polar mountain of accreted material
(dashed contours) and the pinched, flaring, equatorial
magnetic belt (solid contours) are plainly seen.

At first glance, one might expect
such equilibria, with their steep density and magnetic field 
gradients, to be unstable.
Interestingly, this expectation is largely false, as we show in section
\ref{sec:stability}.
We summarise the key equations and notation of magnetic burial here (PM04).

The steady-state ideal-MHD equations for an isothermal
atmosphere ($p = c_{\rm s}^2\rho$, i.e. adiabatic index $\gamma = 1$)
reduce to the force equation
$\nabla p + \rho\nabla\phi - {\mu_0}^{-1}(\nabla\times {\bf B})\times {\bf B} = 0$.
For an axisymmetric configuration in spherical polar coordinates $(r,\theta,\phi)$,
a flux function $\psi(r,\theta)$ generates the magnetic field ${\bf B}$ via
${\bf B} = \nabla\psi(r,\theta)/(r\sin\theta)\times\hat{\bf e}_\phi$.
The flux function satisfies the Grad-Shafranov equation
\begin{equation}
\Delta^2\psi = F^{\prime}(\psi)\exp[-(\phi-\phi_0)/c_{\rm s}^2],
\label{eqn:gradshafranov}
\end{equation}
where
$\rho$, $p$, $\phi$, $\phi_0$, $c_{\rm s}$, and
$\Delta^2$ denote the matter density, pressure,
gravitational potential, surface gravitational potential, sound speed,
and Grad-Shafranov operator respectively (PM04).
In the limit $h_{0} = c_{\rm s}^2 R_{*}^2/(G M_{*}) \ll R_{*}$,
$\phi = {GM_{*}r}/{R_{*}^2}$,
where $M_{*}$ is the mass of the neutron star and $R_{*}$
is its radius, mass-flux conservation in ideal MHD provides the integral
constraint
\begin{equation}
\begin{split}
F(\psi) =&
{c_s^{2}}\frac{dM}{d\psi} \\
	& \times\left\{2\pi\int_{C}ds\, r\sin\theta {|\nabla\psi|}^{-1}\exp[-(\phi-\phi_0)/c_s^2] \right\}^{-1} \, , \\
\label{fpsiparker}
\end{split}
\end{equation}
We prescribe the mass-flux distribution to be
$dM/d\psi = (M_{\rm a}/2\psi_{\rm a})\exp(-\psi/\psi_{\rm a})$,
where
$\psi_{\rm a}~=~\psi_{*}R_{*}/R_{\rm a}$ is the flux enclosed by
the inner edge of the accretion disk at a distance
$R_{\rm a}$ and $\psi_{*}$ is the hemispheric flux.
For the boundary conditions, we fix $\psi$
to be dipolar at $r = R_{*}$, assume north-south symmetry, fix the
$\psi = 0$ field line, and leave the field free at large $r$.

Equations (\ref{eqn:gradshafranov}) and (\ref{fpsiparker}) are solved numerically using an
iterative relaxation scheme \citep{mou74,pay04}.
The mean residual as a function of iteration number
is shown in figure \ref{fig:smalla}(b),
corresponding to the
Grad-Shafranov equilibrium for
$m = \Ma/\Mc = 0.16$
displayed in figure \ref{fig:smalla}(a).
The form of $F(\psi)$, found from (\ref{fpsiparker}), varies
from  $\Ma = 0$ to $\Ma = 0.16\Mc$
in the manner displayed in figure \ref{fig:smalla}(c).

\subsection{Critical accreted mass $\Mc$}
\label{sec:smallma}
According to naive estimates based on hydromagnetic
force balance between
$\mu_{0}^{-1}(\nabla\times\vv{B})\times\vv{B}$ and $\nabla p$
at the polar cap, the magnetic field (and hence $\mu$) is distorted appreciably
away from its initial configuration for
$\Ma \geq \Mc \sim 10^{-8}\Msun$ \citep{bro98,lit01}.
However, self-consistent solutions of
(\ref{eqn:gradshafranov}) and (\ref{fpsiparker}), in which mass does not migrate across flux
surfaces and the
back reaction from equatorial magnetic stresses is included
(PM04), predict a larger value of $\Mc$, given by
\begin{eqnarray}
\label{eq:mcparker}
\frac{\Mc}{\Msun} &=& \pi G M_{*} B_{*}^{2} R_{*}^{2}/(2 \mu_{0} c_{\rm s}^{4}\Msun)
\nonumber \\
    &=& 1.2 \times 10^{-4} \
\left(\frac{M_{*}}{1.4\Msun} \right) \
\left(\frac{R_{*}}{10^{4} {\rm m}} \right)^{2} \nonumber \\
& &\times\left(\frac{B_{*}}{10^{8} {\rm T}}\right)^{2} \
\left(\frac{c_{\rm s}}{10^{6} \, {\rm m \, s^{-1}}}\right)^{-4}.
\end{eqnarray}
In the regime $\Ma\lesssim\Mc$, the
Green function analysis in PM04 gives
\begin{equation} 
\label{eq:psismallma}
\psi(r,\theta) = \psi_{\rm d}(r,\theta)[1 - m b^2 (1 - e^{-x/h_{0}})]\, ,
\end{equation}
with
$m = \Ma/\Mc$ and $b = \psi_{*}/\psi_{\rm a}$.
The distorted field lines develop a large tangential component
as mass is added, such that
$|\vv{B}|$ increases substantially for
$\Ma \gtrsim \Mc/a$, with $a = R_*/h_{0}$.\footnote{
The extra factor of $a$ comes from differentiating $\psi$ with respect to $\theta$.}
The self-consistent density distribution associated with
(\ref{eq:psismallma}) is given by
\begin{equation}
\label{rhosmallMa}
\rho(r,\theta) = \rho_{0} \tilde{F}[\tilde{\psi}(r,\theta)] e^{-x/h_{0}}
\end{equation}
with
\begin{equation}
\label{eq:fpsiparker}
\tilde{F}(\tilde{\psi}) = \frac{b}{2\pi a^2}\exp({-\tilde{\psi}})(1-\tilde{\psi}/b)^{1/2}[J(\tilde{\psi})]^{-1},
\end{equation}
where we write
$\tilde{F} = Fh_{0}^{3}c_{\rm s}^{2}/\Ma$,
$\rho_0 = \Ma/h_0^3$, and
$\tilde{\psi} = \psi/\psi_{\rm a}$,
and the form factor satisfies $J(\tilde\psi)\approx 1$
to better than 10 per cent for all $\theta$ except
near the equator,
$89.5^{\circ}\leq\theta\leq 90^{\circ}$
(see figure A1 of PM04).
In the small-$\Ma$ limit,
$F$ is calculated by
assuming $\psi$ to be dipolar and evaluating the (small) correction
from (\ref{eqn:gradshafranov}) using Green functions.

\subsection{Screening the magnetic dipole}
The magnetic dipole moment is defined in terms of
the radial component of the magnetic field by
\begin{equation}
\label{eq:dipolemoment}
 \mu =
 \frac{3}{4} r^3
 \int^1_{-1} d(\cos\theta)\, \cos\theta
 B_r(r,\theta) \, ,
\end{equation}
assuming axisymmetry.
In the regime $\Ma\lesssim\Mc$ from (\ref{eq:psismallma}), we obtain
\begin{equation}
\label{eq:dipoleinitial}
\mu/\mu_{\rm i} = (1 - M_{\rm a}/M_{\rm c}) \, ,
\end{equation}
where
$\mu_{\rm i} = \psi_* R_*$ is the initial dipole moment.
Equation (\ref{eq:dipoleinitial}) depends only on
$m = \Ma/\Mc$, not $\Ma$ and $\Mc$ individually, just as in
(\ref{eq:psismallma}).
Figure \ref{fig:smalla}(d) shows how $\mu$ decreases as a function of
altitude due to the screening currents within the first few
density scale heights.
It is very important to note that $\psi$ deviates substantially from
its dipole form for $\Ma/\Mc \gtrsim b^{-2}$,
whereas $\mu$ deviates from $\mu_{\rm i}$ for $\Ma/\Mc\gtrsim 1$,
which occurs at a much later stage of accretion
(because $b\gg 1$ usually).

\section{Numerical simulations of magnetic burial}
\label{sec:burialsetup}
Distorted MHD equilibria like the one pictured in
figure \ref{fig:smalla}(a) are notoriously unstable.
In this section, we describe how the astrophysical
MHD code ZEUS-3D can be used to investigate
the stability of our equilibria.  We discuss the set up of
ZEUS-3D in section \ref{sec:usezeus3d} and
appendix \ref{appendix:zeus}, some verification experiments
in section \ref{sec:verifyzeus3d} and appendix
\ref{appendix:zeus}, and
the curvature rescaling required to render the magnetic
burial problem tractable in section
\ref{sec:scalea}.

\subsection{ZEUS-3D} 
\label{sec:usezeus3d}
ZEUS-3D is a multipurpose, time-dependent, ideal-MHD code for astrophysical
fluid dynamics which uses staggered-mesh finite differencing and
operator splitting in three dimensions
\citep{sto92a}.
In this paper, we restrict the dynamics to
two dimensions, disabling the third, but
employ a spherical polar grid, appropriate for 
an axisymmetric stellar atmosphere.
Previous numerical work focused on the magnetic poles
of the accreting star
\citep[][]{bro98,cum01};
here, by contrast, equatorial magnetic stresses are treated fully
by simulating a complete hemisphere.
The density and magnetic field strengths are read
into ZEUS-3D
from the output of our 
Grad-Shafranov code (PM04).
The time-step
$\Delta t_{\rm Z}$
is set by the Courant condition satisfied by the fastest
MHD wave modes.
Details regarding the parameters, initial conditions, boundary conditions,
verification tests, and coordinate choices in the runs are given in appendix
\ref{appendix:zeus}.

\subsection{Verification} 
\label{sec:verifyzeus3d}
Before implementing the burial problem in ZEUS-3D, we
ran a sequence of
simpler verification cases.

First, we reproduced the classical
results for the nonlinear evolution of the
Parker instability of a plane-parallel
field in rectangular geometry
\citep{mou74}.
We achieved agreement on the minimum stable wavelength
$\lambda_{\rm crit} = 4\pi h_0 [B^2/(4\pi p) +1]^{-1/2}$
and reproduced the final, nonlinearly evolved equilibrium
state to an accuracy of 5 per cent.

Second,
to test spherical coordinates in ZEUS-3D, we studied the evolution of
a spherical isothermal atmosphere containing zero magnetic field,
followed by an atmosphere containing a dipolar magnetic field;
both these states are analytic force-free equilibria
[($\nabla\times\vv{B})\times\vv{B} = 0$)].
ZEUS-3D confirmed that these are indeed equilibria; they
do not alter significantly even after several thousand
Alfv\'en times.
The condition at the outer boundary $r = r_{\rm m}$
 is chosen to suit the problem at hand.
The outflow condition leaves the magnetic field free
but allows some mass loss,
which we minimize by keeping $r_{\rm m}$ large,
to prevent the atmosphere from evaporating over long
time-scales.
The inflow condition artificially pins the magnetic field,
thereby introducing a
radial magnetic field at the outer boundary
which increases $\mu$ artificially when integrated at $r_{\rm m}$.
It is used at an intermediate stage in the bootstrap algorithm
(described in section \ref{sec:bootstrap}) when adding mass
in the $\Ma\gg\Mc$ regime.

\subsection{Curvature rescaling} 
\label{sec:scalea}
The characteristic radial ($h_0$) and latitudinal
($R_*$) length scales are very different in a neutron star,
creating numerical difficulties which must be handled by
rescaling the problem.
For a typical neutron star with
$c_s = 10^6$ m s$^{-1}$
\citep{bro98}, we find $h_{0} = 0.54$ m, $a = R_{*}/h_{0} = 1.9\times 10^{4}$,
and $\tau_{0} = h_0/c_{\rm s} = 5.4\times 10^{-7}$ s,
where $\tau_0$ is the sound crossing time over a hydrostatic scale height.

\begin{center}
\begin{figure}
\centering
\includegraphics[height=65mm]{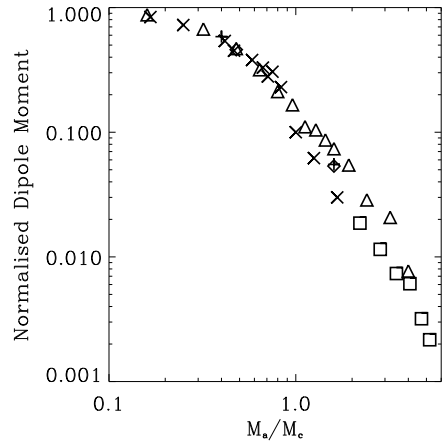}
\caption{\small
Magnetic dipole moment, $\mu$, as a function of accreted mass,
$m = \Ma/\Mc$, for $b = R_{\rm a}/R_* = 3$ and
$a = R_*/h_0 = 1.86\times 10^{4}$ (\emph{crosses}),
$a = 50$ (\emph{triangles}),
$a = 100$ (\emph{pluses}),
$a = 500$ (\emph{diamonds}),
all found with the Grad-Shafranov code,
and $a = 50$ with mass added through the outer
boundary in ZEUS-3D
(\emph{squares}).
}
\label{fig:a50ma}
\end{figure}
\end{center}

Two input parameters are varied in the simulations: $\Ma$ and $b$.
In the regime $\Ma\lesssim\Mc$, where (\ref{eq:psismallma}) holds,
$\psi$ and consequently $\mu$ depend only on
$m = \Ma/\Mc$ and not explicitly on $\Ma$, suggesting that we can
artificially reduce $a = R_{*}/h_{0}$ by
reducing $M_{*}$ and $R_{*}$, as long as we keep
$h_{0}\propto R_{*}^{2}/M_{*}$ fixed.
This has the advantage that
the minimum density increases as $a$ decreases
(because the atmosphere extends further),
decreasing the Alfv\'en speed and hence increasing the ZEUS-3D
time step $\Delta t_{\rm Z}$.
We set $a = 50$, a good compromise that ensures
$a \gg 1$ (as for a realistic star) while 
keeping the computational burden reasonable.
This choice corresponds to a model star with
$M_{*}^{\prime} = 10^{-5}\Msun$ and
$R_{*}^{\prime} = 27$ m.
Our numerical results confirm
that $\mu$ is independent of $a$
(see section \ref{sec:muversusma}),
as predicted analytically.  This is illustrated in 
figure \ref{fig:a50ma}, where $\mu(\Ma)$ is plotted for several
$a$ values;
even for $\Ma < \Mc$, the deviations are less than ten per cent.

Our ZEUS-3D grid reaches an altitude
$x_{\rm m} = r_{\rm m}-R_{*} = 10 h_0$
(cf. $> 10^{3}h_{0}$ in PM04).
We adopt a restricted domain for two reasons:
(i) to maximize grid resolution near the surface of the star,
where gradients are steepest; and
(ii) to stop the time-step, $\Delta t_{\rm Z}$,
from becoming so small that
run time and numerical dissipation become excessive
(as discussed in appendix \ref{appendix:zeus}).
According to
the Courant condition, $\Delta t_{\rm Z}$ scales as
$v_{\rm A, max}^{-1} \propto \rho_{\rm min} \propto e^{-{x}_{\rm m}/h_0}$.
Note that the pole-to-equator sound and Alfv\'en
crossing times
($a h_{0}/c_{\rm s}$ and $a h_{0}/v_{\rm A}$ respectively)
also decrease as $a$ decreases.

\section{Late stages of magnetic burial ($\Ma\gg\Mc$)} 
\label{sec:lateburial}
In this section, we investigate the evolution in time
of the highly distorted Grad-Shafranov equilibria that arise for
$\Ma\gg\Mc \sim 10^{-4}\Msun$.
We aim to answer a question on which the Grad-Shafranov analysis is silent:
what happens when significantly more than $10^{-4}\Msun$ is accreted
at a rate that is slow
compared with the Alfv\'{e}n time and instability oscillation period
(see section \ref{sec:stability})?
To do this,
we employ the bootstrap approach 
described in section
\ref{sec:bootstrap}.

\subsection{$\mu$ versus $\Ma$}
\label{sec:muversusma}
\begin{center}
\begin{figure*}
\begin{tabular}{cc}
 \begin{tabular}{c}
 (a) \\
\includegraphics[height=60mm]{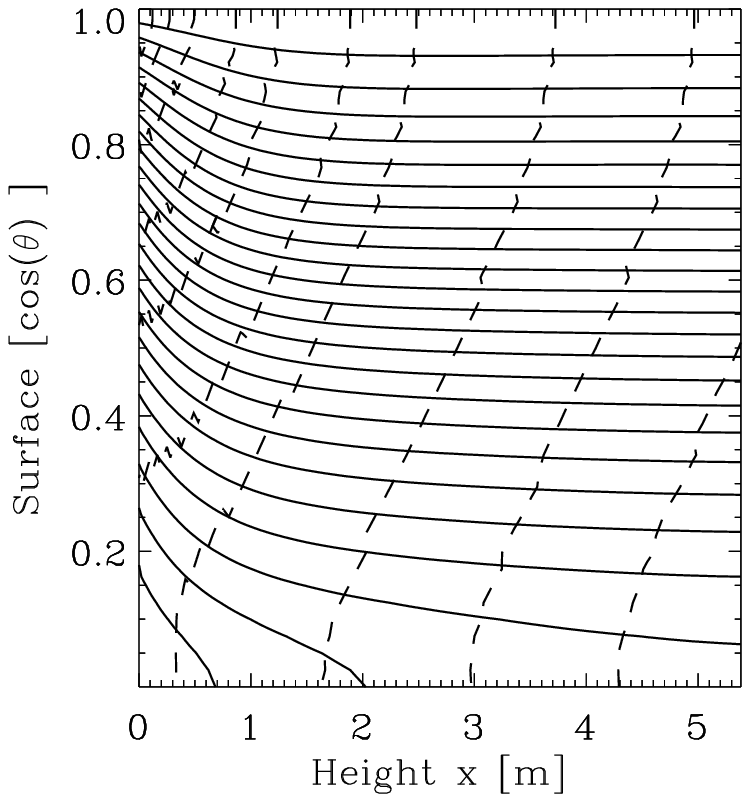}
 \end{tabular}
&
 \begin{tabular}{c}
 (b) \\
\includegraphics[height=60mm]{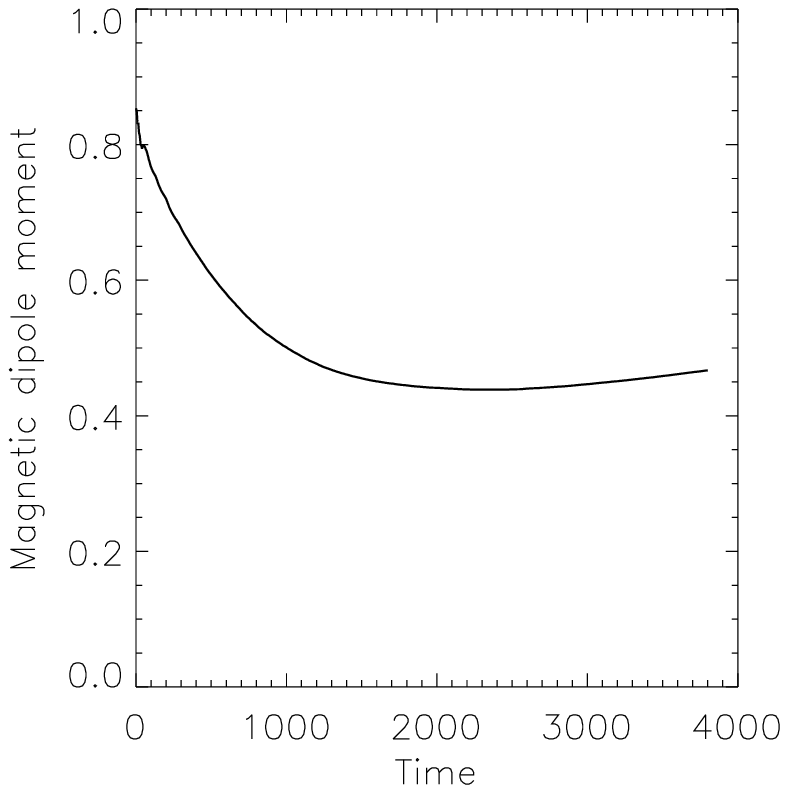}
 \end{tabular}
\\
 \begin{tabular}{c}
 (c) \\
\includegraphics[height=60mm]{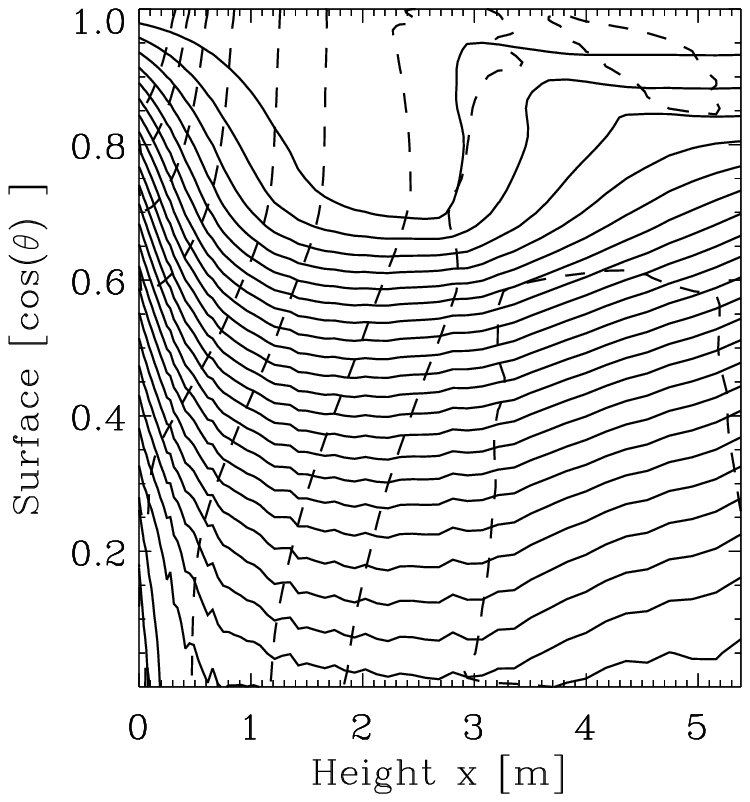}
 \end{tabular}
&
 \begin{tabular}{c}
 (d) \\
\includegraphics[height=60mm]{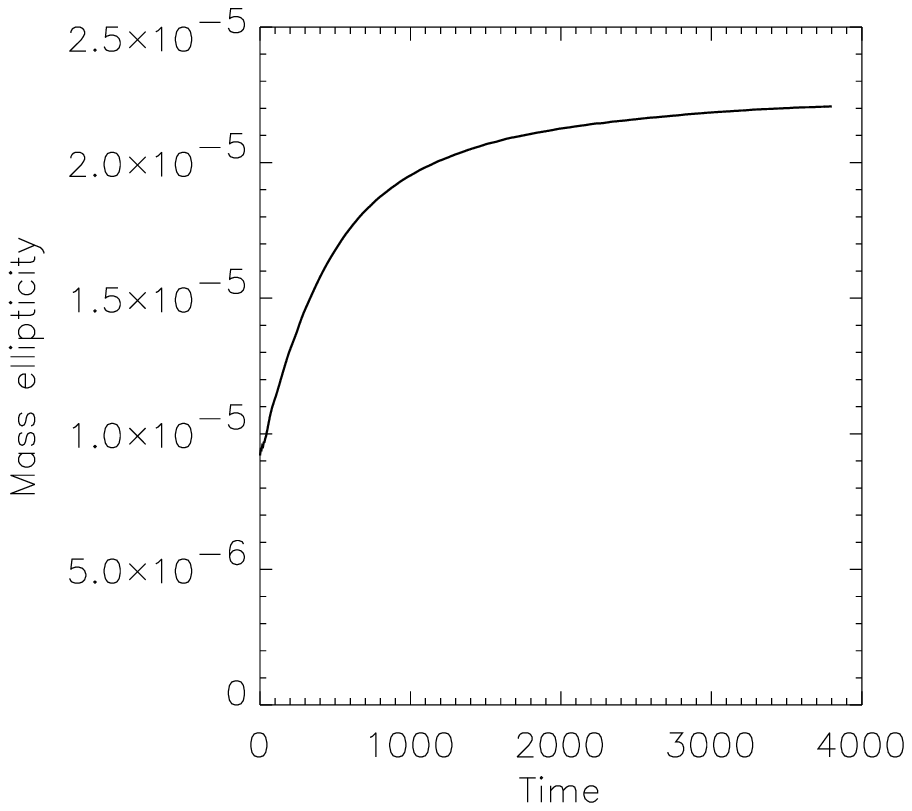}
 \end{tabular}
\\
 \begin{tabular}{c}
 (e) \\
\includegraphics[height=60mm]{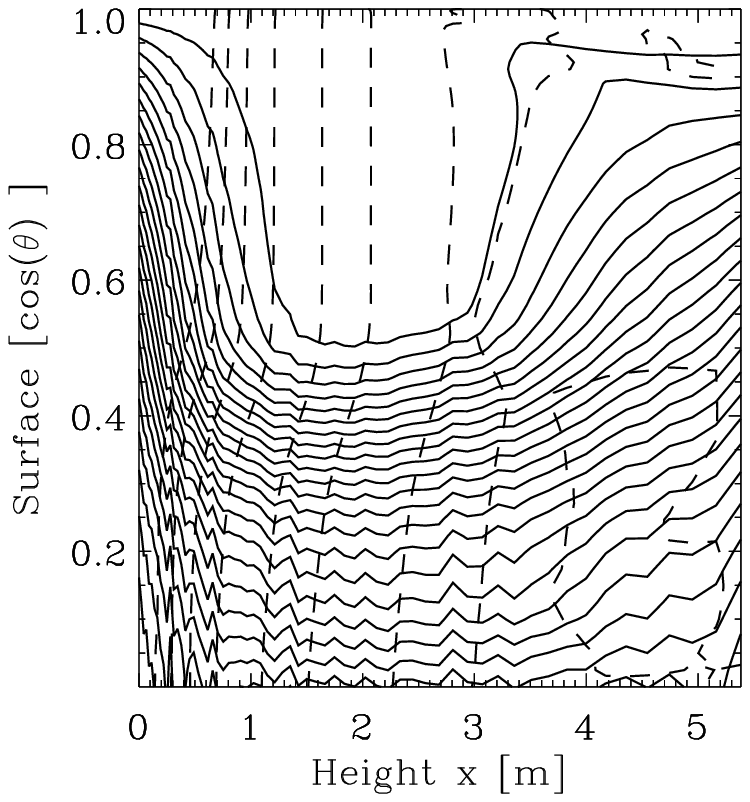}
 \end{tabular}
&
 \begin{tabular}{c}
 (f) \\
\includegraphics[height=60mm]{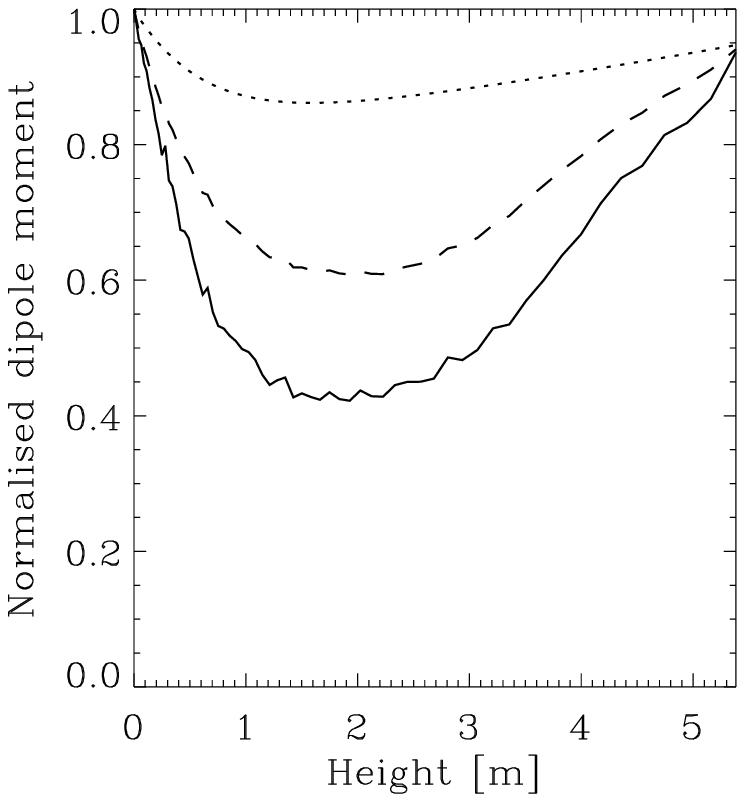}
 \end{tabular}
\end{tabular}
\caption{\small
Magnetic field evolution
with mass added to the outer boundary.
(a), (c), (e):  Magnetic field lines (\emph{solid}) and
density contours (\emph{dashed}) for t = 0, 500, 2000, corresponding to
$m = 0.16, 0.22, 0.4$ respectively.
(b) Dipole moment as a function of time, measured at 3 m above
the stellar surface.
(d) Quadrupole moment as a function of time.
(f) Dipole moment as a function of height for
t = 0 (\emph{dotted}), 500 (\emph{dashed}), and 2000 (\emph{solid}).
}
\label{fig:addmass}
\end{figure*}
\end{center}

Our numerical results confirm
that the magnetic dipole moment is essentially independent of $a = R_{*}/h_{0}$.
Figure \ref{fig:a50ma} displays $\mu(\Ma)$ for several
$a$ values, with $\Ma\ > \Mc$ achieved by adding mass
through the outer boundary of the simulation box (squares).
As predicted analytically, there is negligible dependence on $a$ for
$\Ma < \Mc$.  For $\Ma > \Mc$, deviations of less than
a factor of two occur.
The outer boundary condition
(radial $\vv{B}$)
artificially increases $\mu$, integrated at $x_{\rm m} = 10h_0$,
by about 10 per cent; this is considered further below
[see figure \ref{fig:addmass}(f), where
$\mu$ is plotted as a function of altitude
$x$ (dotted curve) for $m = 0.16$].
Hence the magnetic dipole moment plotted in figure
\ref{fig:a50ma} is an upper bound.

Errors are $\lesssim 1\%$ when compared against other
runs with ${x}_{\rm m} > 10h_0$.
For example, for $m = 0.16$ and $G_{x}=G_{y} = 128$,
the minimum dipole moment is
$0.8695$ at $\tilde{x}_{\rm m} = 10$ and $0.8555$ at
$\tilde{x}_{\rm m} = 20$.
(A $64\times 64$ grid gives similar results, e.g.
$\mu = 0.875$ at $\tilde{x}_{\rm m}= 10$.)

\subsection{Bootstrap accretion: $\Ma\gg\Mc$} 
\label{sec:bootstrap}
The data in figure \ref{fig:a50ma} extend to
$\Ma \approx 5\Mc$, yet
the Grad-Shafranov code only produces equilibria for
$\Ma \lesssim 10^{-4}\Msun$, and breaks down
when magnetic bubbles first appear at
$\Ma\gtrsim \Mc b^{-1}$ (section \ref{sec:bubble}).
This state of affairs is unsatisfactory because,
in many real accreting systems (e.g. LMXBs), $\Ma$ exceeds
substantially the critical mass for bubbles to form
\citep{taa86,van95}.
For such systems, the previous calculations are useful for the
early stages of accretion, at
$\tau_{\rm a} = \Ma/\dot{\Ma} \leq 10^{4} b^{-1}$ yr, 
but teach us little about the final stages of accretion
and hence the relic magnetic structure once accretion stops.

In this section, we perform a numerical experiment to address this issue.
We begin with a hydromagnetic equilibrium configuration for
$\rho$ and $\psi$ where $\Ma$ is just below the critical value for bubbles,
calculated using the previous Grad-Shafranov method
(PM04).
We load this configuration into ZEUS-3D.
We then add mass on the polar flux tube
$0 \leq \psi \leq \psi_{\rm a}$,
at a rate slower than all the hydromagnetic time-scales involved
in the process of coming to equilibrium (including the Alfv\'en
time-scale and the period of the global Parker oscillations
discussed in section \ref{sec:stability}), but at a rate faster than the true
accretion rate $\dot{\Ma}$ (so as to make the computation
tractable).
We then see what final magnetic structure we get after a realistic
amount of mass is accreted, e.g. $\Ma \sim 0.1 \Msun$ for LMXBs.

This approach does not track properly any processes
that operate on time-scales between $\tau_{\rm A}$
and $\tau_{\rm a}$,
such as Hall drift \citep{cum04} and
Ohmic diffusion \citep{rom90}.
However, it does model all time-dependent ideal-MHD 
effects in the context of polar
magnetic burial for the first time.
Note that this experiment would be completely
impractical without the Grad-Shafranov equilibria previously
computed for
$\Ma b \lesssim 10^{-4}\Msun$ (PM04),
because the separation of the hydromagnetic and accretion
time-scales is too great if one starts from
$\Ma = 0$, even when the accretion is accelerated
artificially.

\begin{center}
\begin{figure*}
\centering
\begin{tabular}{cc}
 \begin{tabular}{c}
 (a) \\
\includegraphics[height=60mm]{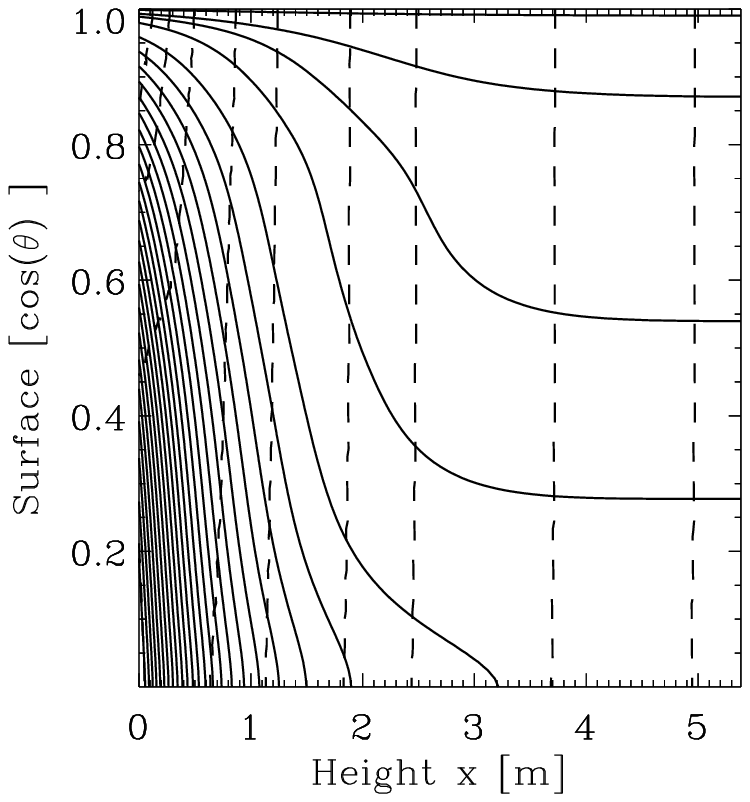}
 \end{tabular}
&
 \begin{tabular}{c}
 (b) \\
\includegraphics[height=60mm]{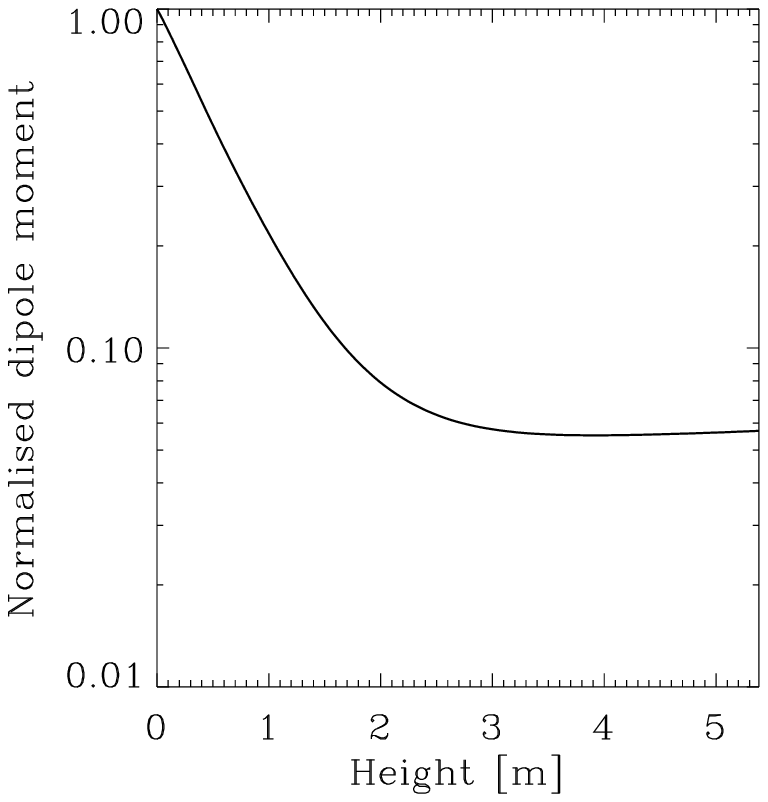}
 \end{tabular}
\\
 \begin{tabular}{c}
 (c) \\
\includegraphics[height=60mm]{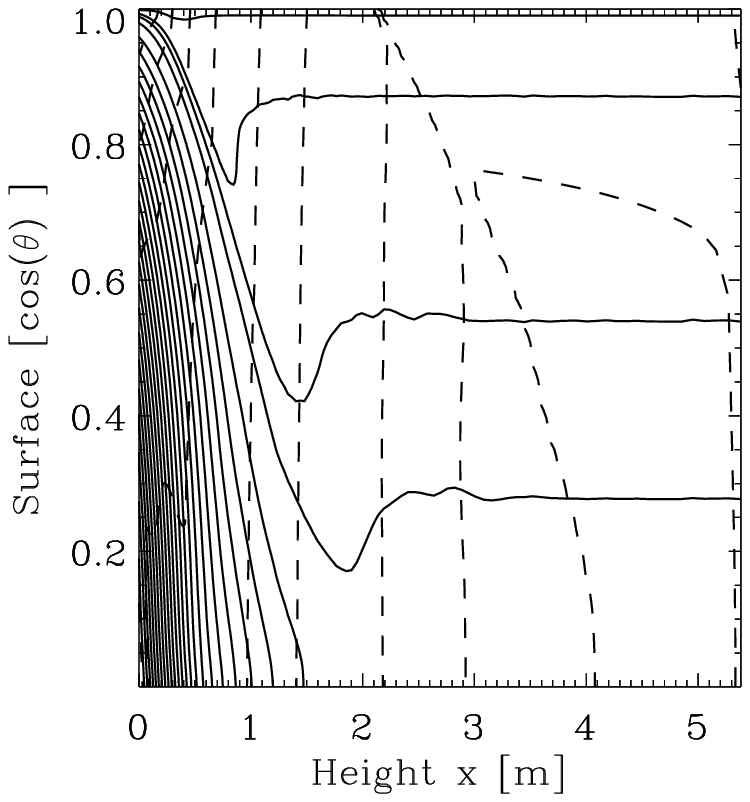} 
 \end{tabular}
&
 \begin{tabular}{c}
 (d) \\
\includegraphics[height=60mm]{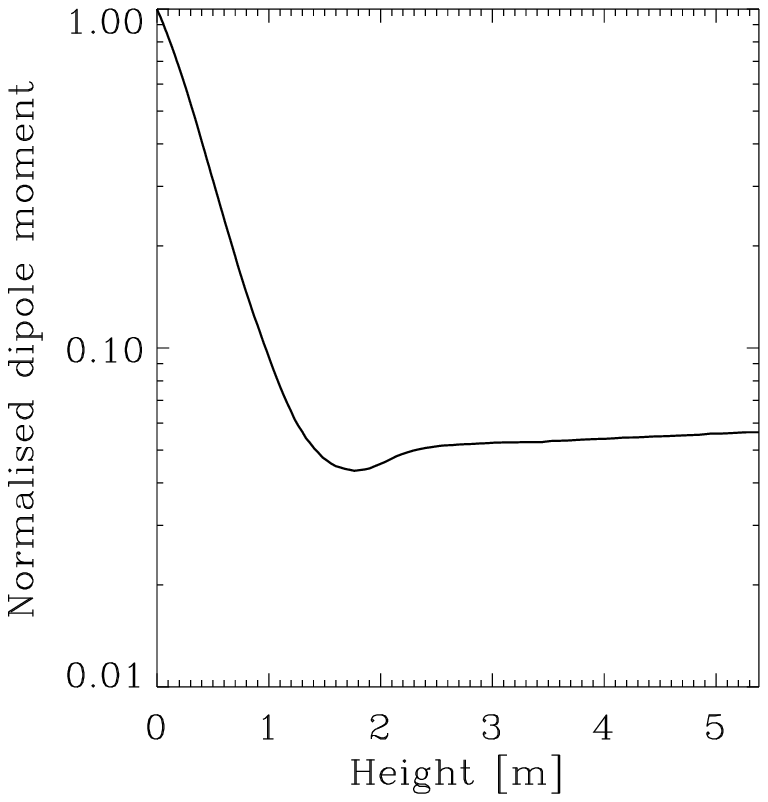}
 \end{tabular}
\\
 \begin{tabular}{c}
 (e) \\
\includegraphics[height=60mm]{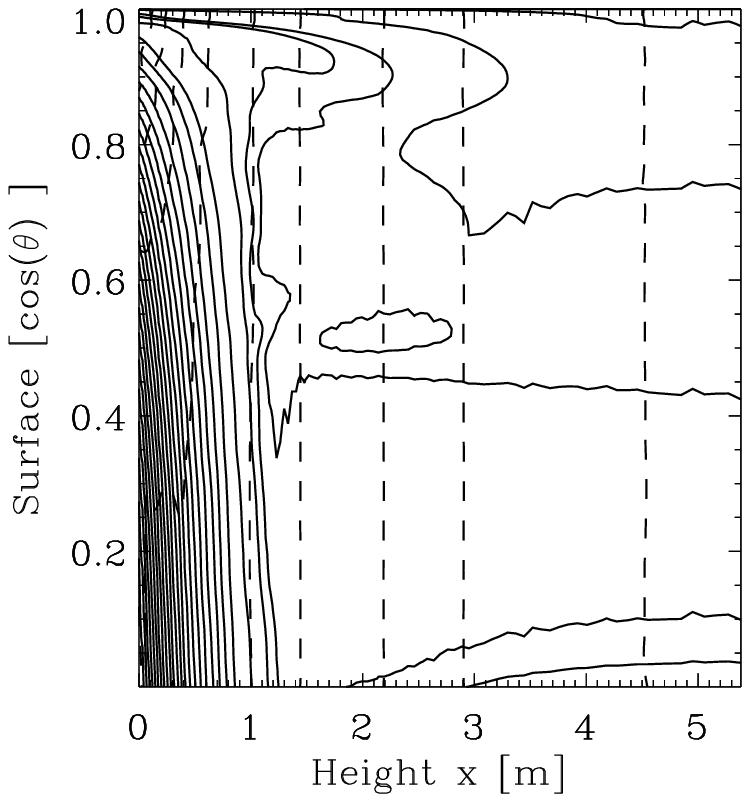}
 \end{tabular}
&
 \begin{tabular}{c}
 (f) \\
\includegraphics[height=60mm]{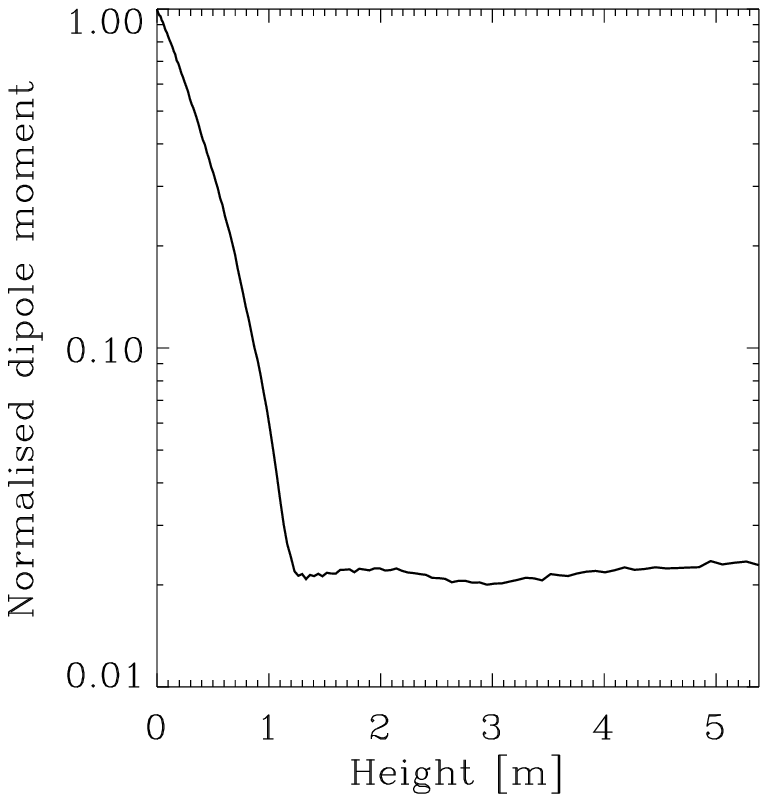}
 \end{tabular}
\end{tabular}
\caption{\small
Illustration of one step in the bootstrapping algorithm.
Magnetic field lines (\emph{solid}) and density contours (\emph{dashed})
(\emph{left}) and the magnetic dipole moment as a function of height
(\emph{right}) for
$m = 1.6$ at $t = 0$ (\emph{top}),
$m = 2.2$ at $t = 100$ (\emph{middle}), reached by adding mass
at the outer boundary, and $m = 4.1$ at
$t = 400$ (\emph{bottom}).
}
\label{fig:add10more}
\end{figure*}
\end{center}

\begin{center}
\begin{figure*}
\includegraphics[height=65mm]{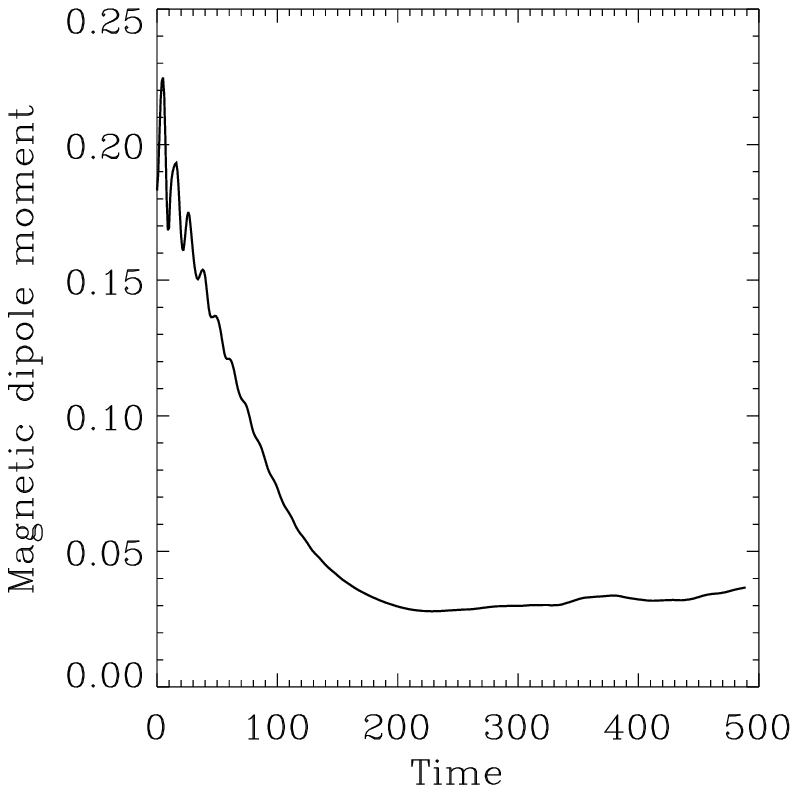}
\includegraphics[height=65mm]{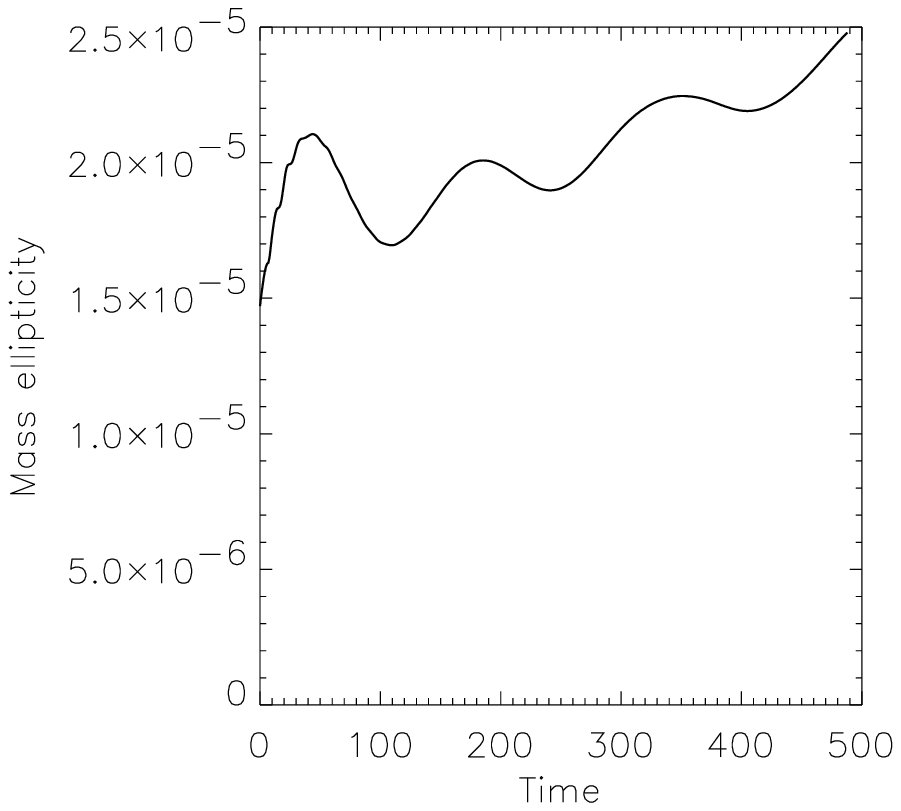}
\caption{\small
Evolution of the magnetic dipole moment, as measured 1.5 m above the
stellar surface (\emph{left}), and the mass ellipticity (\emph{right}),
when mass is added to a $m = 0.16$ equilibrium.
}
\label{fig:addmass10}
\end{figure*}
\end{center}

Figure \ref{fig:add10more} illustrates one bootstrapping cycle; it
is also a good illustration of the dynamics of magnetic burial
for $\Ma\gg\Mc$.
Starting from a steady state equilibrium with
$m = 1.6$,
we add mass to the outer boundary, with inflow boundary conditions.
The magnetic field lines are tied to the outer boundary
by this condition in ZEUS-3D,
preventing $\mu$ from decreasing there and
raising $\mu$ artificially closer to the star.
To overcome this, we change
the outer boundary condition
from inflow ($\vv{B}$ pinned) to outflow ($\vv{B}$ free)
after adding progressively more mass
through the polar flux tube, with
$\dot{M}_{\rm a}(\theta) = \dot{M}_{\rm a, max}e^{-b\sin^2\theta}$
and
$\dot{M}_{\rm a, max} \approx 2\times 10^{-6}\Msun/\tau_0$
when scaled up from the model ($a=50$) to a realistic neutron star.
Mass falls along field lines
[figure \ref{fig:add10more}(a)]
towards the stellar surface and
flattens the equatorial `tutu', until the magnetic tension matches the
hydrostatic pressure.
Matter subsequently flows equatorward, dragging magnetic
field lines with it [figure \ref{fig:add10more}(c)].
Superposed on this process are sound and Alfv\'en oscillations, which are
discussed in section \ref{sec:stability}; they cause wiggles
in the field lines, visible above the spreading matter
(2 -- 3 m above the surface).
The oscillations show up clearly in the time evolution
of the magnetic dipole moment,
and mass quadrupole moment, plotted in figure \ref{fig:addmass10}.
[Further discussion of the mass distribution appears in \citet{mel05}.]
Finally, the mass inflow is stopped and the outer boundary
condition is switched to outflow, allowing the magnetic
field lines to relax [figure \ref{fig:add10more}(e)].
The configuration in figure \ref{fig:add10more}(e) becomes
the initial state for further mass to be added.

\begin{center}
\begin{figure}
\centering
\includegraphics[height=65mm]{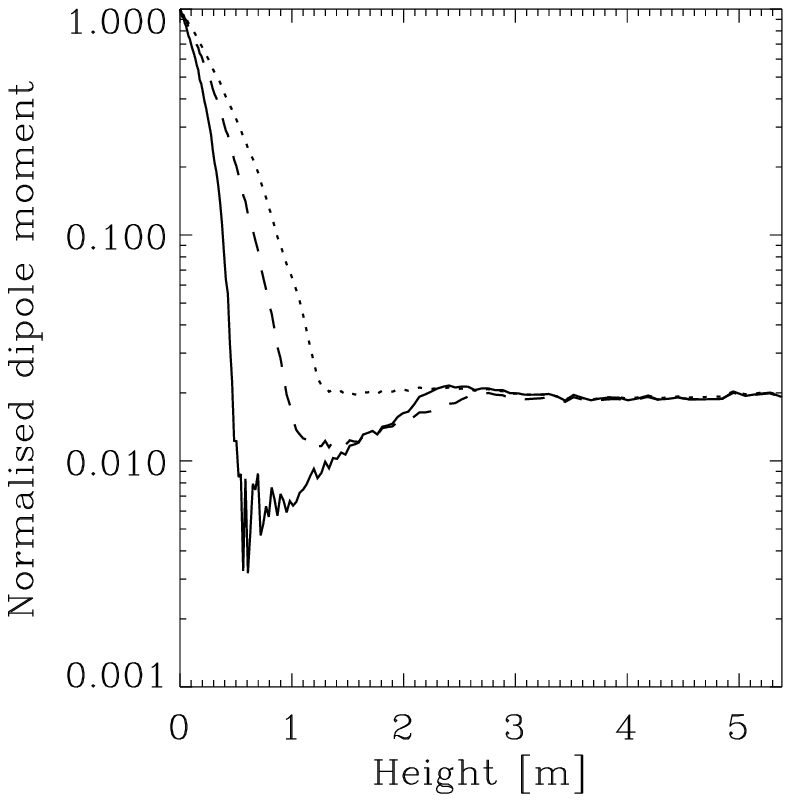}
\caption{\small
Dipole moment versus altitude when mass
is added to the configuration in figure
\ref{fig:add10more}(c), at
$t = 0$ (\emph{dotted}), $t = 100$ (\emph{dashed}), and
$t = 500$ (\emph{solid}),
corresponding to
$m = 2.23, 2.86$, and $4.76$ respectively.
}
\label{fig:add10more11}
\end{figure}
\end{center}

The dipole moment attained after adding a total of
$3\times 10^{-4}\Msun$
on a $128 \times 128$ grid
is shown in
figure \ref{fig:add10more11}
as a function of altitude.  We find that
a minimum dipole moment of
$\approx 3\times 10^{-3}$ times the surface value
is obtained for $m = 4.76$ at an altitude $\sim h_0$.
Smaller dipole moments are expected
for larger $\Ma$,
but numerical constraints prevented us from improving
the grid resolution nearer the equator sufficiently to see this.

\begin{center}
\begin{figure*}
\includegraphics[height=65mm]{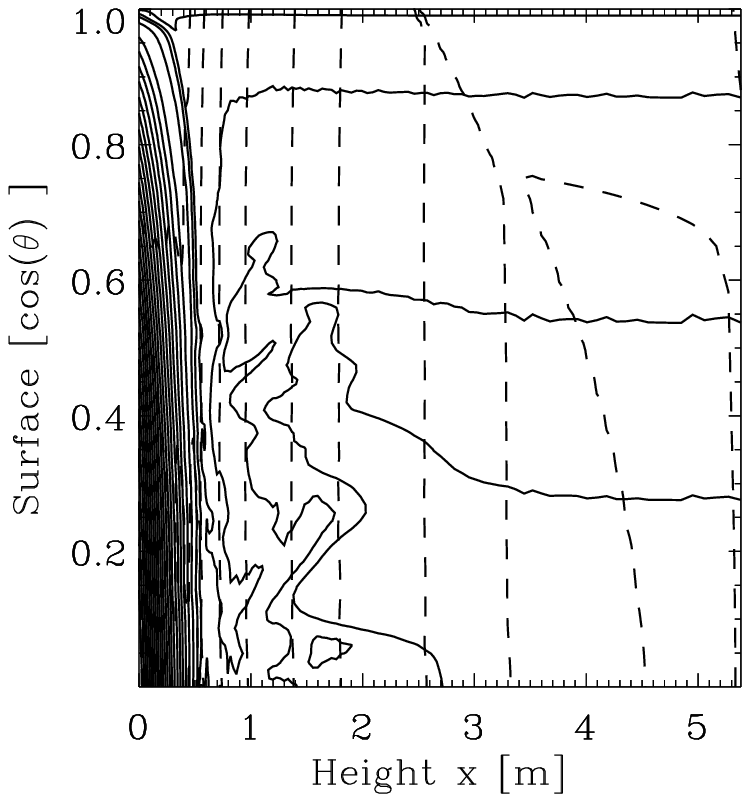}
\includegraphics[height=65mm]{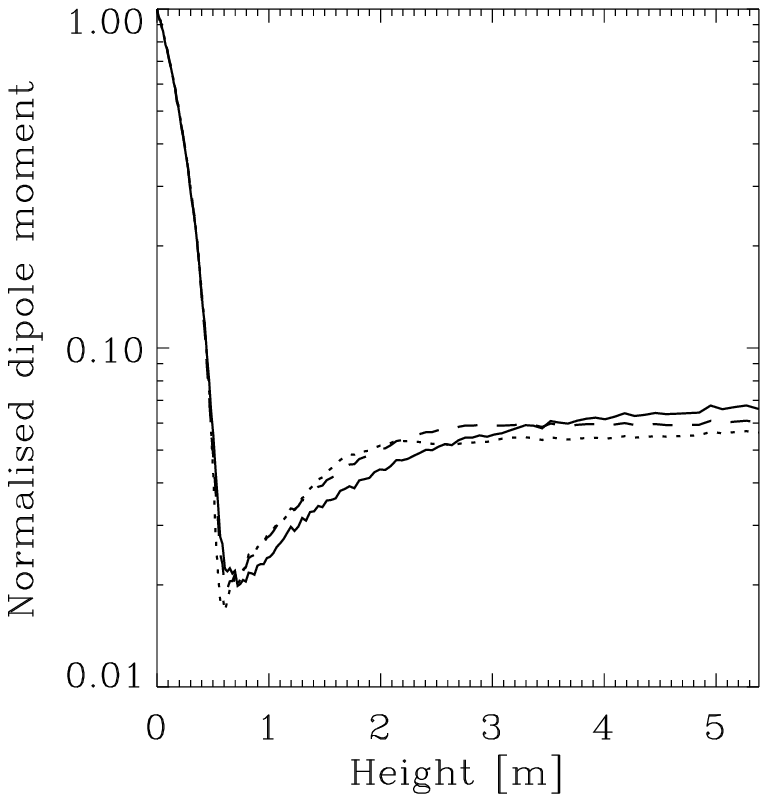}
\caption{\small
(\emph{Left}.)
Magnetic field lines (\emph{solid}) and density contours
(\emph{dashed}) when mass is added to a
$m = 0.16$ equilibrium, reaching $m = 4.1$ at $t = 400$.
(\emph{Right}.)
Magnetic dipole moment as a function of altitude
when the $m = 4.1$ configuration at $t = 0$ (\emph{dotted})
relaxes in the presence of an outflow outer boundary condition,
at $t =500$ (\emph{dashed}) and
$t = 900$ (\emph{solid}).
}
\label{fig:addmass400}
\end{figure*}
\end{center}

One might be tempted to ask whether the intermediate step of relaxing
the field at $r_{\rm m}$ is really necessary.
It is, and figures \ref{fig:addmass} and \ref{fig:add10more}
illustrate why.
If too much mass is added, it is found numerically that the
field lines break off the underlying magnetic dipole and,
when allowed to relax, remain pinned to the outer boundary.
Figure \ref{fig:addmass10} shows $\mu$ as a
function of time,
measured $\sim 1.5$ m above the
stellar surface where the spreading mostly occurs.
$\mu$ reaches a minimum after $\sim 100\tau_0$,
even when more mass is added, because of 
line-tying at $r_{\rm m}$.
Adding more mass is ineffective
because closed magnetic loops form.
They are apparent
in figure \ref{fig:addmass400}, which shows the
configuration at $t = 400$
(starting from the $m = 1.6$ equilibrium).
This configuration cannot be used as the initial state for
the next bootstrapping iteration because
the magnetic field lines remain fixed at the outer boundary.

To validate the bootstrapping method, we take an equilibrium that is free
of magnetic bubbles (i.e. $m \ll 1$) and add mass up to a value
of $\Ma$
for which the Grad-Shafranov equilibrium is already known.
Generally speaking, we reproduce the Grad-Shafranov value of
$\mu$ to an accuracy of 5 per cent.

Figure \ref{fig:addmass} shows the results of such a test,
beginning with $m = 0.16$
and adding mass at a rate
$\dot{M}_{\rm a}(\theta) = \dot{M}_{\rm a, max}e^{-b\sin^2\theta}$,
with $\dot{M}_{\rm a, max} \approx 7\times 10^{-8}\Msun/\tau_0$.
We point out several important features.
(i) The initial $\mu$ in figure \ref{fig:addmass}(f)
rises towards the outer boundary (see section \ref{sec:verifyzeus3d}).
(ii) $\mu$ is unchanged at the outer boundary,
because the magnetic field lines are tied there
by the flux-freezing condition, to accomodate the inflowing mass.
The magnetic field lines bend
towards the equator, and $\mu$ decreases
closer to the stellar surface.
(iii) Closer to the surface, $\mu$ reaches a minimum
within 5 per cent of that obtained from
the Grad-Shafranov equilibrium.
(iv) The kinks apparent in the magnetic field lines are a result
of numerical dissipation on the grid scale.

\subsection{Bubbles} 
\label{sec:bubble}
When the Grad-Shafranov equation is solved
with $\psi$ free (Neumann) at the outer boundary,
closed {bubbles} of magnetic field, disconnected from the inner
(Dirichlet) boundary, can arise (PM04)
for $\Ma \geq 1.6 \Mc b^{-1}$.
Physically, they are generated when the toroidal screening current
exceeds a threshold.
However, the field lines are deformed continuously from a
simply connected initial state 
(e.g. a dipole, or something close to it is the best guess
for a typical neutron star) in ideal MHD.
Therefore, the bubbles are not realizable in an accreting neutron star
even though they are admissible mathematical solutions to the
steady-state boundary-value problem. 
Instead,
the bubbles point to a loss of equilibrium,
analogous
to that which occurs during
eruptive solar phenomena \citep{kli89},
where no simply connected hydromagnetic
equilibrium exists.
In the language of the Grad-Shafranov analysis in section \ref{sec:stability2},
the source term
$\propto F^{\prime}(\psi)$ in (\ref{eqn:gradshafranov})
increases with $M_{\rm a}b$, boosting $\Delta^2\psi$ and
creating flux surfaces with
$\psi < 0$ or $\psi > \psi_{*}$, which cannot
connect to the star and
either form closed loops or are anchored
at infinity (here, the accretion disk).

\begin{figure*}
\begin{tabular}{cc}
 \begin{tabular}{c}
 (a) \\
\includegraphics[height=60mm]{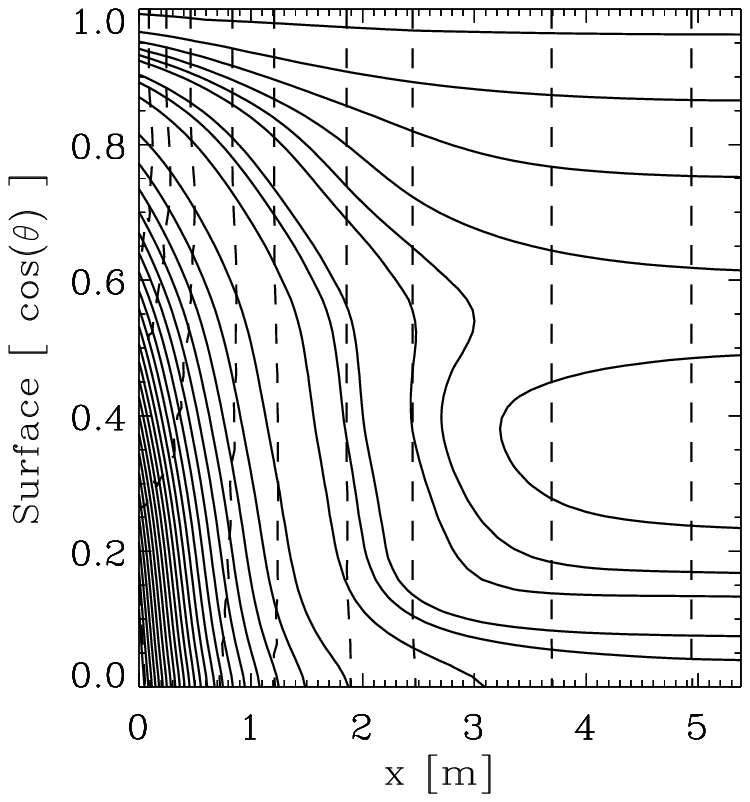}
 \end{tabular}
&
 \begin{tabular}{c}
 (b) \\
\includegraphics[height=60mm]{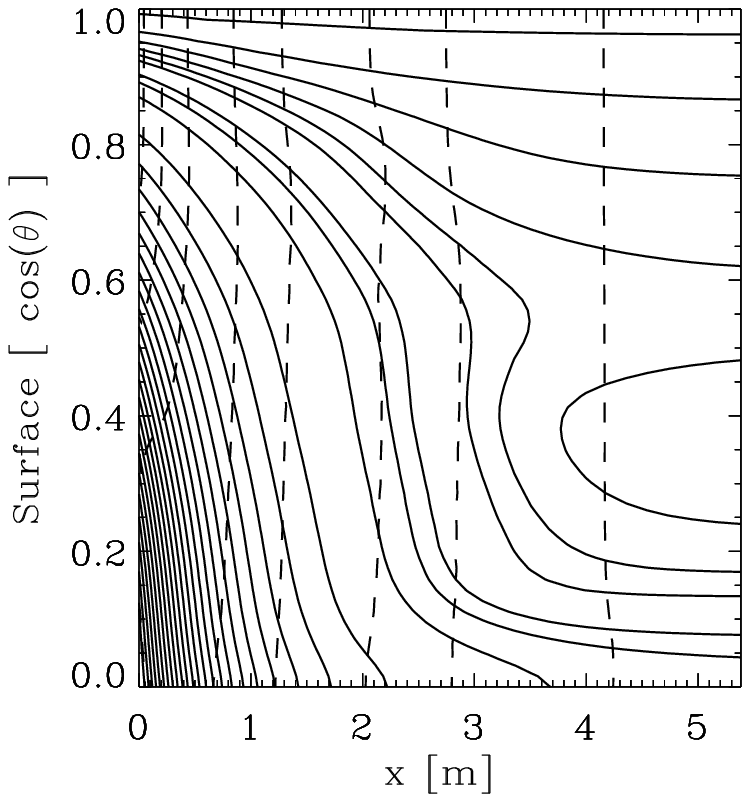}
 \end{tabular}
\\
 \begin{tabular}{c}
 (c) \\
\includegraphics[height=60mm]{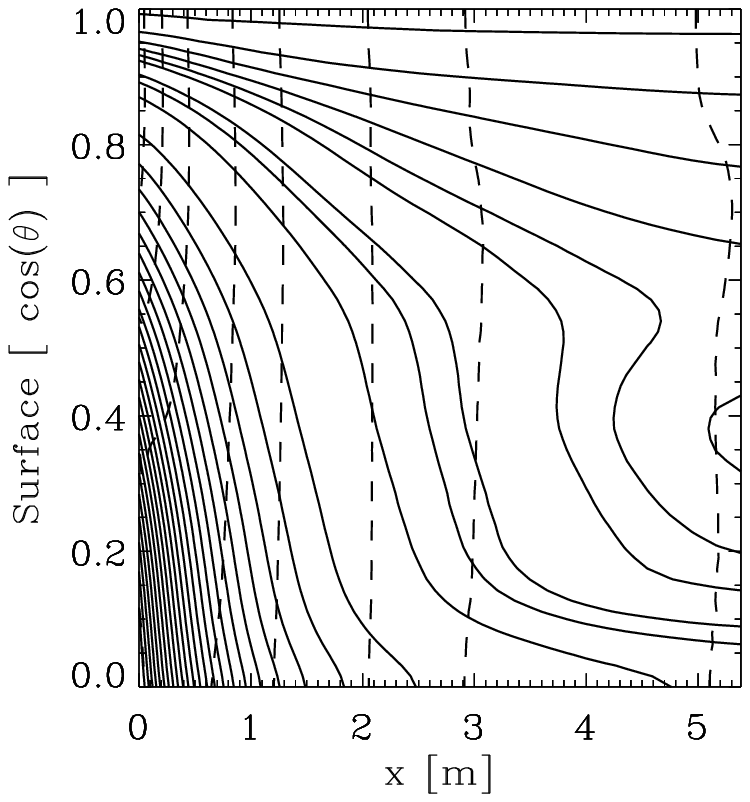}
 \end{tabular}
&
 \begin{tabular}{c}
 (d) \\
\includegraphics[height=60mm]{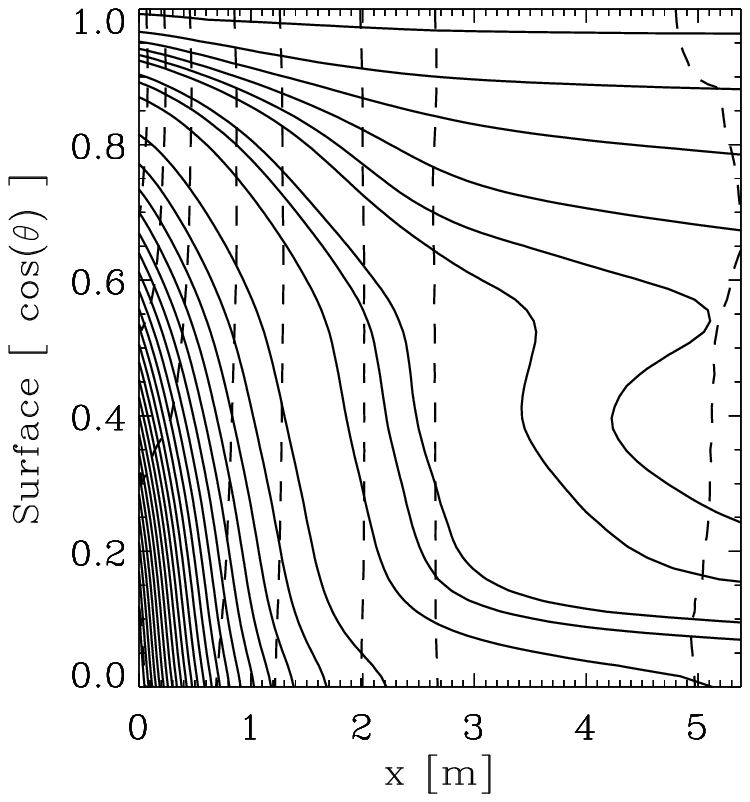}
 \end{tabular}
\\
\end{tabular}
\centering
\caption{\small
Evolution of a bubble
[found initially in the region $3\leq x \leq 5$ m, $0.2\leq \cos(\theta)\leq 0.5$ in (a)]
with $m = 1.6$ initially, after the density is increased uniformly
to $m = 3.2$ at $t = 0$. Snapshots are
plotted at (a) $t = 0$, (b) $t=3$, (c) $t=6$, (d) $t=9$.
}
\label{fig:bubble}
\end{figure*}

How does a bubble evolve in ZEUS-3D?
We import a Grad-Shafranov equilibrium
for $\Ma = 1.6\Mc$ into ZEUS-3D, retaining
the self-consistent density $\rho = F[\psi(r,\theta)] e^{-x/h_0}$
in the region
$0 \leq\psi\leq\psi_{*}$ but replacing it with
an isothermal atmosphere
inside the bubble
($\psi <0$ and $\psi > \psi_*$),
where strict flux-freezing would imply
$\rho = 0$ (matter cannot enter without crossing flux surfaces).
Figure \ref{fig:bubble} shows part of the evolution for $m = 1.6$.
The bubble rises to $r_{\rm m}$,
during the first Alfv\'en oscillation of the magnetosphere, at
the Alfv\'en speed.
During the bubble's rise,
0.5 per cent of the accreted mass
($4.4\times 10^{-7}\Msun$ when converted to realistic
neutron star parameters; see appendix \ref{sec:appendix:convert})
is ejected through the outer boundary in one
Alfv\'en time.
This compares with thermal evaporation of
less than 0.001 per cent in the same time,
which occurs naturally given
outflow boundary conditions (see appendix \ref{appendix:zeus}).
For $m > 1.6$, the bubble oscillates about an equilibrium point
rather than rising buoyantly.

\subsection{Uniform density increase} 
\label{sec:densityincrease}
\begin{center}
\begin{figure*}
\begin{tabular}{cc}
\includegraphics[height=60mm]{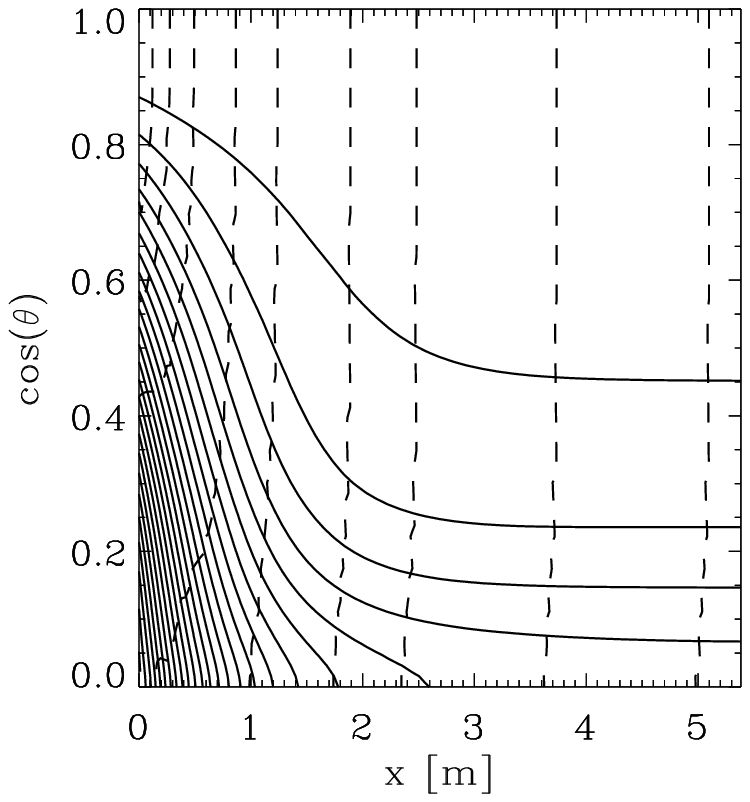}
&
\includegraphics[height=60mm]{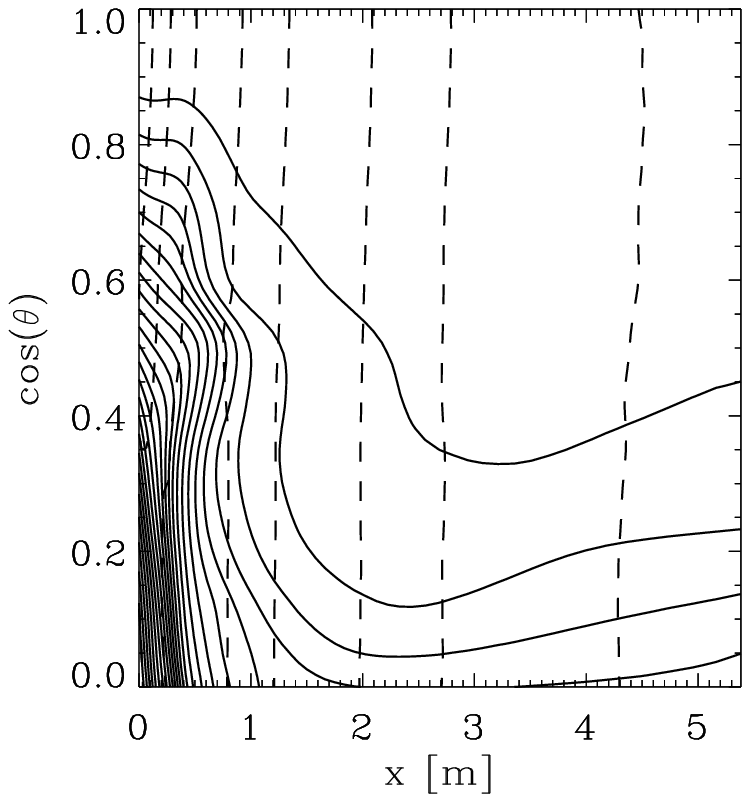}
\\
\includegraphics[height=60mm]{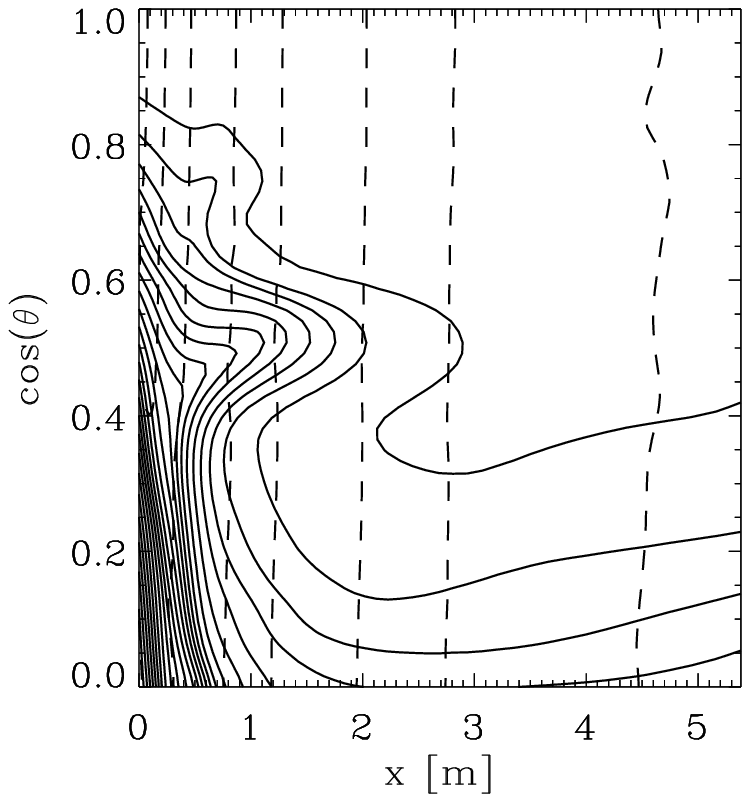}
&
\includegraphics[height=60mm]{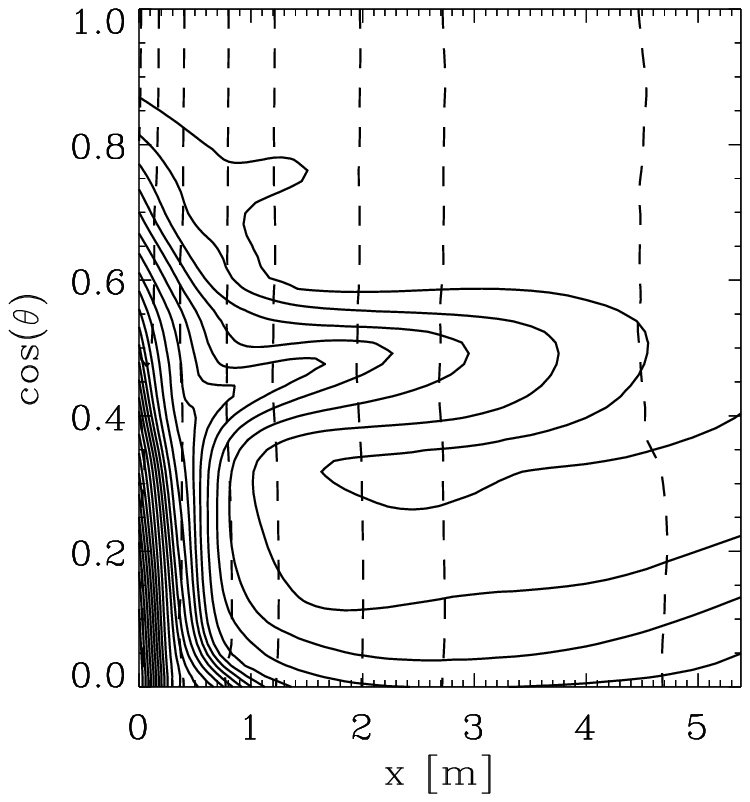}
\\
\includegraphics[height=60mm]{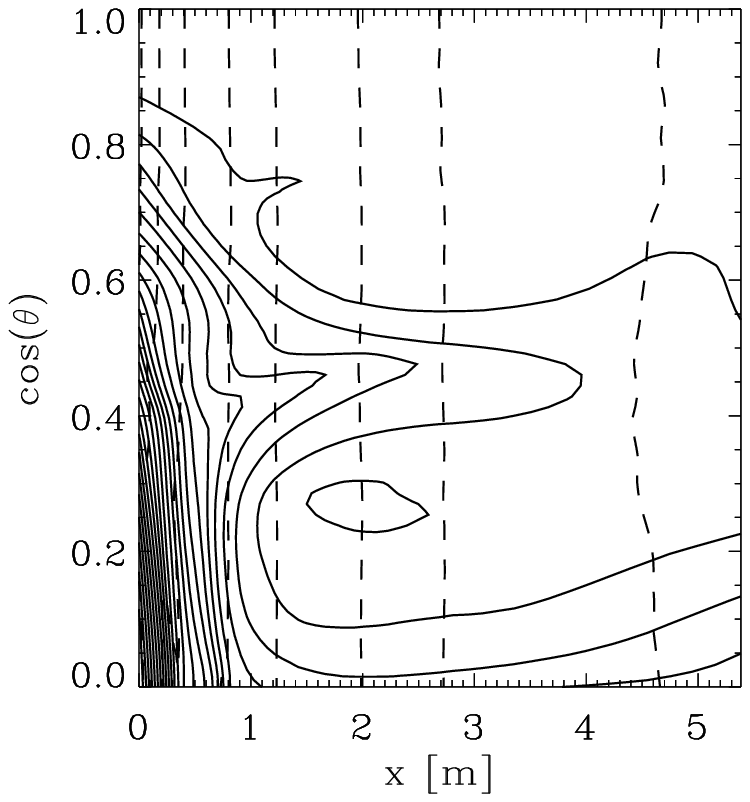}
&
\includegraphics[height=60mm]{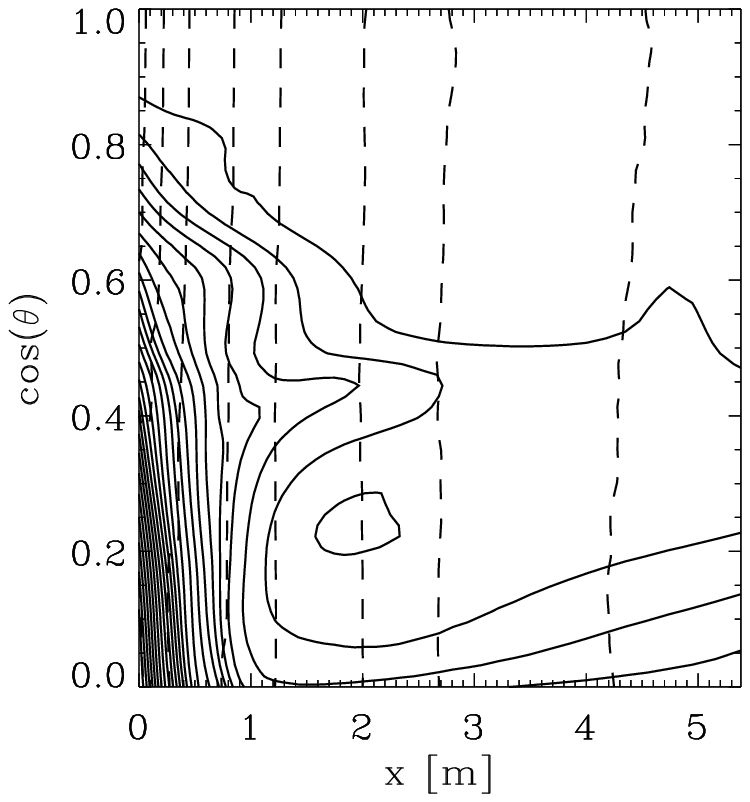}
\end{tabular}
\caption{\small
Evolution of the Parker instability for
$\Ma = 1.12 \Mc$ with the density increased uniformly across the grid
by a factor of 5.  Snapshots of the magnetic field lines are shown
at 
$t = 0, 120, 160, 200, 240$, and 280 Alfv\'en times
(top left to bottom right).
}
\label{fig:parkerevol}
\end{figure*}
\end{center}

Another route to computing equilibria with
$\Ma > 10^{-4}\Msun$ is to start from a bubble-free
Grad-Shafranov equilibrium with $\Ma < \Mc$ and
increase the density
uniformly across the grid by some factor (usually $\gg 1$),
while leaving the magnetic field unchanged.
If we start from the initial dipole ($\Ma = 0$),
this kind of numerical experiment is badly controlled;
it leads to excessive numerical dissipation and mass loss through $r = r_{\rm m}$.
However, if we start from $\Ma \sim 10^{-5}\Msun$, the readjustment is gentler.
During the early stages of the readjustment,
some classic instances of the Parker
instability are observed.  For example,
figure \ref{fig:parkerevol} shows how the initial state for
$m = 1.12$ evolves after the density is increased five-fold uniformly 
to $m = 5.6$.
The magnetic field component $B_{\theta}$ is compressed in the $r$ direction
on an Alfv\'en time-scale
--- ripe conditions for
the Parker instability.
The blistering becomes clear after 160 Alfv\'en times (third frame).
After 280 Alfv\'en times, the material settles down to a new equilibrium.
Importantly, negligible mass and magnetic flux
(less than 1 \% of the total)
is lost through this blistering.

\begin{center}
\begin{figure}
\centering
\includegraphics[height=65mm]{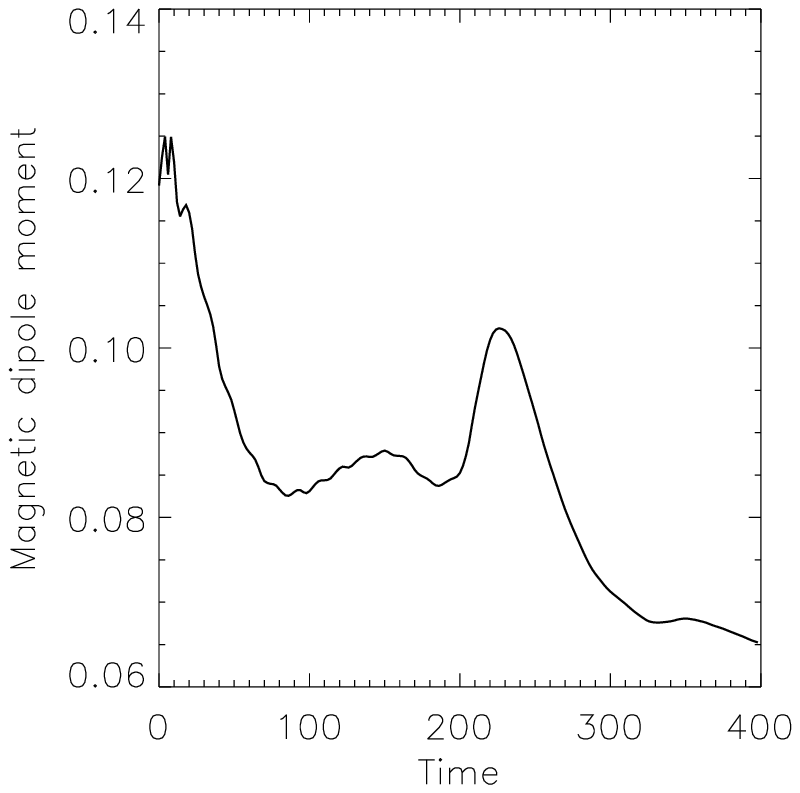} 
\caption{\small
Magnetic dipole moment as a function of time during the blistering
of the Parker instability displayed in
figure \ref{fig:parkerevol}.
}
\label{fig:parkerdipole}
\end{figure}
\end{center}

The magnetic dipole moment
$\mu$, plotted in figure \ref{fig:parkerdipole},
settles down to $\approx 0.06\mu_{\rm i}$.
This is larger than we expect for $m=5.6$, given that
we found
$\mu(m=4.7) < 0.01$ using the bootstrapping method.
Two factors explain this:
(i) the sudden density increase does not allow the magnetic field
and matter to gradually find their equilibria, and (ii)
flux-freezing is violated by increasing the density as above.
This method can be used in conjunction with bootstrapping with
less severe density increases (factor $\ll 5$).

\subsection{Size of the polar cap} 
In our numerical experiments,
we choose the polar cap radius to be 
relatively large because
(i) it helps the Grad-Shafranov solution converge numerically
while still capturing the essential idea of a polar mountain,
(ii) it raises the minimum density $\rho_{\rm min}\propto e^{-b}$
on the grid, preventing $\Delta t_{\rm Z}$ from becoming
too small via the Courant condition, and
(iii) in a real star, the polar plasma flow can spread due to Rayleigh-Taylor
and Kelvin-Helmholtz instabilities \citep{aro84}.
We have not managed to get ZEUS-3D running for $b \gg 3$.
For $b = 3$ and $b = 5$,
$\mu$ reaches an equilibrium value in a few hundred Alfv\'en times, 
and the oscillation frequencies are roughly the same, with
$\mu/\mu_{\rm i} \rightarrow 0.85, 0.82$ for
$b = 3,5$ as $t \rightarrow\inf$.


\subsection{Equatorial magnetic belt} 
\citet{mel01} predicted that the magnetic field at the equator
intensifies during magnetic burial,
if magnetic flux is conserved.
The results presented here confirm this.
As seen in figure \ref{fig:smalla}(a), the magnetic field
is ``combed" away from the pole, and flattened against the stellar
surface towards the equator, increasing $B_{\theta}$ at the expense of $B_{r}$.
The maximum magnetic field strength, $B_{\rm max}$, is 
given as a function of $\Ma$ in \citet{pay06}
Note that
the analytic approximation in the limit of small $\Ma$ captures
the equatorial belt provided that
$J(\psi)$ is calculated properly (PM04).

\section{Stability} 
\label{sec:stability}
In this section, we investigate whether the Grad-Shafranov
equilibria of magnetically confined mountains calculated
in this paper ($\Ma \gtrsim \Mc$) and PM04 ($\Ma\lesssim\Mc$)
are stable to small and large perturbations.
To do this, we load
the numerical output from
our Grad-Shafranov code into ZEUS-3D,
evolve it forward in time, and report on the nature of
any instabilities observed.
In ideal MHD, any instabilities manifest themselves as 
slow, Alfv\'en, or fast magnetosonic waves,
modified by buoyancy effects in a stratified medium.

We explore the stability of the system in the context of
three numerical experiments:
(i) we test how the
small perturbation evolves, which arises from the numerical error
in converting from the Grad-Shafranov to the ZEUS-3D grid
(section \ref{sec:linstability}, \ref{sec:globaloscillate}
and \ref{sec:transientparker});
(ii) we consider the fate of large perturbations
(section \ref{sec:largeperturb}); and
(iii) we compare the evolution in ZEUS-3D with the convergence of
the Grad-Shafranov algorithm.


\subsection{Linear stability: numerical experiment} 
\label{sec:linstability}
Equilibria of the kind depicted in figure \ref{fig:smalla}(a),
where the magnetic field is markedly distorted,
are normally expected to be unstable.
To assess this here,
we track separately the evolution of
$W_{\rm g}$, $W_{\rm k}$, and $W_{B}$,
the gravitational, kinetic, and magnetic energies
respectively,
as the magnetic mountain evolves in ZEUS-3D.
The total energy is given by
\[ W = W_{B} + W_{\rm g} + W_{\rm k}, \]
where we define
\[ W_{\rm g} = \int  d^3\vv{x}\, \rho\phi ,\quad
W_{\rm k} = \int  d^3\vv{x}\, (\rho v^2/2) ,\]
and
\[W_{B} = \int  d^3\vv{x}\, (B^2/2\mu_0) .\]
Key observables of the system, such as the
magnetic dipole moment $\mu$,
defined in (\ref{eq:dipolemoment}), and the mass ellipticity
$\epsilon$, are also tracked.
We do not plot the total thermal energy
$W_{\rm p} = \int d^3\vv{x}\, P \log P$ in what follows
because ZEUS-3D holds it constant during isothermal
(but not adiabatic) runs.

\begin{center}
\begin{figure}
\centering
\includegraphics[height=65mm]{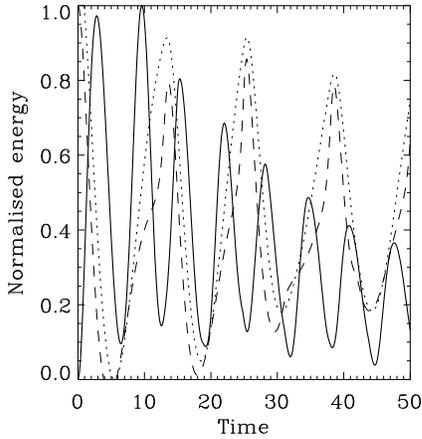}
\caption{\small
Kinetic (\emph{solid}), magnetic (\emph{dashed}), and
gravitational potential (\emph{dotted}) energies,
normalized to their maximum values as a function of time
for $m = 1.6$, $G_{x}=G_{y} = 128$.
}
\label{fig:energywave}
\end{figure}
\end{center}

We begin, as an example, with the equilibrium for $m = 1.6$
[figure \ref{fig:add10more}(a)],
our starting point for adding mass through the outer boundary 
in section \ref{sec:bootstrap}.
Figure \ref{fig:energywave} displays
$W_{\rm g}$, $W_{\rm k}$ and $W_{B}$
a function of $t$ for this case.
The energies are normalized by their maximum values so
that the energy exchanges are clear, because their absolute
values differ by several orders of magnitude:
$W_{\rm g}$ dominates, with
$W_{B} = 3.7\times 10^{-3} W_{\rm g}$ and
$W_{\rm k} = 4.2\times 10^{-5} W_{\rm g}$ initially.
After $300\tau_{\rm A}$, we obtain
$W_{B} = 3.4\times 10^{-3} W_{\rm g}$ and
$W_{\rm k} = 1.2\times 10^{-5} W_{\rm g}$.

The evolution proceeds as follows.
The equilibrium state imported into ZEUS-3D 
is not exact due to grid resolution, imperfect
convergence in the Grad-Shafranov code, and numerical discrepancies
when translating between the Grad-Shafranov and ZEUS-3D grids
(see appendix \ref{appendix:zeus}).
Initially, during the first oscillation,
$W_{\rm g}$ and $W_{B}$ are converted to $W_{\rm k}$.
When the oscillation overshoots, the energy flow
reverses direction.
In ideal MHD, where there is no dissipation, the oscillation
persists.  In ZEUS-3D, where there is some
numerical dissipation and energy is radiated by hydromagnetic waves
through the outer boundary,
the oscillations are damped, as in figure \ref{fig:energywave}.
However, this damping is slow.
(Experiments demonstrating this numerical dissipation are reported
in appendix \ref{appendix:zeus}.)
Note that $W_{\rm k}$ oscillates at twice the frequency
of $W_{\rm g}$ and $W_{B}$,
because the leading term in $W_{\rm k}$ is quadratic
in perturbed quantities ($\propto \delta v^{2}$)
whereas $W_{\rm g}$ ($\propto \delta\rho$) and
$W_{B}$ ($\propto B\delta B$) are linear.

\begin{center}
\begin{figure}
\centering
\includegraphics[height=65mm]{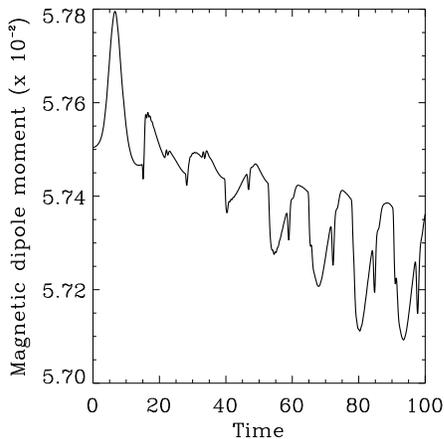}
\caption{\small
Magnetic dipole moment
normalized to its surface value,
plotted as a function of time
for $m = 1.6$, $G_{x}=G_{y} = 128$.
}
\label{fig:dipoledetail}
\end{figure}
\end{center}

The dipole moment, normalized to its value at the stellar surface,
is plotted in figure \ref{fig:dipoledetail} for $m = 1.6$.
The oscillation period equals that of $W_{B}$
for the reason above.
The kinks in the curve occur when
the magnetic field reflects
off the boundaries at the pole and equator.
The oscillations are identified as acoustic modes
(see section \ref{sec:globaloscillate}).

\subsection{Global MHD oscillations} 
\label{sec:globaloscillate}
We explore here the physical nature of the oscillations
resulting from small perturbations to the hydromagnetic equilibria
and how their amplitude and period depend on $\Ma$.

\begin{center}
\begin{figure}
\centering
\includegraphics[height=65mm]{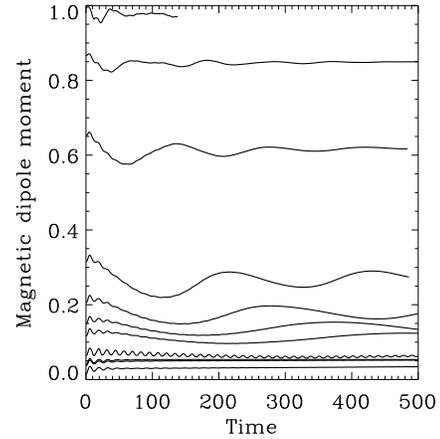}
\includegraphics[height=65mm]{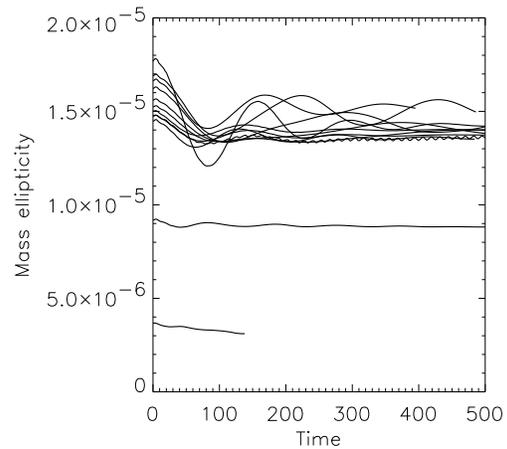}
\caption{\small
Magnetic dipole moment (\emph{top}) and mass quadrupole moment (\emph{bottom})
as a function of time for
$\Ma/\Mc =$ 0.053,\, 0.16,\, 0.32,\, 0.48,\, 0.64,\, 0.8,\,
\noindent
0.96,\, 1.12,\, 1.6,\, 2.4,\, 3.2, and 4.0
(top to bottom for $\mu$, and bottom to top for the quadrupole moment).
Grid resolution: $G_{x,y} = 128$,
with logarithmic scaling in altitude.
}
\label{fig:dipoleall}
\end{figure}
\end{center}

\begin{center}
\begin{figure}
\centering
\includegraphics[height=65mm]{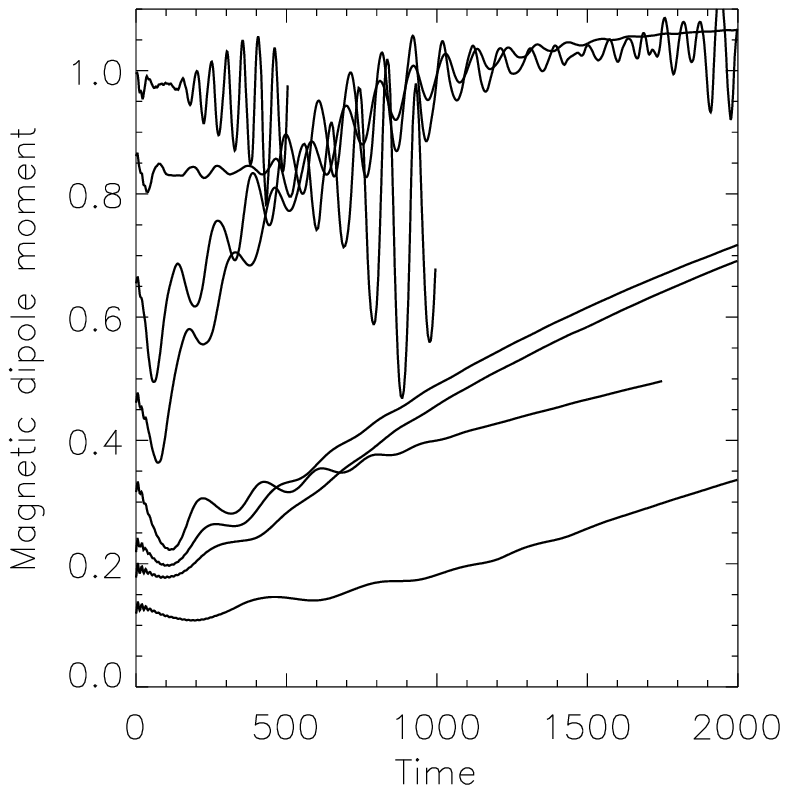}
\includegraphics[height=65mm]{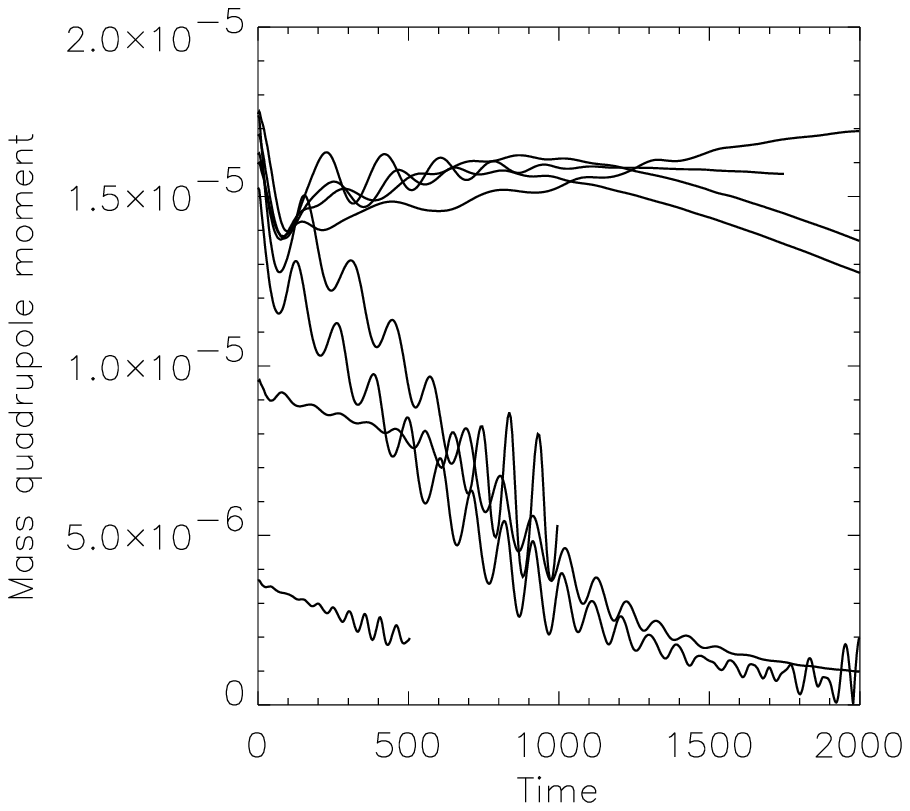}
\caption{\small
Magnetic dipole moment (\emph{top}) and mass quadrupole moment (\emph{bottom})
as a function of time for
$\Ma/\Mc =$ 0.053,\, 0.16,\, 0.32,\, 0.48,\, 0.64,\, 0.96,\, 1.12
(top to bottom for $\mu$, and bottom to top for the quadrupole moment).
Grid resolution: $G_{x,y} = 64$,
with linear scaling in altitude. 
}
\label{fig:dipoleall2}
\end{figure}
\end{center}

\begin{center}
\begin{figure}
\centering
\includegraphics[height=65mm]{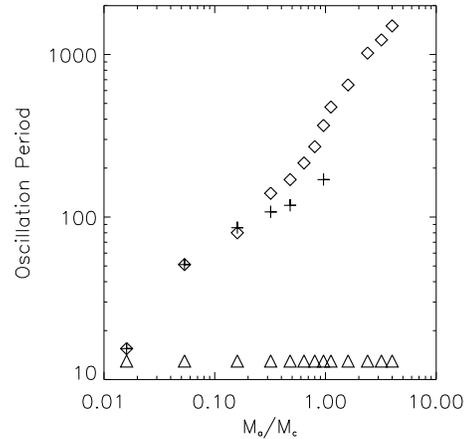} 
\caption{\small
Oscillation period (in units of the Alfv\'en time) as a function of $\Ma$,
with linear (\emph{crosses}) and logarithmic (\emph{diamonds}) grid sampling
in altitude.  The Alfv\'en period is well fit by
$300\tau_0 (\Ma/\Mc)^{1/2}$.
The period of the sound mode (\emph{triangles}) is also plotted.
Note that $\tau_{0}$ is defined at a particular value of $\Ma$.
}
\label{fig:maoscill}
\end{figure}
\end{center}

Figure \ref{fig:dipoleall} shows the time evolution of $\mu$
and mass ellipticity
for several $\Ma$ values and logarithmic scaling of the
altitude.  For reference and numerical comparison,
the results for a linear scaling in altitude are shown in
figure \ref{fig:dipoleall2}.
\emph{The key result is that the equilibria are marginally stable:
the buried field is not disrupted significantly,
but the configuration oscillates about its equilibrium state.}
Two modes are clearly present:
(i) a short period oscillation, with fixed period $13\tau_{0}$
for all $\Ma$ (and thus $\rho$),
which is an acoustic mode with velocity $c_{\rm s}$; and
(ii) a longer period oscillation, with period increasing
with $\Ma$ as displayed in figure \ref{fig:maoscill},
which is an Alfv\'en mode.
The Alfv\'en oscillation frequency is fitted by
\begin{equation}
\label{eq:alfvenfreq}
f_{\rm A} \approx 0.003\tau_0^{-1} (\Ma/\Mc)^{-1/2} {\rm Hz}
\end{equation}
for $a = 50$, which, when
scaled to a realistic neutron star, yields
\begin{equation}
\label{eq:alfvenfreq2}
f_{\rm A} \approx 17 (\Ma/\Mc)^{-1/2} {\rm Hz}.
\end{equation}

There are a few numerical considerations.
The magnetic dipole moment artificially rises
from its equilibrium value towards
the surface value for $0.2 \lesssim m \lesssim 1.6$.
under the following conditions.
(i) There is insufficient resolution close to the stellar surface,
e.g. a linear grid in $r$ with $G_{x} = 128$ is not good enough.
A comparison of figures
\ref{fig:dipoleall} (logarithmic $r$ grid) and
\ref{fig:dipoleall2}  (linear $r$ grid) brings out this point.
$\mu$ and the quadrupole moment exhibit oscillations,
whose amplitude grows and whose mean varies
for a linear grid, whereas
the amplitude is damped and the mean remains constant for
a logarithmic grid.
(ii) The outer boundary condition is set to \emph{outflow},
leading to a small amount of mass loss.
While the mass loss amounts to $\lesssim$ 2\% by $t = 10^3$,
it allows the field lines at the magnetic equator to
flare up towards the pole and destabilize the equilibrium.
These problems are overcome by
(i) using a logarithmic scale in $r$ to increase the
grid resolution near the stellar surface,
(ii) managing carefully the translation from logarithmic scaling
in the Grad-Shafranov grid to the ZEUS-3D grid
(see appendix \ref{appendix:zeus}), and
(iii) setting the outer boundary condition to \emph{inflow}.

\subsection{Identifying MHD modes} 
To classify the oscillation modes discovered in the previous
section, we need the dispersion relation of small-amplitude
MHD waves in the limit (verified by the simulations)
where the wavelength of the $\rho$ and $\vv{B}$
perturbations is small compared with
$|\rho|/|\nabla\rho|$ and $|\psi|/|\nabla\psi|$.
The force equation governs stability:
intuitively, if a fluid (Lagrangian) displacement {\boldmath{$\xi$}}
creates a force $\vv{F}$ in the opposite
direction, equilibrium tends to be restored and the system
is stable.
Upon analysing the perturbation in Fourier modes,
{\boldmath{$\xi$}}$(\vv{r},t) \propto$
{\boldmath{$\xi$}}$(\vv{k},\omega) e^{-i(\vv{k}\cdot\vv{r} - \omega t)}$,
the force equation reads
\begin{equation}
\label{eq:xifourier}
\rho_{0}\omega^{2}{\bm \xi} = - c_{\rm s}^{2}\rho_{0} \vv{k}
\vv{k}\cdot{\bm \xi}(\vv{k},\omega) + \
\mu_{0}^{-1}\{[\vv{k}\times(\vv{k}\times({\bm \xi}\times\vv{B}_{0}))]\times\vv{B}_{0}\}\, .
\end{equation}

In the WKB limit $k \gg |\nabla\psi|/|\psi|$,
the dispersion relation (\ref{eq:xifourier}) supports two modes:
shear Alfv\'en waves, with
\begin{equation}
\label{eq:shearmode}
\omega^{2} = k_{\parallel}^{2} v_{\rm A}^{2} \, ,
\end{equation}
and fast ($+$) and slow ($-$) magnetosonic waves, with
\begin{equation}
\label{eq:magnetosonic}
\omega^{2} = {\tfrac{1}{2}} k^{2}(c_{\rm s}^{2} + v_{\rm A}^{2}) \
[1 \pm (1-\delta)^{1/2}]
\end{equation}
and
\begin{equation}
\delta = \
\frac{4 k_{\parallel}^{2}c_{\rm s}^{2} v_{\rm A}^{2}}{k^2(c_{\rm s}^{2} + v_{\rm A}^{2})^{2}}\, .
\end{equation}
The shear Alfv\'en wave is incompressible ($\vv{k}\cdot{\bm \xi} = 0$).
It propagates analogously to a transverse wave along a string under tension,
the magnetic field lines playing the role of the string.

Let us denote the right-hand side of (\ref{eq:xifourier})
by $\vv{F}({\bm \xi})$.
The operator $\rho^{-1}\vv{F}$ is linear.
It can be shown [see p 244 of \citet{goe04} for a proof] that in ideal MHD,
the operator $\rho^{-1}\vv{F}$ is also self-adjoint, so that the eigenvalues
$\omega^{2}$ in (\ref{eq:xifourier}) are real (making $\omega$ either real or purely imaginary).
Two classes of solution occur:
(i) stable pure waves ($\omega^{2} > 0$), and
(ii) exponentially growing instabilities ($\omega^{2} < 0$).
There are no damped oscillations in ideal MHD, even in a nonuniform background;
any dissipation observed is numerical.


%


A full theoretical calculation of the two-dimensional stability
is beyond the scope of this paper.
However, we attempt to interpret the numerical
results displayed in equations (\ref{eq:alfvenfreq}) and (\ref{eq:alfvenfreq2})
in terms of known results for a gravitating, magnetized slab \citep{goe04}.
We assume that the slab is infinite
and homogeneous in the $y$- and $z$-directions, contained between
two planes at $x = x_1$ and $x = x_2$; that the equilibrium
$\rho$ and $\psi$ vary in the $x$-direction; and that the
excited modes satisfy the boundary conditions
$\xi_{x}(x_1) = \xi_{x}(x_2) = 0$.
The results for a slab provide insight
into the oscillation modes of a neutron star's magnetosphere,
where the above boundary conditions are well satisfied due to
a combination of line tying at $r = R_{*}$ and stratification
at all $r$ (the component of {\boldmath{$\xi$}} perpendicular
to equipotential surfaces is small).
Working with coordinates in a field-line projection,
the Alfv\'en and slow continuum frequencies are estimated to be \citep{goe04}
\begin{equation}
\omega_{\rm A}^2 = 4\pi\mu_{0}^{-1}\rho^{-1}F^2, \quad\quad
\omega_{\rm S}^2 = 4\pi\mu_{0}^{-1}\frac{\gamma p}{\gamma p +4\pi\mu_{0}^{-1}B^2}\rho^{-1}F^2
\end{equation}
with
$F = -i \vv{B}\cdot\nabla$.
From our numerical simulations, we find empirically that $\approx 1/2$
wavelength of the dominant oscillation mode fits within one quadrant,
yielding radial and latitudinal wavenumbers $\sim 2/R_*$.  This gives
$F\sim 2 B_{\rm max}/R_{*}$ and thus
$\omega_{\rm A}^{2} = 4 B_{\rm max}^{2}/(R_{*}^{2} \rho_{\rm max})$,
using the maximum magnetic field strength and density as characteristic
values.
Empirically, from our simulations, we find
$B_{\rm max} \approx 10^{9}$ T for $\Ma = \Mc$,
$\rho_{\rm max} \approx 10^{13}(\Ma/\Mc) {\rm \, kg \, m}^{-3}$,
and hence
$f_{\rm A} = \omega_{\rm A}/(2\pi) = 32(\Ma/\Mc)^{-1/2}$ Hz.
This scales the same way as the numerical result
$f_{\rm A} = 17(\Ma/\Mc)^{-1/2}$ Hz but is larger, due to the
crudeness of our slab approximation and numerical estimates.
The plasma varies with density (and $\Ma$)
from being high-$\beta$ at the stellar surface
to low-$\beta$ far from the star, given that
$c_{\rm s}$ is uniform where the mass is concentrated
and the waves are launched.

Figure \ref{fig:maoscill} shows the oscillation period as a function of $m$.
When analyzing the MHD evolution, we extract
the period of the waves by Fourier analysis
in order to distinguish which modes are excited.
For the Alfv\'en mode,
as the average density increases (with $m$), $v_{\rm A}\propto\Ma^{-1/2}$
decreases, and thus the oscillation period increases.
The Alfv\'en period (in units of the Alfv\'en time) is fitted by
$300\tau_0(\Ma/\Mc)^{-1/2}$.
The period of the acoustic mode remains constant throughout all the simulations
we perform.

\subsection{Transient Parker instability} 
\label{sec:transientparker}
In the experiments described in section \ref{sec:densityincrease},
it is observed that
increasing the density uniformly by a factor $\gtrsim 4$
causes Parker instabilities to occur.
Figure \ref{fig:parkerevol} consists of a sequence of frames illustrating the
evolution of the Parker instability for an equilibrium state whose
density is uniformly increased five-fold.
The wavelength $\lambda$ of the instability subtends
$\approx 1$ radian,
i.e. $\lambda\approx 0.3 R_{*}$.
The radial density profile is exponential, while the magnetic field
is mostly confined to a layer of thickness $\sim h_0$,
directed perpendicular to $\nabla P$.
This is the classic situation in which Parker instabilities are
expected \citep{mou74}.

The evolution of $\mu$, as measured at ${x} = 10h_0$,
is displayed in figure \ref{fig:parkerdipole}.
Note that, after the Parker blister subsides,
$\mu$ returns to its original value.
The equilibrium is not disrupted permanently;
indeed, less than $\sim 1\%$ of the accreted layer
and frozen-in magnetic flux are expelled from the
simulation domain.
In this respect, the instability can be considered transient.

The Grad-Shafranov equilibria imported from PM04
are generated by an algorithm that can
follow, by successive relaxation, the full nonlinear evolution of
the Parker instability \citep{mou74,par66}.
Therefore it is
unsurprising that the equilibria are stable;
the relaxation algorithm evolves the magnetic field
quasistatically through a sequence of intermediate
``states" quite close to those that the real
solution to the time-dependent MHD equations would pass through.
True, the final state is distorted, and one might ask
why the buoyancy of the compressed magnetic flux does not drive
long-wavelength, slow MHD modes
that overturn the accreted matter on the Alfv\'en
time-scale --- something that does not occur in the ZEUS-3D runs
(except for the transient in figure \ref{fig:parkerevol}).
The explanation is that the equatorial magnetic `tutu'
represents the end state of a Parker instability that occurs
quasistatically
as we add material.
The tutu is the analogue of the stable magnetic `blisters' which form when
the plane-parallel Parker instability saturates, while
the polar mountain is the analogue of the material that drains into
the magnetic valleys.

To test the assertion (PM04) that the iterative
solution of the Grad-Shafranov equation coupled to a
flux-freezing condition provides
a good proxy for the true time-dependent behaviour,
we begin with a magnetic dipole with excess mass loaded
into the polar
flux tube in both the Grad-Shafranov code and in ZEUS-3D.
In ZEUS-3D, this is effectively the same as applying a
uniform density increase (section \ref{sec:densityincrease})
to the $m = 0$ equilibrium,
ending up with $m = 1.6$ in this example.
We plot $\mu$ as a function of
time (ZEUS-3D) and iteration number (Grad-Shafranov)
in figure \ref{fig:dipolestatic}.
An underrelaxation parameter
$\Theta = 0.99$ is selected by trial and error
to give a comparable rate of convergence in the two codes.
The initial progress to equilibrium is similar in both cases.
However, after 100 iterations
(or equivalently $t\geq 100$), $\mu$
fluctuates more in ZEUS-3D than
in the converging Grad-Shafranov code.
In ZEUS-3D, the added mass initially squashes the field into a layer
$\sim 1/4$ the thickness of the equilibrium layer
[figure \ref{fig:add10more}(a)].
The field then bounces back and transient Parker-like instabilities
occur, as we expect ($\rho$ increases by a factor $\gtrsim 4$).
This is best illustrated in figure \ref{fig:psistatic}, which
shows a snapshot of the configurations at $t = 100$
(in ZEUS-3D) and iteration 100 (in the Grad-Shafranov
code).
Notice how the field lines are bent by the transient instabilities
in ZEUS-3D but remain smooth in the Grad-Shafranov code.
This comparison also illustrates why
Grad-Shafranov equilibria are
a necessary starting point for ZEUS-3D when aiming to reach stable
equilibria for large $\Ma \gtrsim 10^{-4}\Msun$.
Note that the convergence-matching shown here is not a
proof of the stability of the equilibria
\citep{ass78}.


\subsection{Large perturbations} 
\label{sec:largeperturb}
Next, we investigate what happens when the Grad-Shafranov equilibria
are perturbed far from equilibrium,
i.e. $\delta B/B \sim 1$;
the perturbations considered thus far are small,
arising mainly from imperfect translation between
the grids of the two codes.

\begin{figure}
\centering
\includegraphics[height=65mm]{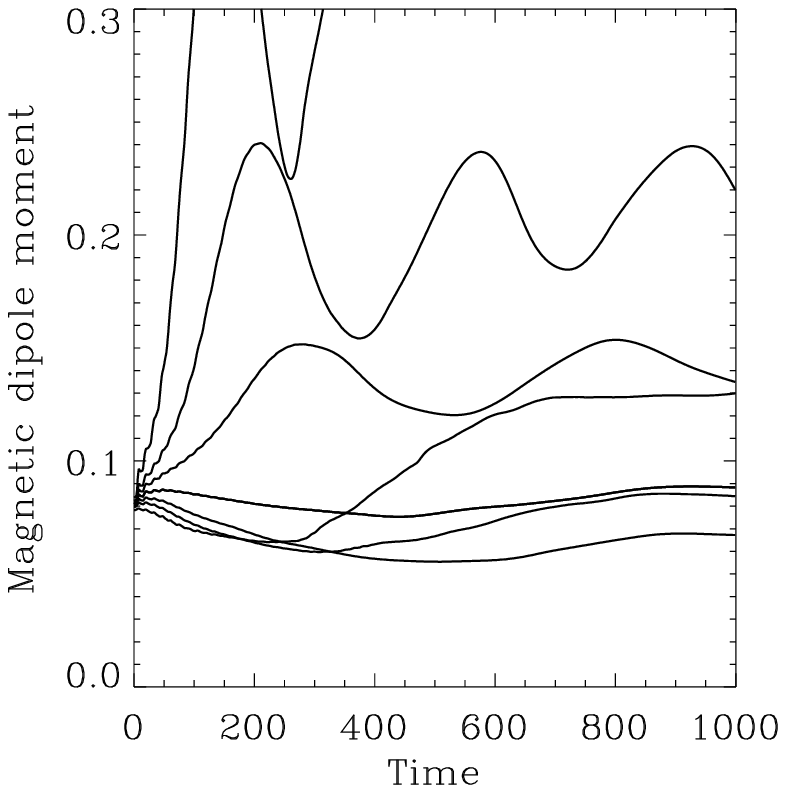} 
\includegraphics[height=65mm]{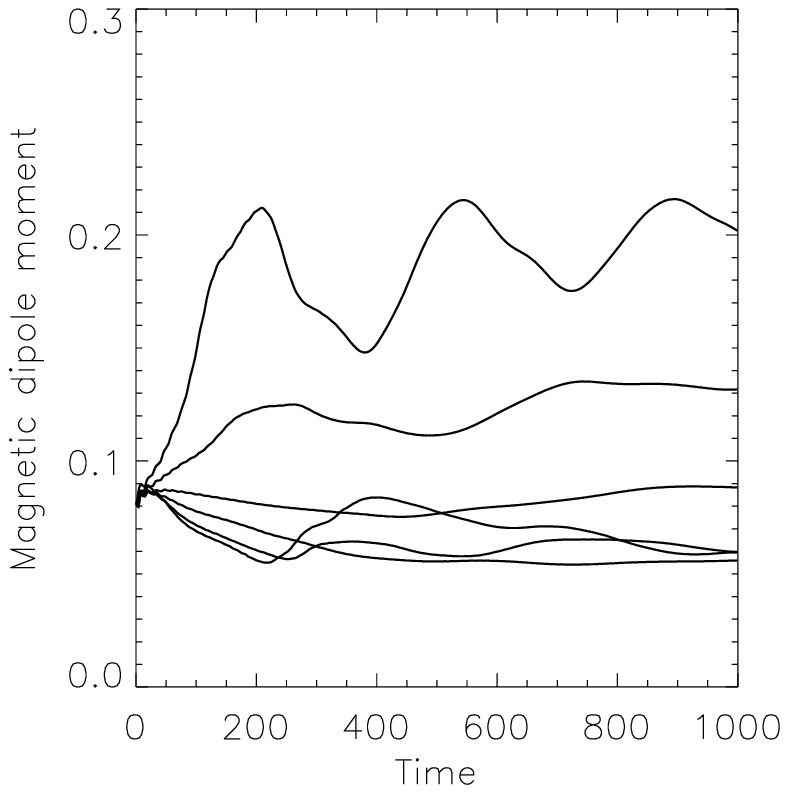} 
\caption{\small
(\emph{Top}.) Magnetic dipole moment as a function of time for
perturbation amplitudes
$\delta = -1.0,\, -0.5,\, -0.25,\, 0,\, 0.25,\, 0.5,\, 1.0$
(\emph{bottom to top}),
when the magnetic field is perturbed as described in
(\ref{eq:bigperturbmag}) but $\rho$ is not altered.
(\emph{Bottom}.)
As above, but with the magnetic field perturbed via
(\ref{eq:bigperturbpsi}) and then iterated once
through the Grad-Shafranov code to obtain the self-consistent $\rho$,
for
$\delta = -0.2,\, -0.1,\, 0,\, 0.1,\, 0.25,\, 0.5$
(\emph{bottom to top}).
}
\label{fig:perturb}
\end{figure}

\begin{center}
\begin{figure}
\centering
\includegraphics[height=65mm]{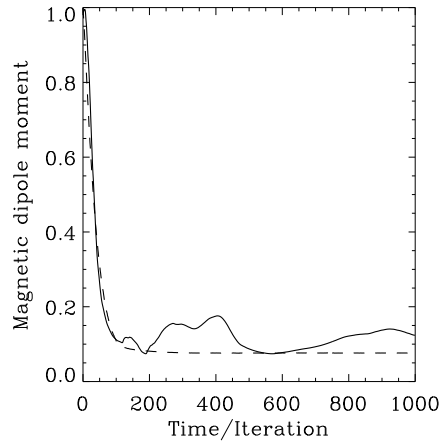}
\caption{\small
Magnetic dipole moment as a function of time
in ZEUS-3D (\emph{solid})
and as a function of iteration number in the Grad-Shafranov code
with underrelaxation parameter $\Theta = 0.99$
(\emph{dashed}).
}
\label{fig:dipolestatic}
\end{figure}
\end{center}

\begin{center}
\begin{figure*}
\centering
\begin{tabular}{cc}
 \begin{tabular}{c}
 (a) \\
\includegraphics[height=60mm]{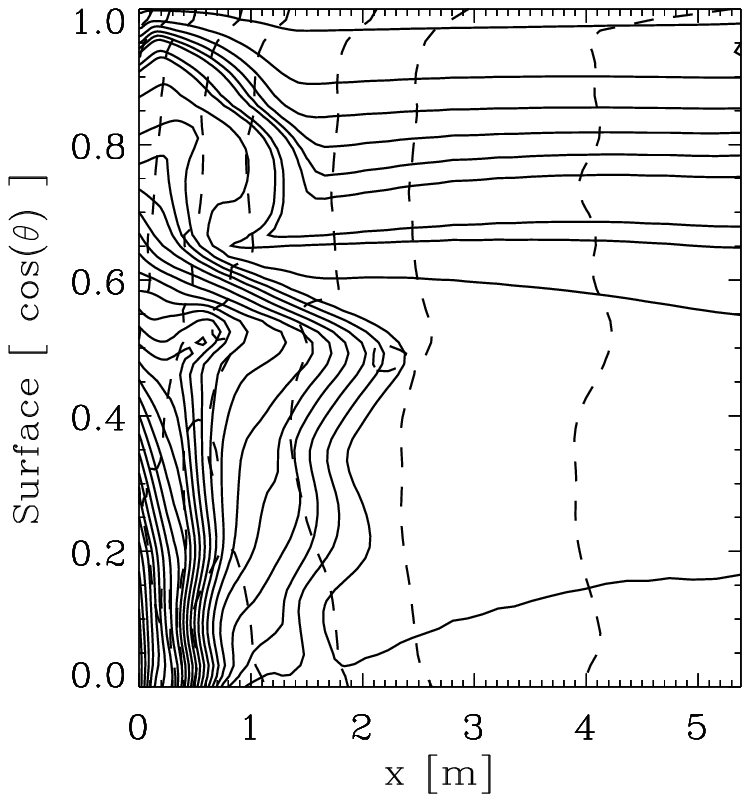}
 \end{tabular}
& 
 \begin{tabular}{c}
 (b) \\
\includegraphics[height=60mm]{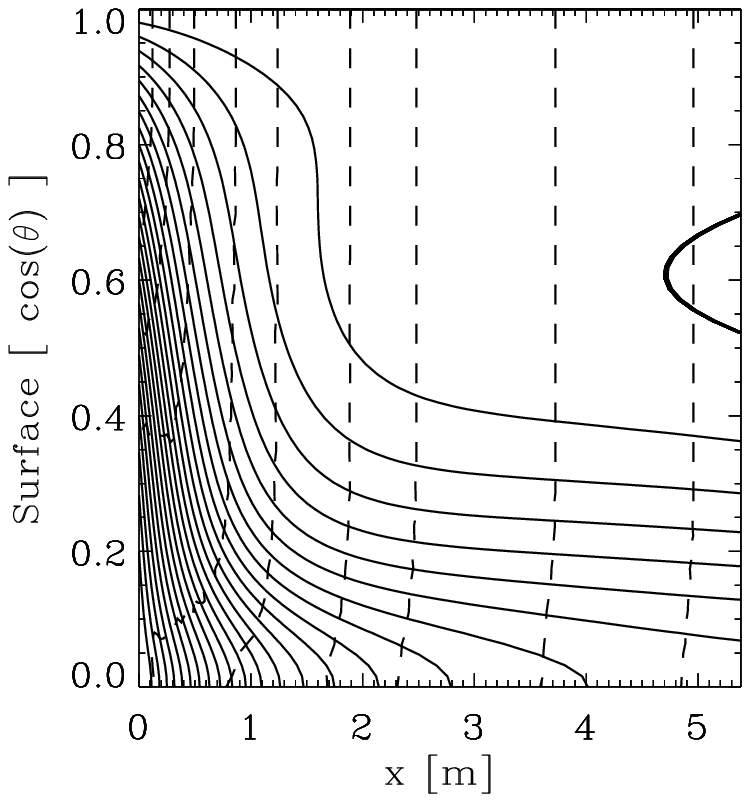}
 \end{tabular}
\end{tabular}
\caption{\small
Magnetic field lines (\emph{solid}) and density contours
(\emph{dashed}) for $t = 100$ in ZEUS-3D (\emph{left}) and
the $100^{\rm th}$ iteration in the Grad-Shafranov code
(\emph{right}).
The initial state is a magnetic dipole with excess mass
($m = 1.6$) in the polar flux tubes
($\psi_{*}/\psi_{\rm a} = 3$).
}
\label{fig:psistatic}
\end{figure*}
\end{center}

We perturb the magnetic field, while simultaneously
respecting the boundary conditions, by setting
\begin{equation}
\label{eq:bigperturbmag}
\vv{B} \mapsto \vv{B}\{1 + \delta \sin[\pi(r-R_{*})/(r_{\rm m}-R_{*})]\sin\theta\}\, .
\end{equation}
Figure \ref{fig:perturb}(a) shows the fate of this perturbation
for different amplitudes $\delta$.
Note that the perturbation does not strictly respect
flux-freezing because it changes $dM/d\psi$ slightly.
\emph{Nonetheless, even for significant perturbations,
the equilibria are not disrupted --- a significant and robust result.}
For $\delta < 0.2$, $\mu$ settles back to within 20 per cent
of its initial value.
For $\delta > 0.2$, $\mu$ does not settle back to its initial value,
but it does settle down to some steady value.
Again, line-tying plays a key role in conferring such exceptional
stability on the system.
The evolution also depends on the sign of $\delta$.
The perturbation is a low-order spatial mode (in $r$ and $\theta$)
and either compresses ($\delta > 0$)
or relaxes ($\delta < 0$) the magnetic field, with
compression causing a back reaction which increases $\mu$
[as can be seen in figure \ref{fig:perturb}].

The above perturbation (\ref{eq:bigperturbmag})
induces an effective redistribution of mass in flux tubes.
To avoid this,
we perturb the magnetic potential $\psi$ in the Grad-Shafranov code
according to
\begin{equation}
\label{eq:bigperturbpsi}
\psi \mapsto \psi\{1 + \delta \sin[\pi(r-R_{*})/(r_{\rm m}-R_{*})]\sin\theta\}
\end{equation}
and iterate once to calculate $\rho$ self-consistently without
changing $dM/d\psi$, so that the above mass redistribution is 
performed self-consistently.
We then load the \emph{self-consistent}
perturbed equilibrium into ZEUS-3D as before.
The results are shown in figure \ref{fig:perturb}(b).
For a given $\delta$, the amplitude of the oscillations is
reduced, confirming that the mass redistribution contributes
to the back reaction.
The main effect is to bring the final value of $\mu$ closer to
its initial value for $\delta \geq 0.5$.  The 
evolution still depends on the sign of $\delta$,
suggesting that mass-flux redistribution
remains somewhat important.

\section{Conclusions}
\label{sec:stabilityconclusion}
We find that, when
the MHD equilibria computed in PM04 are perturbed, they
are marginally stable.
The magnetic field remains essentially undisrupted,
with the exception of transients that expel a tiny
fraction of flux and matter.
MHD oscillations, including both
acoustic and Alfv\'en modes, cause $\mu$ and $\epsilon$ to
oscillate about their mean values.
Using ZEUS-3D, we extend the results of PM04 to larger $\Ma$ in
two ways.
(i) We add mass through the outer boundary at an
artificially accelerated rate which is still slow compared
to the MHD equilibration time-scale and oscillation period
(as in a real neutron star).
We use a bootstrapping method in which the mass is
progressively added then
allowed to settle to equilibrium.
(ii) We increase the density instantaneously uniformly, and then allow
the configuration to settle to equilibrium.
The equilibria we obtain are stable to  Parker modes,
as one might expect given that they are
the output of a Grad-Shafranov code which follows the
full nonlinear evolution of Parker instabilities.

The existence of stable magnetic mountains with persistent
(albeit oscillating) mass quadrupole moments
($\epsilon \lesssim 10^{-6}$) raises the prospect that
accreting millisecond pulsars \citep{wij98}
are sources of gravitational radiation.
This application is pursued in \citet{mel05} and \citet{pay06b}.
Likewise, the existence of a stable equatorial magnetic
belt of intense magnetic field, with its ability to 
impede thermal transport,
has interesting implications for the persistence of
millisecond oscillations in the tails of type I
thermonuclear X-ray bursts in LMXBs \citep{mun02}.
This application is pursued in \citet{pay06}.


\subsection{Limitations}

The work presented above and in PM04 can be generalised by
relaxing certain key assumptions.
Some areas to explore include resistive effects
(e.g. Ohmic diffusion) \citep[e.g.][]{kon97},
sinking of accreted material through a ``soft" stellar surface
\citep[e.g.][]{kon04},
Hall currents \citep[e.g.][]{rhe04}, nonaxisymmetry,
and other equations of state.
For example, by assuming a hard surface, we end up with
densities at the base of the accreted layer that
can be unphysically high.
And it is well known that MHD equilibria can be stable in two
dimensions, yet unstable in three dimensions,
even with line tying.
The magnetization of the accreted plasma also warrants inclusion;
it has been ignored by all authors except \citet{uch81}.

The main bottleneck hindering progress towards
Grad-Shafranov equilibria for
$\Ma \gtrsim 10^{-3}\Msun$ is grid resolution at the magnetic equator.
At present, we believe that the only way to alleviate this is by
appropriate logarithmic gridding in $\theta$.


Ohmic dissipation
due to electron-phonon and electron-impurity scattering
\citep[e.g.][]{bha95}
is not modelled in this paper,
to keep the problem manageable.
Ultimately, it should be.
It is important for magnetic structures
$\sim 1$ m in size, which develop for $\Ma > 10^{-2}\Msun$
\citep{mel05,bro98,cum01}.
Note, however, that nonideal effects are already present
at some level in our calculations accidentally:
the grid introduces
numerical dissipation, because
a field line is defined by linear interpolation between grid points
and the error in this interpolation effectively allows matter to
move across field lines. 

\subsection{Instabilities in three dimensions}
This paper must be generalised to three dimensions before the
stability question is truly settled.
As a trivial example, by restricting ourselves to two-dimensional
equilibria with $B_{\phi} = 0$,
we suppress unstable modes involving
the toroidal magnetic field.
The Parker instability in the galactic disc has been studied in
three dimensions by \citet{kim98}, and many
simulations of toroidal and nonaxisymmetric field structures appear
in the plasma physics literature.
These can be studied in ZEUS-3D and will be explored in a future paper.

Line-tying boundary conditions, as distinct from periodic boundary
conditions, change the character of the basic MHD waves
\citep{goe94}.
Line-tying boundary conditions generally conflict with the phase
relationships between the components of the displacement vector
{\boldmath{$\xi$}} of the three pure modes.
MHD waves of a mixed nature occur as a result.
Stability is enhanced, because it takes extra energy to bend the
longitudinal component of the magnetic field.
We do not compare line-tying and periodic boundary conditions
quantitatively here but point out that line tying enhances
stability even in three dimensions
\citep{goe94}.

Short-wavelength ballooning instabilities are stabilized by line-tying
until the overpressure at the top of the stellar crust exceeds the
magnetic pressure by a factor $8 R_{*}\arcsin(b^{-1/2})/h_0$
\citep{lit01}.  While they are predicted to occur within $h_0$ of the surface,
we do not observe them in these simulations.
Resistive ballooning \citep{vel86} and resistive Rayleigh-Taylor modes \citep{kha93}
are not considered
by virtue of our restriction to ideal MHD,
but they are known to be common in distorted axisymmetric
equilibria similar to those in Figure \ref{fig:smalla}(a).
These instabilities can allow plasma slippage
faster than Ohmic diffusion, but slower than
the Alfven time.\footnote{Arons, J., private communication}
To properly model these instabilities globally,
large toroidal quantum numbers need to be considered
\citep{hua06}.




\section{Acknowledgements}
This research was supported in part by 
an Australian Postgraduate Award and a David Hay write-up
award.


\bibliographystyle{mn2e}

\newcommand{\prd}{Phys. Rev. D} \newcommand{\apj}{ApJ} \newcommand{\apjs}{ApJS}
  \newcommand{\prb}{Phys. Rev. B} \newcommand{\nat}{Nature}
  \newcommand{\mnras}{MNRAS}
  \newcommand{\aap}{A\&A}\newcommand{\aaps}{A\&AS}\newcommand{\aj}{AJ}
  \newcommand{\apjl}{ApJL} \newcommand{\sva}{SvA} \newcommand{\solphys}{Sol.
  Phys.} \newcommand{\physscr}{Physica. Scripta.}
\begin{thebibliography}{}

\bibitem[\protect\citeauthoryear{{Arons}, {Burnard}, {Klein}, {McKee},
  {Pudritz} \& {Lea}}{{Arons} et~al.}{1984}]{aro84}
{Arons} J.,  {Burnard} D.~J.,  {Klein} R.~I.,  {McKee} C.~F.,  {Pudritz} R.~E.,
     {Lea} S.~M.,  1984, in {Woosley} S.~E.,  ed., High Energy Transients in
  Astrophysics Accretion onto magnetized neutron stars - magnetospheric
  structure and stability.
p.~215

\bibitem[\protect\citeauthoryear{{Asseo}, {Cesarsky}, {Lachieze-Rey} \&
  {Pellat}}{{Asseo} et~al.}{1978}]{ass78}
{Asseo} E.,  {Cesarsky} C.~J.,  {Lachieze-Rey} M.,    {Pellat} R.,  1978,
  \apjl, 225, L21

\bibitem[\protect\citeauthoryear{{Bhattacharya} \& Srinivasan}{{Bhattacharya}
  \& Srinivasan}{1999}]{bha95}
{Bhattacharya} D.,  Srinivasan G.,  1999, in in {Lewin}, W.~H.~G. and {van
  Paradijs}, J. and {van den Heuvel}, E.~P.~J. eds, X-ray Binaries Evolution of
  neutron star magnetic fields.
p.~235

\bibitem[\protect\citeauthoryear{{Bisnovatyi-Kogan} \&
  {Komberg}}{{Bisnovatyi-Kogan} \& {Komberg}}{1974}]{bis74}
{Bisnovatyi-Kogan} G.~S.,  {Komberg} B.~V.,  1974, Soviet Astronomy, 18, 217

\bibitem[\protect\citeauthoryear{{Brown} \& {Bildsten}}{{Brown} \&
  {Bildsten}}{1998}]{bro98}
{Brown} E.~F.,  {Bildsten} L.,  1998, \apj, 496, 915

\bibitem[\protect\citeauthoryear{{Burderi}, {Possenti}, {Colpi}, {di Salvo} \&
  {D'Amico}}{{Burderi} et~al.}{1999}]{bur99}
{Burderi} L.,  {Possenti} A.,  {Colpi} M.,  {di Salvo} T.,    {D'Amico} N.,
  1999, \apj, 519, 285

\bibitem[\protect\citeauthoryear{{Cumming}, {Arras} \& {Zweibel}}{{Cumming}
  et~al.}{2004}]{cum04}
{Cumming} A.,  {Arras} P.,    {Zweibel} E.,  2004, \apj, 609, 999

\bibitem[\protect\citeauthoryear{{Cumming}, {Zweibel} \& {Bildsten}}{{Cumming}
  et~al.}{2001}]{cum01}
{Cumming} A.,  {Zweibel} E.,    {Bildsten} L.,  2001, \apj, 557, 958

\bibitem[\protect\citeauthoryear{{Goedbloed} \& {Halberstadt}}{{Goedbloed} \&
  {Halberstadt}}{1994}]{goe94}
{Goedbloed} J.~P.,  {Halberstadt} G.,  1994, \aap, 286, 275

\bibitem[\protect\citeauthoryear{{Goedbloed} \& {Poedts}}{{Goedbloed} \&
  {Poedts}}{2004}]{goe04}
{Goedbloed} J.~P.~H.,  {Poedts} S.,  2004, {Principles of
  Magnetohydrodynamics}.
Cambridge University Press

\bibitem[\protect\citeauthoryear{{Hameury}, {Bonazzola}, {Heyvaerts} \&
  {Lasota}}{{Hameury} et~al.}{1983}]{ham83}
{Hameury} J.~M.,  {Bonazzola} S.,  {Heyvaerts} J.,    {Lasota} J.~P.,  1983,
  \aap, 128, 369

\bibitem[\protect\citeauthoryear{{Huang}, {Zweibel} \& {Sovinec}}{{Huang}
  et~al.}{2006}]{hua06}
{Huang} Y.-M.,  {Zweibel} E.~G.,    {Sovinec} C.~R.,  2006, Physics of Plasmas,
  13, 2102

\bibitem[\protect\citeauthoryear{{Khan} \& {Bhatia}}{{Khan} \&
  {Bhatia}}{1993}]{kha93}
{Khan} A.,  {Bhatia} P.~K.,  1993, \physscr, 48, 607

\bibitem[\protect\citeauthoryear{{Kim}, {Hong}, {Ryu} \& {Jones}}{{Kim}
  et~al.}{1998}]{kim98}
{Kim} J.,  {Hong} S.~S.,  {Ryu} D.,    {Jones} T.~W.,  1998, \apjl, 506, L139

\bibitem[\protect\citeauthoryear{{Klimchuk} \& {Sturrock}}{{Klimchuk} \&
  {Sturrock}}{1989}]{kli89}
{Klimchuk} J.~A.,  {Sturrock} P.~A.,  1989, \apj, 345, 1034

\bibitem[\protect\citeauthoryear{{Konar} \& {Bhattacharya}}{{Konar} \&
  {Bhattacharya}}{1997}]{kon97}
{Konar} S.,  {Bhattacharya} D.,  1997, \mnras, 284, 311

\bibitem[\protect\citeauthoryear{{Konar} \& {Choudhuri}}{{Konar} \&
  {Choudhuri}}{2004}]{kon04}
{Konar} S.,  {Choudhuri} A.~R.,  2004, \mnras, 348, 661

\bibitem[\protect\citeauthoryear{{Litwin}, {Brown} \& {Rosner}}{{Litwin}
  et~al.}{2001}]{lit01}
{Litwin} C.,  {Brown} E.~F.,    {Rosner} R.,  2001, \apj, 553, 788

\bibitem[\protect\citeauthoryear{{Melatos} \& {Payne}}{{Melatos} \&
  {Payne}}{2005}]{mel05}
{Melatos} A.,  {Payne} D.~J.~B.,  2005, \apj, 623, 1044

\bibitem[\protect\citeauthoryear{{Melatos} \& {Phinney}}{{Melatos} \&
  {Phinney}}{2001}]{mel01}
{Melatos} A.,  {Phinney} E.~S.,  2001, PASA, 18, 421

\bibitem[\protect\citeauthoryear{{Mouschovias}}{{Mouschovias}}{1974}]{mou74}
{Mouschovias} T.,  1974, \apj, 192, 37

\bibitem[\protect\citeauthoryear{{Mouschovias}}{{Mouschovias}}{1996}]{mou96}
{Mouschovias} T.,  1996, in Solar and Astrophysical Magnetohydrodynamic Flows,
  ed. K. C. Tsinganos (NATO ASI Ser. C, 481) (Dordrecht: Kluwer), p.~475

\bibitem[\protect\citeauthoryear{{Muno}, {{\" O}zel} \& {Chakrabarty}}{{Muno}
  et~al.}{2002}]{mun02}
{Muno} M.~P.,  {{\" O}zel} F.,    {Chakrabarty} D.,  2002, \apj, 581, 550

\bibitem[\protect\citeauthoryear{{Muslimov} \& {Tsygan}}{{Muslimov} \&
  {Tsygan}}{1985}]{mus85}
{Muslimov} A.~G.,  {Tsygan} A.~I.,  1985, SvAL, 11, 80

\bibitem[\protect\citeauthoryear{{Parker}}{{Parker}}{1966}]{par66}
{Parker} E.,  1966, \apj, 145, 811

\bibitem[\protect\citeauthoryear{{Payne} \& {Melatos}}{{Payne} \&
  {Melatos}}{2004}]{pay04}
{Payne} D.~J.~B.,  {Melatos} A.,  2004, \mnras, 351, 569

\bibitem[\protect\citeauthoryear{{Payne} \& {Melatos}}{{Payne} \&
  {Melatos}}{2006a}]{pay06b}
{Payne} D. J.~B.,  {Melatos} A.,  2006a, ApJ, 641, 471

\bibitem[\protect\citeauthoryear{{Payne} \& {Melatos}}{{Payne} \&
  {Melatos}}{2006b}]{pay06}
{Payne} D. J.~B.,  {Melatos} A.,  2006b, ApJ, 652, 597

\bibitem[\protect\citeauthoryear{{Rheinhardt}, {Konenkov} \&
  {Geppert}}{{Rheinhardt} et~al.}{2004}]{rhe04}
{Rheinhardt} M.,  {Konenkov} D.,    {Geppert} U.,  2004, \aap, 420, 631

\bibitem[\protect\citeauthoryear{{Romani}}{{Romani}}{1990}]{rom90}
{Romani} R.~W.,  1990, \nat, 347, 741

\bibitem[\protect\citeauthoryear{{Shibazaki}, {Murakami}, {Shaham} \&
  {Nomoto}}{{Shibazaki} et~al.}{1989}]{shi89}
{Shibazaki} N.,  {Murakami} T.,  {Shaham} J.,    {Nomoto} K.,  1989, \nat, 342,
  656

\bibitem[\protect\citeauthoryear{{Srinivasan}, {Bhattacharya}, {Muslimov} \&
  {Tsygan}}{{Srinivasan} et~al.}{1990}]{sri90}
{Srinivasan} G.,  {Bhattacharya} D.,  {Muslimov} A.,    {Tsygan} A.,  1990,
  Curr. Sci., 59, 31

\bibitem[\protect\citeauthoryear{{Stone} \& {Norman}}{{Stone} \&
  {Norman}}{1992a}]{sto92a}
{Stone} J.~M.,  {Norman} M.~L.,  1992a, \apjs, 80, 753

\bibitem[\protect\citeauthoryear{{Stone} \& {Norman}}{{Stone} \&
  {Norman}}{1992b}]{sto92b}
{Stone} J.~M.,  {Norman} M.~L.,  1992b, \apjs, 80, 791

\bibitem[\protect\citeauthoryear{{Strohmayer} \& {Bildsten}}{{Strohmayer} \&
  {Bildsten}}{2006}]{str03}
{Strohmayer} T.~E.,  {Bildsten} L.,  2006, in Compact Stellar X-Ray Sources,
  eds. W.H.G. Lewin and M. van der Klis, Cambridge University Press, p.~113

\bibitem[\protect\citeauthoryear{{Taam} \& {van den Heuvel}}{{Taam} \& {van den
  Heuvel}}{1986}]{taa86}
{Taam} R.~E.,  {van den Heuvel} E.~P.~J.,  1986, \apj, 305, 235

\bibitem[\protect\citeauthoryear{{Tauris}, {van den Heuvel} \&
  {Savonije}}{{Tauris} et~al.}{2000}]{tau00}
{Tauris} T.~M.,  {van den Heuvel} E.~P.~J.,    {Savonije} G.~J.,  2000, \apjl,
  530, L93

\bibitem[\protect\citeauthoryear{{Uchida} \& {Low}}{{Uchida} \&
  {Low}}{1981}]{uch81}
{Uchida} Y.,  {Low} B.~C.,  1981, JA\&A, 2, 405

\bibitem[\protect\citeauthoryear{{Urpin} \& {Geppert}}{{Urpin} \&
  {Geppert}}{1995}]{urp95}
{Urpin} V.,  {Geppert} U.,  1995, \mnras, 275, 1117

\bibitem[\protect\citeauthoryear{{Urpin} \& {Konenkov}}{{Urpin} \&
  {Konenkov}}{1997}]{urp97}
{Urpin} V.,  {Konenkov} D.,  1997, \mnras, 284, 741

\bibitem[\protect\citeauthoryear{{van den Heuvel} \& {Bitzaraki}}{{van den
  Heuvel} \& {Bitzaraki}}{1995}]{van95}
{van den Heuvel} E.~P.~J.,  {Bitzaraki} O.,  1995, \aap, 297, L41

\bibitem[\protect\citeauthoryear{{Velli} \& {Hood}}{{Velli} \&
  {Hood}}{1986}]{vel86}
{Velli} M.,  {Hood} A.~W.,  1986, \solphys, 106, 353

\bibitem[\protect\citeauthoryear{{Wijers}}{{Wijers}}{1997}]{wij97}
{Wijers} R.~A.~M.~J.,  1997, \mnras, 287, 607

\bibitem[\protect\citeauthoryear{{Wijnands} \& {van der Klis}}{{Wijnands} \&
  {van der Klis}}{1998}]{wij98}
{Wijnands} R.,  {van der Klis} M.,  1998, \nat, 394, 344

\end{thebibliography}

\newcommand{\prd}{Phys. Rev. D} \newcommand{\apj}{ApJ} \newcommand{\apjs}{ApJS}
  \newcommand{\prb}{Phys. Rev. B} \newcommand{\nat}{Nature}
  \newcommand{\mnras}{MNRAS}
  \newcommand{\aap}{A\&A}\newcommand{\aaps}{A\&AS}\newcommand{\aj}{AJ}
  \newcommand{\apjl}{ApJL} \newcommand{\sva}{SvA} \newcommand{\solphys}{Sol.
  Phys.} \newcommand{\physscr}{Physica. Scripta.}

\appendix
\section{Magnetic burial in ZEUS-3D}
\label{appendix:zeus}
In this appendix, we explain briefly how to
simulate the problem of magnetic burial in ZEUS-3D,
so that the reader can reproduce and generalize our results.\footnote{
This work also serves as a prototype of a simulation pipeline
being developed by the Australian Virtual Observatory.
We have demonstrated a preliminary version of this pipeline in which
hydromagnetic equilibria generated by the time-independent,
Grad-Shafranov code described in PM04 are output in a standard format
(VOTable) and fed into ZEUS-3D to be evolved in time.}
For details on the general design and operation of ZEUS-3D,
please consult
\citet{sto92a, sto92b}.

After discussing the control variables in ZEUS-3D,
we provide several test cases for reference.
The Parker instability in rectangular coordinates,
whose nonlinear evolution was computed by
\citet{mou74,mou96},
tests the basic functionality of ZEUS-3D.
An unmagnetized, isothermal atmosphere in spherical coordinates
tests the boundary conditions on $\rho$ and $\vv{v}$.
A magnetized, isothermal atmosphere in spherical coordinates
tests the boundary conditions on $\vv{B}$ and yields the
equilibrium state for
$\Ma = 0$.

\subsection{Control variables}
\label{sec:appendix:control}
The number of active grid zones in each coordinate,
{\tt ggen1:nbl} and {\tt ggen2:nbl} for {\tt i} and {\tt j} respectively,
is $G_{x} \times G_{y}$ (typically  $128 \times 128$).
In addition,
there are two ghost zones at each edge of the grid, for setting
boundary conditions.
The zones in, say, the {\tt i} direction
($r$ in spherical coordinates)
can scale geometrically using {\tt ggen1:x1rat}, such that the width of
zone {\tt (i+1)} is {\tt x1rat} times zone {\tt i}.
This is how we implement logarithmic coordinates to
increase grid resolution near the stellar surface, and to match
input data from the Grad-Shafranov code in PM04 (section \ref{sec:logscale}).
There are several conditional compilation switches which control
the geometry.
To implement two dimensions,
enable {\tt KSYM}, and set the number of zones
in the third ordinate, {\tt ggen3:nbl} to one.
For Cartesian coordinates, enable {\tt XYZ};
for spherical polar coordinates, enable {\tt RTP}.
When printing out the grids, we use {\tt nxpx} = {\tt nypx} = 400 pixels
and suppress the third dimension ($z$ or $\phi$) by setting
{\tt pixcon:ipixdir} = 3.

The results in this paper assume an isothermal equation of state
for simplicity, thus {\tt ISO} is enabled.
Self-gravity, {\tt GRAV}, is disabled.
Dimensionless quantities are chosen as follows:
$\tilde{\mu}_{0} = 1$ (automatic in ZEUS-3D),
$\tilde{G} = 1$ (set {\tt grvcon:g}=1),
$\tilde{c}_{\rm s} = 1$ (set {\tt eqos:ciso}=1), and
$\tilde{h}_{0} = 1$
($h_{0} = c_{\rm s}^{2} R_{*}^{2}/G M_{*}$).
With these choices, the basic units of mass, magnetic field
and time become
$M_{0} = h_0 c_{\rm s}^{2}/G$,
$B_{0} = [\mu_{0} c_{\rm s}^{4}/(G h_{0}^{2})]^{1/2}$,
and
$\tau_{0} = h_{0}/c_{\rm s}$.
In ideal MHD,
the evolution is controlled by the geometry as well as
the ratio of magnetic to thermal pressure,
$\alpha = B^{2}/(\mu_{0} c_{\rm s}^{2} \rho)$;
the physical units of $B$ and $\rho$ do not enter separately.
Varying $\tilde{G}$, set to one by default,
changes the unit conversions but not the physics.
ZEUS-3D may be run as a hydrodynamic code without magnetic fields
by disabling {\tt MHD}, reducing the run time.

The time-step $\Delta t_{\rm Z}$ is adjusted adaptively so that
the Courant-Friedrichs-Lewy stability condition
is satisfied
in every zone, with the tolerance determined by the Courant number
({\tt hycon:courno} $= 0.5$ by default).
In the case of an isothermal atmosphere and subsonic fluid motions,
it is set by
$\max(c_{\rm s},v_{\rm A})$.
Since one has
$v_{\rm A}\propto 1/\rho\propto e^{\tilde{x}}$
in a typical isothermal atmosphere, the code is limited
in the range of altitudes and densities that it can faithfully simulate.
The grid is dumped at times separated by {\tt pixcon:dtpix}
and the code runs for a maximum time {\tt pcon:tlim}.

\subsection{Verifying the Parker instability in a rectangular geometry}
A test case which is easy to implement
(and relevant to the problem of magnetic burial) is
the Parker or magnetic Rayleigh-Taylor instability
\citep{par66,mou74,mou96}.
The initial equilibrium state is a semi-infinite,
exponentially
stratified atmosphere in a uniform gravitational field,
with a magnetic field perpendicular
to the gravitational force.
This configuration is unstable to overturn (slow MHD) modes
longer than a critical wavelength $\lambda_{\rm crit}$.

We simulate 
the Parker instability in the region
$-X \leq \tilde{x} \leq X$,
$0 \leq \tilde{y} \leq Y$.
A uniform gravitational acceleration $g_{y}$ in the $y$ direction is implemented
in ZEUS-3D
by placing a point mass far from the simulation box, at
$\tilde{y} = \tilde{R}_{\rm eff} \gg Y$.
The boundary conditions are set to be periodic at $\tilde{x} = \pm X$
({\tt niib, noib} = 4) and
reflecting at $\tilde{y} = 0, Y$
({\tt nijb} = {\tt nojb} = 1).

In dimensionless units, the equilibrium initial conditions are
\begin{equation}
\label{eq:pkrinitrho}
\tilde{\rho}(\tilde{x},\tilde{y}) = 0.5 \exp[{-\tilde{y}/(1+\alpha)}]
\end{equation}
and
\begin{equation}
\label{eq:pkrinitmag}
\tilde{A}_{z}(\tilde{x},\tilde{y}) = 2(1+\alpha) \exp[{-\tilde{y}/2(1+\alpha)}] \, ,
\end{equation}
with
$\tilde{B}_{x} = (\nabla\times\tilde{A})_{x} = - e^{-\tilde{y}/2(1+\alpha)}$.
The initial velocity perturbation is taken as
$\tilde{v}_{y} = \epsilon \sin(\pi \tilde{y}/Y) \cos(\pi \tilde{x}/X)$,
where, for $\epsilon = 0.3$, the kinetic energy equals
$0.23 \%$ of the magnetic energy;
cf. \citet{mou96}.

The stability condition is tested by importing a stable equilibrium state, with
$X < \lambda_{\rm crit}/2$,
and checking that the kinetic energy remains negligible compared to the
magnetic and gravitational potential energies.
The ratio of the kinetic energy to the total energy is given
in table \ref{table:pkrgrav},
scaling as $\tilde{R}_{\rm eff}^{-2}$ for
$\tilde{R}_{\rm eff} \lesssim 10^{5}$.
We choose $\tilde{R}_{\rm eff} = 7.6\times 10^{4}$ for most runs.
This test also confirms that the equilibrium is stable for
$(X,Y) = (7,25)$, just below the critical value
$\lambda_{\rm crit}/2 = 7.26$ \citep{mou74}, and is unstable for
$(X,Y) = (8,25)$.

\begin{table}
\begin{center}
\begin{tabular}{cccc}
\hline
$\tilde{R}_{\rm eff}$ & $\tilde{W}_{\rm g}$ & $\tilde{W}_{B}$ & $\tilde{W}_{\rm k}$ \\
\hline
$7.6\times 10^{2}$  & $-1.03\times 10^{3}$ & 12.9   &  $1.03\times 10^{-1}$    \\
$7.6\times 10^{2}$  & $-1.06\times 10^{4}$ & 14.0   &  $1.75\times 10^{-3}$    \\
$7.6\times 10^{3}$  & $-1.06\times 10^{5}$ & 14.0   &  $1.32\times 10^{-5}$    \\
$7.6\times 10^{4}$  & $-1.06\times 10^{6}$ & 14.0   &  $1.22\times 10^{-6}$    \\
$7.6\times 10^{5}$  & $-1.06\times 10^{7}$ & 14.0   &  $1.69\times 10^{-6}$    \\
$7.6\times 10^{6}$  & $-1.06\times 10^{8}$ & 14.0   &  $1.75\times 10^{-6}$    \\
\hline
\end{tabular}
\caption{Gravitational, magnetic and kinetic energies
(dimensionless units)
resulting from a point mass located at $R_{\rm eff}$ such that
$\tilde{M}/\tilde{R}_{\rm eff}^{2} = 1$.  The initial state is
the equilibrium given by (\ref{eq:pkrinitrho})
and (\ref{eq:pkrinitmag}),
with $\tilde{\tau} = 10$ and $(X,Y) = (7,25)$,
perturbed as described in the text.
}
\label{table:pkrgrav}
\end{center}
\end{table}

\begin{center}
\begin{figure}
\centering
\includegraphics[height=65mm]{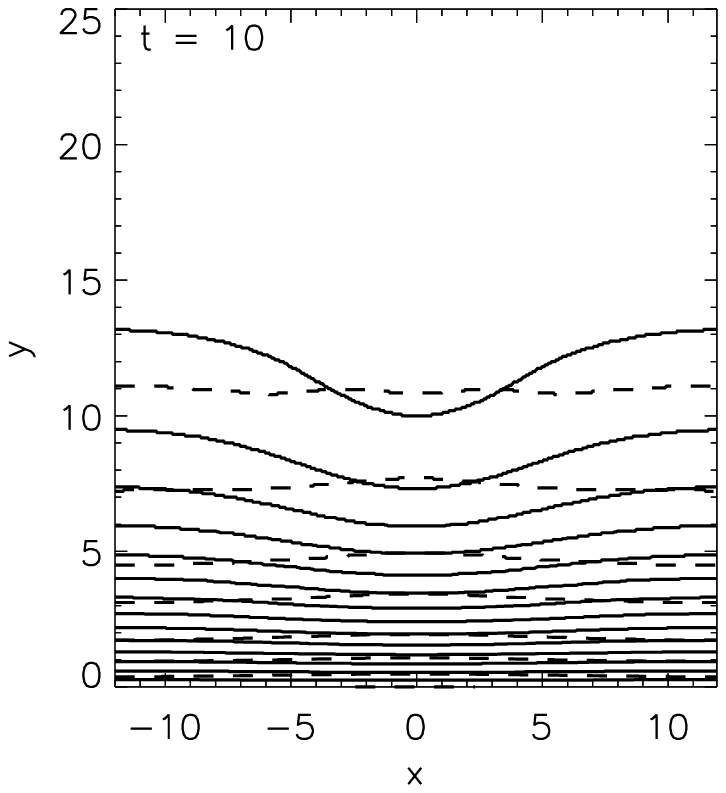} 
\includegraphics[height=65mm]{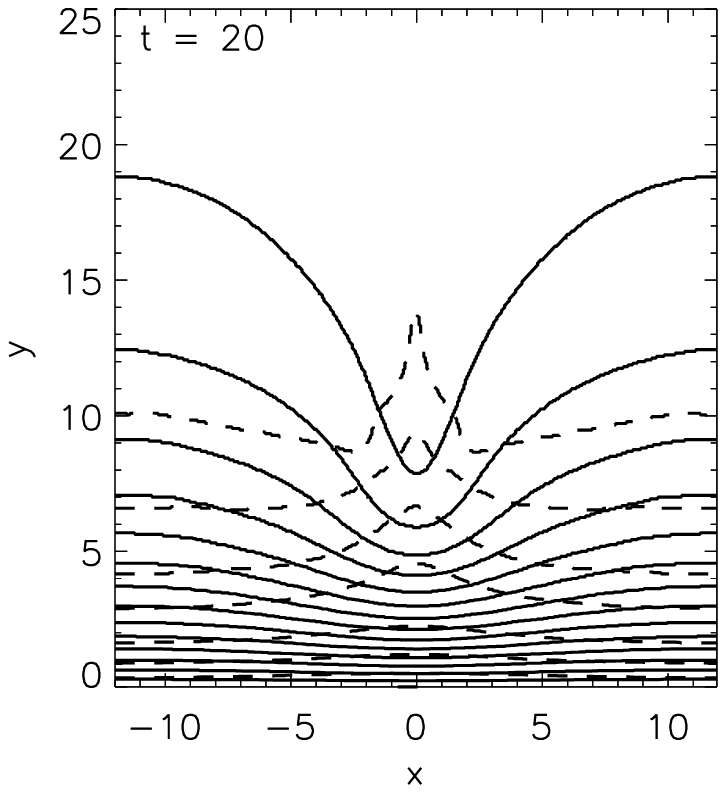} 
\includegraphics[height=65mm]{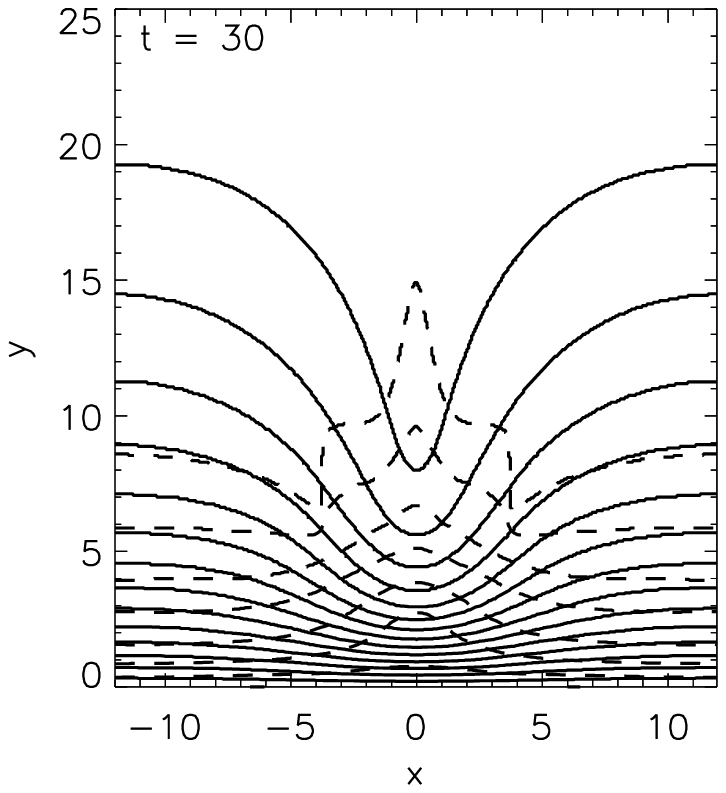} 
\caption{\small
Nonlinear evolution of the Parker instability, showing
magnetic flux surfaces (\emph{solid}) and isodensity contours
(\emph{dashed}) for $(X,Y) = (12,25)$, $\tilde{\tau} = 10,\, 20,\, 30$, and
$G_{x} = G_{y} = 64$
}
\label{fig:pkr}
\end{figure}
\end{center}

We also set (X,Y) = (12,25) and
verify the output against the results given in figure 3
of \citet{mou96}.
We choose the same units, namely
$c_{\rm s} = 6.2$ km s$^{-1}$,
$h_{0} =  1.3 \times 10^{18}$ m,
$\tau_{0} = r_{0}/c_{\rm s} = 6.5 \times 10^{6}$ yr,
$M_{0} = 5.8 \times 10^{17}$ kg,
$\rho_{0} = 2.9 \times 10^{-37}$ kg \, m$^{-3}$,
$B_{0} = 3.7 \times 10^{-18}$ T.
(To have exactly the same units as \citet{mou96}
choose $\tilde{G} = 6.67 \times 10^{-17}$ in ZEUS-3D.)
Figure \ref{fig:pkr} shows the ZEUS-3D results at the three
times displayed in figure 3 of \citet{mou96}.  The results
are in good accord, differing by less than 5 per cent.
A mountain is clearly present where the magnetic field bends down
in the negative $y$ direction.

\subsection{Spherical, isothermal atmosphere, zero magnetic field}
\label{sec:zeusapp3}
Having verified a known solution, we now move closer to our goal
and consider spherical geometry
({\tt RTP}).
We consider first an isothermal atmosphere with no magnetic field
to test the effect of the boundary conditions.
In spherical geometry, the curvature of the star,
characterized by $a = R_{*}/h_{0} \sim 2\times 10^{4}$,
enters the problem.
To cope with 
this large value,
we scale coordinates in ZEUS-3D logarithmically
(section \ref{sec:appendix:control}).
However, the code halts prematurely when
the time-step $\Delta t_{\rm Z}$ becomes too small;
for $a > 10^{3}$, it does not run at all.
The problem is most severe far from the surface, where
$\rho$ is small, $v_{\rm A}$ is large, and $\Delta t_{\rm Z}$ is small.
As an initial test case in spherical coordinates,
we set up a spherical isothermal atmosphere
with a negligible magnetic field ($\alpha = 4\times 10^{-8}$)
in 2D (single quadrant).
The initial condition is an exponential isothermal atmosphere
\begin{equation}
\label{eq:earthinitrho}
\tilde{\rho}(\tilde{r},\theta) = \exp[{-a(\tilde{r}-a)/\tilde{r}}]\approx\exp(-\tilde{x})
\end{equation}
threaded by a dipolar magnetic vector potential
\begin{equation}
\label{eq:earthinitmag}
\tilde{A}_{\phi}(\tilde{r},\theta) = 10^{-10} \tilde{r}^{-2} \sin\theta
\end{equation}
The boundary conditions are
reflecting at the pole ($B_{\theta} = 0$, {\tt nijb = 1}) and
at the equator ($B_{r} = 0$, {\tt nijb = 5}),
dipolar at the surface
({\tt niib = 3}, i.e. inward flow at zero velocity, fixed density,
dipolar magnetic field),
and free at the outer boundary ($\tilde{r} = \tilde{r}_{\rm m}$)
({\tt niob = 2}, outward flow).
Note that, in ZEUS-3D, a fixed magnetic field must be accompanied
by inflow, even if the inflow is at zero speed.

The experiment described in this section aims to calibrate the
role of the outer boundary condition.
Equations (\ref{eq:earthinitrho}) and (\ref{eq:earthinitmag})
almost describe a force-free equilibrium except that
matter can evaporate through the outer boundary.
However, alternative boundary conditions offered in ZEUS-3D are not
appropriate, e.g. inflow artificially pins the magnetic field lines
at the boundary.
The amount of mass loss is monitored and $\tilde{r}_{\rm m}$
is chosen large to minimize it.
There is a trade off between minimizing
mass loss as well as simulation run time
(determined primarily by $\Delta t_{\rm Z}$),
as illustrated in table \ref{table:varya}.
Logarithmic grid scaling has little effect on the amount of mass which
evaporates through the outer boundary.
The outer boundary condition artificially increases the dipole moment
at large $r$, as discussed in the main text and illustrated in
figure \ref{fig:addmass}(f).
The run time is reduced if the magnetic calculations are switched off
by disabling {\tt MHD};
the results are indistinguishable.
The results in table \ref{table:varya} are for
$R_{*} = 6.4\times 10^{6}$ m,
$M_{*} = 6.0\times 10^{24}$ kg,
$c_{\rm s} = 290$ m s$^{-1}$,
$h_{0} = 8.4\times 10^{3}$ m.
With these choices, the basic units become
$M_{0} = c_{\rm s}^{2}/G = 1.2 \times 10^{15}$ kg,
$\rho_{0} = M_{0}/r_{0}^{3} = 2.1 \times 10^{3}$ kg \,  m$^{-3}$,
$B_{0} = [\mu_{0} c_{\rm s}^{4}/(G r_{0}^{3})]^{1/2} = 15$ T ,
$\tau_{0} = r_{0}/c_{\rm s} = 29$ s;
the alert reader will note that these parameters describe the
Earth!

\subsection{Spherical, isothermal atmosphere, dipole magnetic field}
As a prelude to testing the stability of the Grad-Shafranov
equilibria found by PM04, we consider the simpler situation of the
spherical, isothermal atmosphere (section \ref{sec:zeusapp3}) threaded by a substantial dipolar
magnetic field.
Note that this is a valid force-free equilibrium state
[$(\nabla\times\vv{B})\times\vv{B}=0$ in
(\ref{eqn:gradshafranov}) except at $r = 0$],
except for evaporation.
This configuration corresponds to the pre-accretion
neutron star with $\Ma = 0$.
It serves to calibrate ZEUS-3D and estimate the magnitude of
numerical errors.


The relevant neutron star parameters are
$R_{*} = 10^{4}$ m,
$M_{*} = 1.4\Msun$,
$c_{\rm s} = 10^{6}$ m s$^{-1}$.
With these choices, the basic units become
$M_{0} = c_{\rm s}^{2}/G = 1.5 \times 10^{22}$ kg,
$\rho_{0} = M_{0}/r_{0}^{3} = 9.5 \times 10^{22}$ kg \, m$^{-3}$,
$B_{0} = [\mu_{0} c_{\rm s}^{4}/(G r_{0}^{3})]^{1/2} = 3.5 \times 10^{14}$ T ,
$\tau_{0} = r_{0}/c_{\rm s} = 5.4 \times 10^{-7}$ s.
However, as discussed in section \ref{sec:scalea},
we consider a scaled version of the real star with $a = 50$,
leaving the physical scale height,
$h_{0} = c_{\rm s}^{2} R_{*}^{2}/(G M_{*}) = 0.54$ m, unchanged.

\begin{figure*}
\begin{tabular}{cc}
\includegraphics[height=65mm]{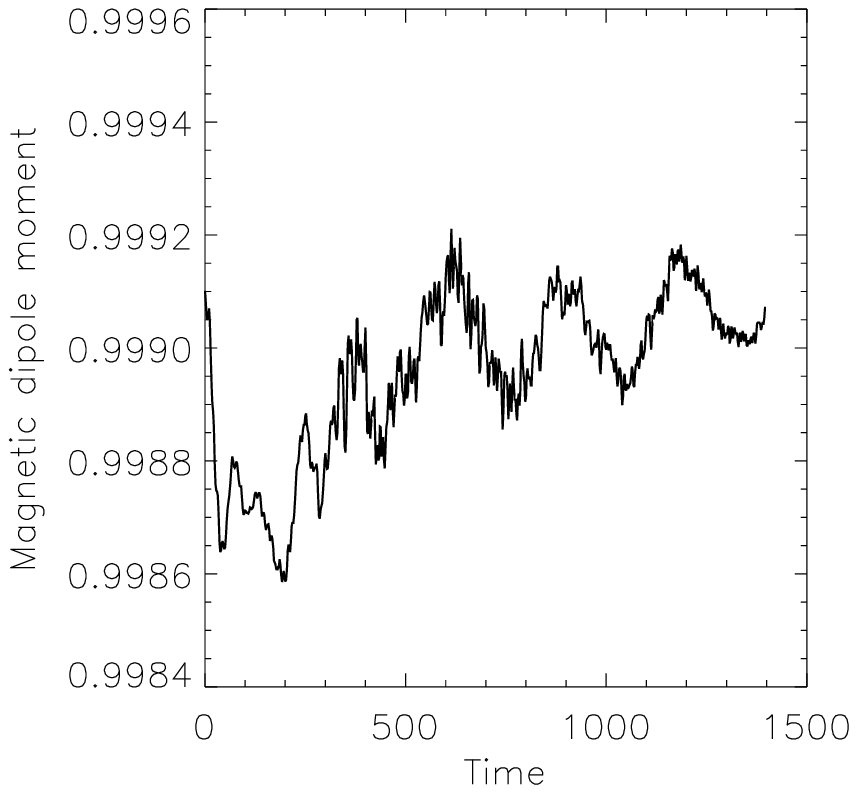} 
&
\includegraphics[height=65mm]{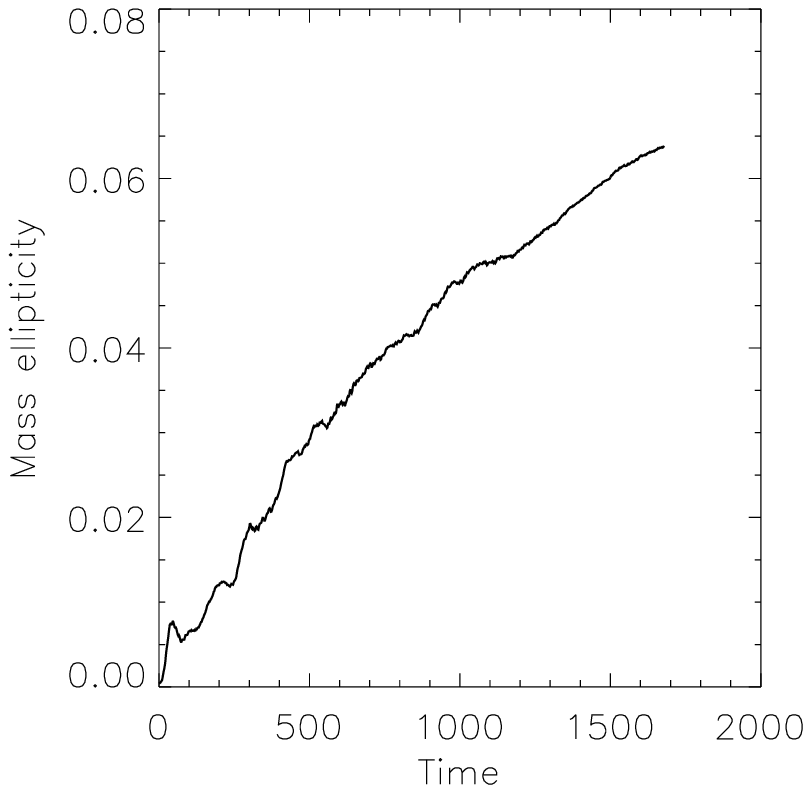}
\\
\includegraphics[height=65mm]{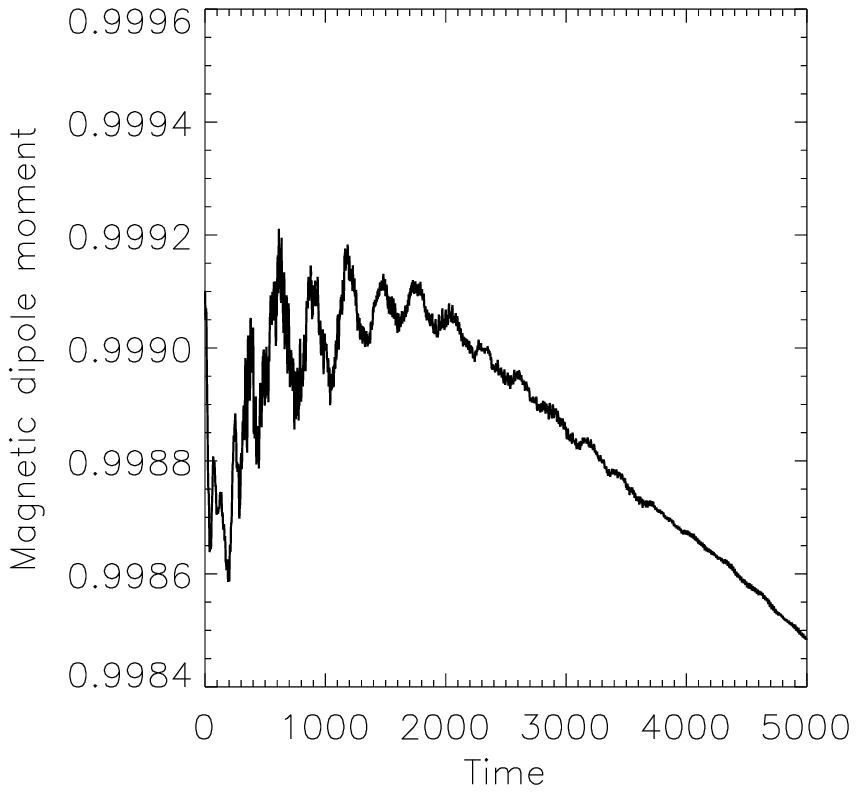} 
&
\includegraphics[height=65mm]{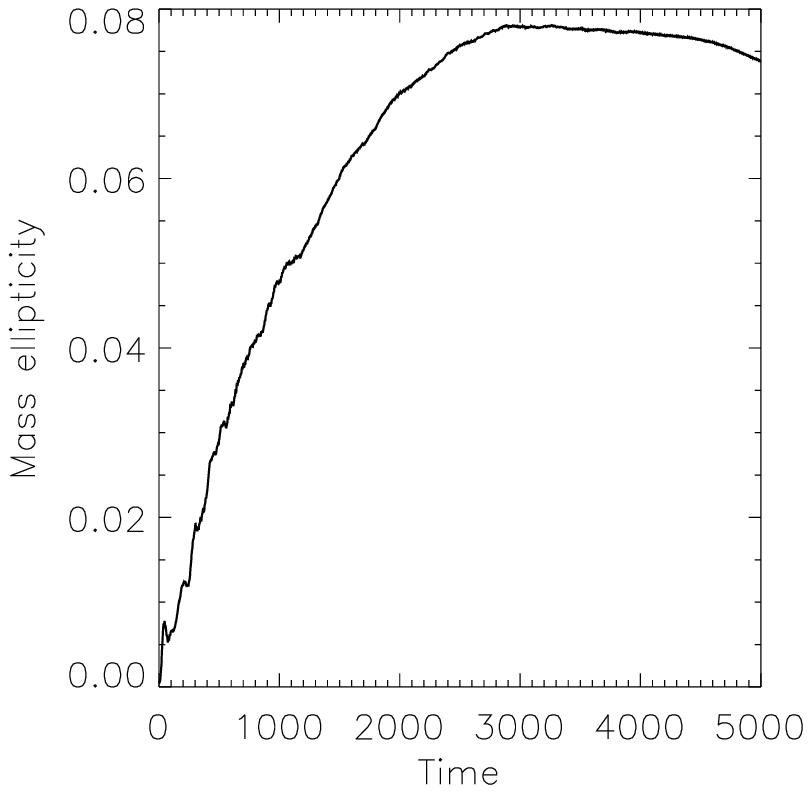}
\\
\end{tabular}
\caption{\small
The dipole (\emph{left}) and mass ellipticity (\emph{right})
as a function of time for an isothermal atmosphere with
$a = 100$ and
$\tilde{r}_{\rm m} = 110$ in a $64\times 64$ grid.
Parameters:
$\rho_{0} = 1$ and
$B_{0} = 0.1$,
{\tt x1rat = 1.03}.
}
\label{fig:isodipole}
\end{figure*}

To estimate the errors in the code, we calculate the
magnetic dipole moment and mass quadrupole moment
and compare to the theoretical values
$\mu = 0.5 a^3 \tilde{B}_{*}$ and
$\epsilon = 0$ in table \ref{table:muerror}.
For $b = 1$, we obtain $\epsilon = 1.02619$;
for $b = 3$, $\epsilon = 1.23373$.
The errors are less than 1 per cent.
Figure \ref{fig:isodipole} shows how the dipole and mass ellipticity
vary with time for these cases.
The dipole moment is accurate to $< 0.2\%$.
The minor variations are caused by numerical jitter
(also see figure \ref{fig:dissipate2}) and evaporation at the
outer boundary (section \ref{sec:transientparker}.
The dependence of the error on grid size is tabulated in table
\ref{table:muerror}.

\begin{center}
\begin{figure}
\centering
\includegraphics[height=65mm]{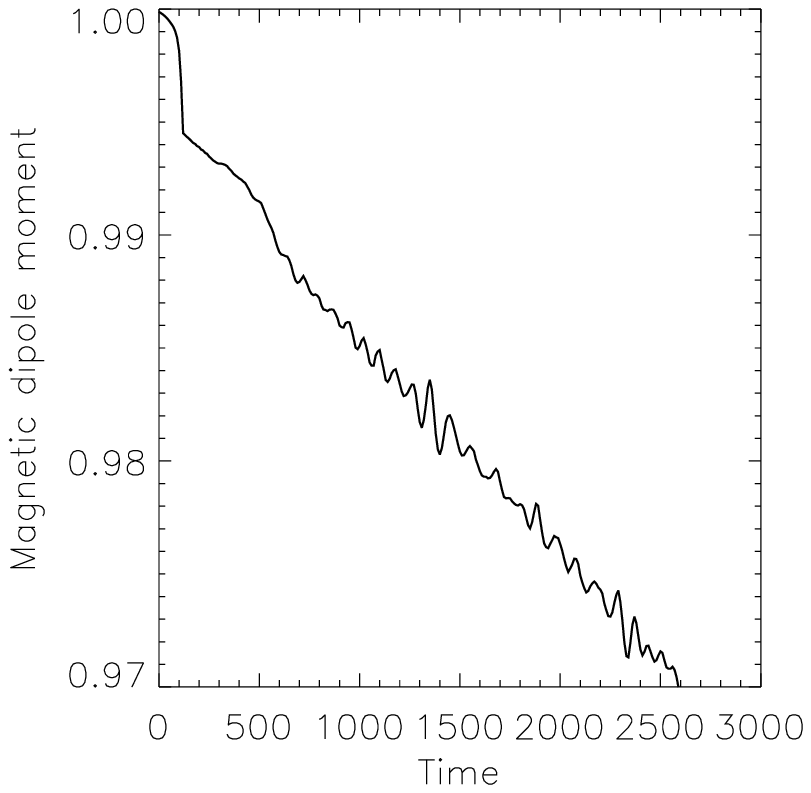} 
\includegraphics[height=65mm]{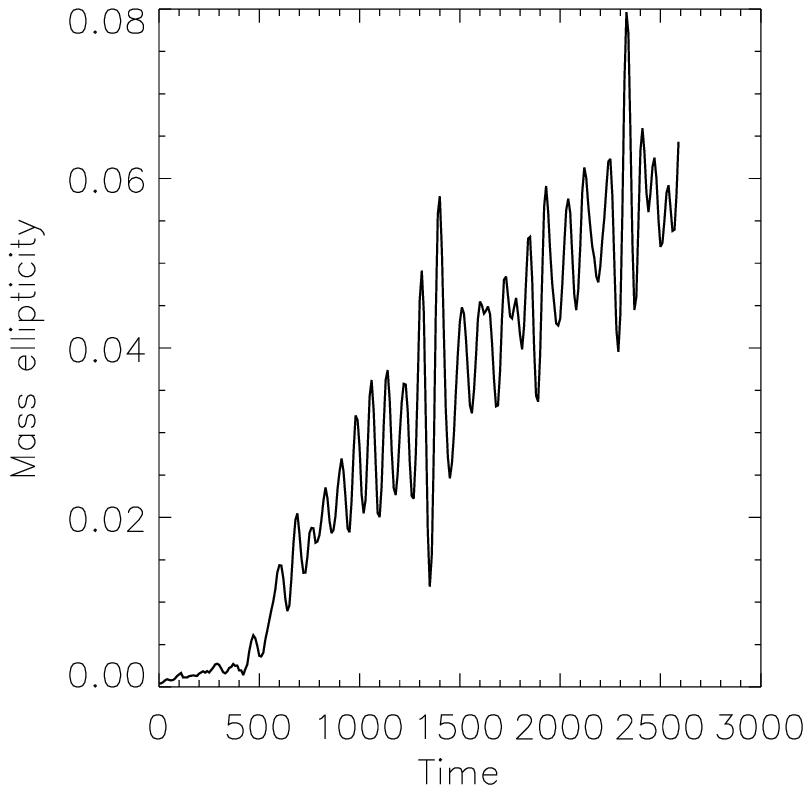}
\caption{\small
The dipole (\emph{left}) and mass ellipticity (\emph{right})
as a function of time for an isothermal atmosphere with
$a = 500$ and
$\tilde{r}_{\rm m} = 510$ in a $64\times 64$ grid.
Parameters:
$\rho_{0} = 1$ and
$B_{0} = 0.1$,
{\tt x1rat = 1.03}.
}
\label{fig:isodipole2}
\end{figure}
\end{center}

\begin{center}
\begin{figure}
\centering
\includegraphics[height=65mm]{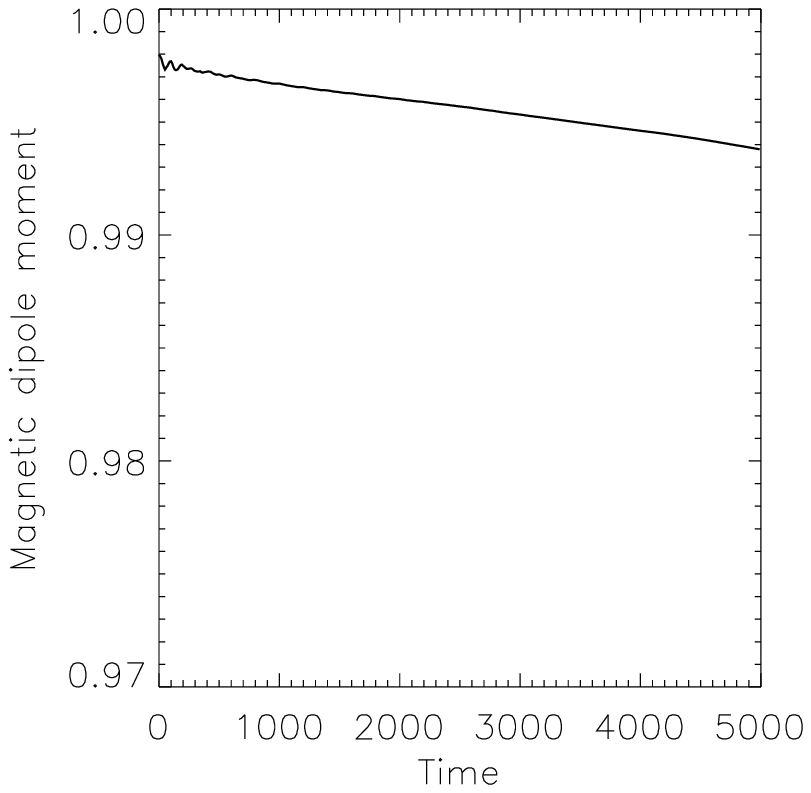} 
\includegraphics[height=65mm]{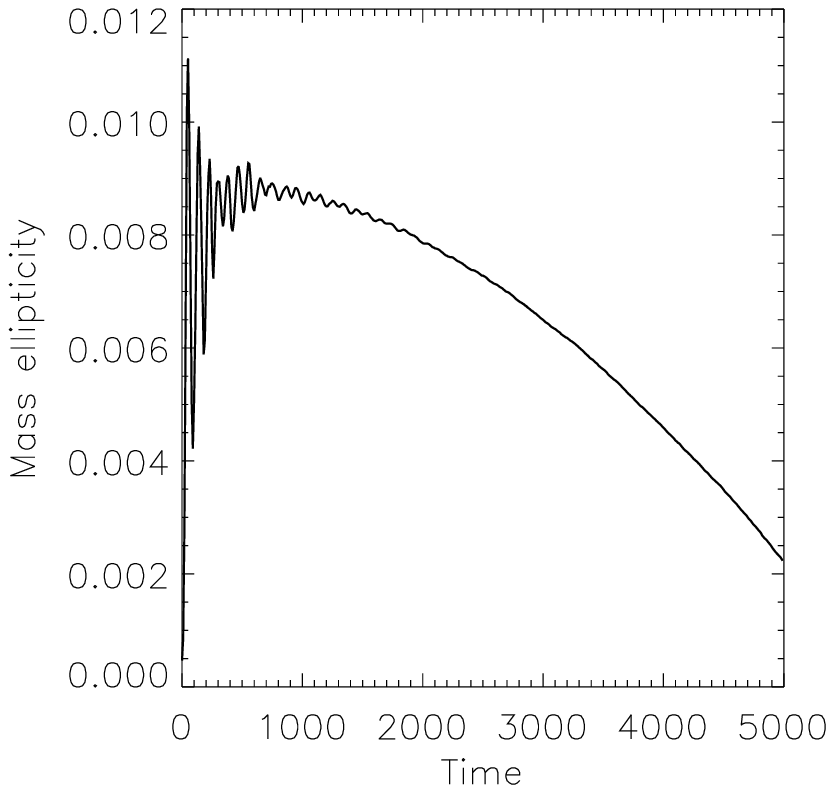}
\caption{\small
The dipole (\emph{left}) and mass ellipticity (\emph{right})
as a function of time for an isothermal atmosphere with
$a = 40$ and
$\tilde{r}_{\rm m} = 50$ in a $64\times 64$ grid.
Parameters:
$\rho_{0} = 1$ and
$B_{0} = 0.1$,
{\tt x1rat = 1.03}.
}
\label{fig:isodipole3}
\end{figure}
\end{center}

\begin{center}
\begin{figure}
\centering
\includegraphics[height=65mm]{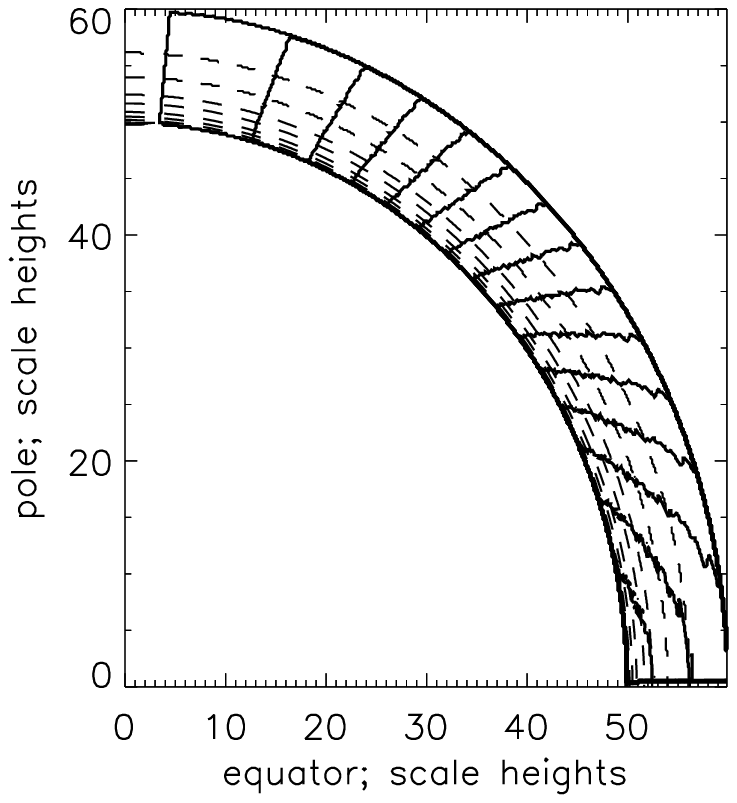} 
\includegraphics[height=65mm]{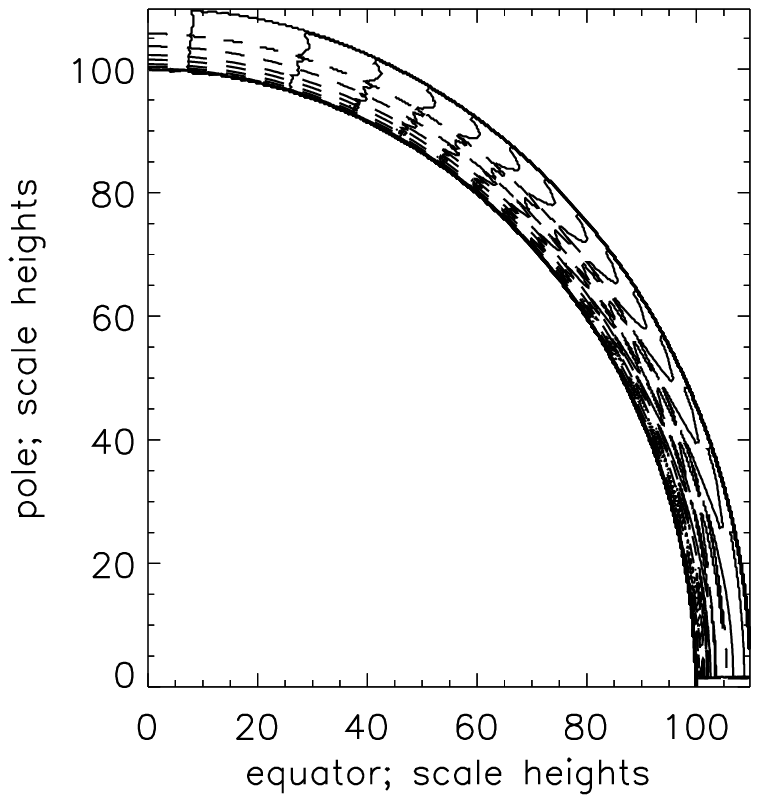} 
\caption{\small
Isothermal atmosphere with dipolar magnetic field.
The kinks in the magnetic field lines (\emph{solid}) are due to
numerical dissipation.
Parameters:
$a = 50, \tilde{r}_{\rm m} = 60, \tilde{\tau} = 66, 150\times 150$ grid
(\emph{left});
$a = 100, \tilde{r}_{\rm m} = 110, \tilde{\tau} = 66, 50\times 100$ grid
{\tt x1rat = 1.1}
(\emph{right}).
}
\label{fig:dissipate2}
\end{figure}
\end{center}

\subsection{Outer boundary}
The outflow condition at the outer boundary has
minimal effect on the evolution if the
outer boundary is sufficiently far away.
To test this, we try several values of $\tilde{r}_{\rm m}$,
keeping $\tilde{r}_{\rm m}$ small enough so that
the Alfv\'en time-step, set by the Courant condition with
$v_{\rm A} \propto B^{2}/\rho$, is not too short.

Numerical dissipation occurs in ZEUS-3D for runs longer than
$\gtrsim 10^{5}$ time-steps.
Kinks in the magnetic field appear at grid points,
as illustrated in figure \ref{fig:dissipate2}.
If the outer boundary is set too many scale heights away, the
minimum density across the grid is very low, the maximum Alfv\'en
velocity is very high, and the time-step, $\Delta t_{\rm Z}$,
determined by the Courant condition,
is very short.  This means that
(i) the simulation takes a very long time to run, and
(ii) the numerical dissipation causes the simulation to stop.
For $a = 50$ and
$\tilde{r}_{\rm m} = 70$, the code stops at $\tilde{\tau} < 2$; for
$\tilde{r}_{\rm m} = 80$, it stops at $\tilde{\tau} < 0.2$.
For $a = 100$ and 
$\tilde{r}_{\rm m} = 110$, the code stops at $\tilde{\tau} = 10.26$,
but with {\tt ggen1:x1rat = 1.03} instead of 1, it continues until
$\tilde{\tau} = 27.22$.
For $a = 1000$ and 
$\tilde{r}_{\rm m} = 1010$, the code stops immediately.
Throughout the paper, we use
$\tilde{r}_{\rm m} = 60$, so that the runs are long enough
to give sufficient time to assess the stability of the equilibria.
Table \ref{table:x1rat} shows how the grid ratio affects the
time to halt.


\begin{table}
\begin{center}
\begin{tabular}{cccc}
\hline
$a$             & $\tilde{r}_{\rm m}$      & \% mass loss & run time (min) \\ 
\hline
50     & 60     & 0.5  & 5\\ 
50     & 55     & 14.1 & 5\\ 
50     & 52     & 17.0 & 5\\ 
20     & 30     & 1.72 & 3\\ 
40     & 50     & 1.2  & 4\\ 
100    & 110    & 0.3  & 10\\ 
500    & 510    & 0.1  & 20\\ 
\hline
\end{tabular}
\caption{ 
Proportion of the initial mass lost from the simulation volume, and
run time for
$\tilde{t} = 100$, as  functions of the dimensionless hydrostatic
scale height $a^{-1}$.
The grid size used was $50 \times 100$.
Naturally, the relative run times are more useful than the absolute.}
\label{table:varya}
\end{center}
\end{table}

\begin{table}
\begin{center}
\begin{tabular}{ccc}
\hline
grid size           &$\mu/\mu_{\rm t}$ & $\epsilon$ \\
\hline
$64\times 64$   & 0.999101      & $ 4.32268\times 10^{-4}$     \\
$50\times 100$  & 1.003032      & $ 2.63789\times 10^{-4}$     \\
$150\times 150$ & 1.001214      & $ 4.52317\times 10^{-4}$     \\
$300\times 300$ & 1.000302      & $ 1.26503\times 10^{-4}$     \\
\hline
\end{tabular}
\caption{ 
Errors as a function of grid size as measured by the
deviation of $\mu$ and $\epsilon$ from theoretical values for $\mu_{\rm t}$
and $\epsilon_{\rm t} = 0$ for
$a = 100$, $\tilde{r}_{\rm m} = 110$ and $\tilde{t} = 0$.
}
\label{table:muerror}
\end{center}
\end{table}

\begin{table}
\begin{center}
\begin{tabular}{cccc}
\hline
{\tt x1rat}     & $\Delta t_{\rm A}$ & $\Delta t_{\rm s}$   & time to halt ($\tau_{0}$) \\ 
\hline
1               & $2.93714\times 10^{-3}$ & 0.20 & 10.26 \\ 
1.02            & $4.64149\times 10^{-3}$ & 0.31 & 17.28 \\ 
1.03            & $5.77838\times 10^{-3}$ & 0.38 & 27.22 \\ 
1.05            & $8.12472\times 10^{-3}$ & 0.52 & 55.62 \\ 
1.10            & $9.67961\times 10^{-3}$ & 0.01 & 66.71 \\ 
1.102           & $9.63428\times 10^{-3}$ & 0.01 & 43.48 \\ 
\hline
\end{tabular}
\caption{
The time taken for numerical dissipation to halt the simulation for
$a = 100$
$\tilde{r}_{\rm m} = 110$, and
a $50 \times 100$ grid,
as a function of {\tt ggen1:x1rat},
the geometric ratio in the rescaled coordinate $r$.
}
\label{table:x1rat}
\end{center}
\end{table}



\subsection{Converting to realistic $a$ values and between codes}
\label{sec:appendix:convert}
Here we give the formulae for converting between the Grad-Shafranov code
and ZEUS-3D.
Furthermore, we detail the effects of converting from a small
($a = 50$) to a realistic ($a \approx 10^{4}$) star.
With $h_0$ fixed, but allowing $a$ to vary, we have
$R_{*} = a h_0 = 27 (a/50)$ m,
$M_{*}/\Msun = a^2 h_0 c_{\rm s}^{2}/(G\Msun) = 1.0138\times 10^{-5}(a/50)^2$,
and
\begin{equation}
\label{eq:mca50dim}
\frac{M_{\rm c}}{\Msun} = \
6.1\times 10^{-15} \
\left(\frac{a}{50} \right)^{4} \
\left(\frac{B_{*}}{10^{8} {\rm T}}\right)^{2} \
\left(\frac{c_{\rm s}}{10^{6} \, {\rm m \, s^{-1}}}\right)^{-4} \, .
\end{equation}
$h_0$ is the dimensionless unit of length.
The dimensionless units in the Grad-Shafranov code (subscript `G') are
$\rho_{0,{\rm G}} = \Ma/h_0^{3}$
$= 9.8\times 10^{-16} (m/0.16)(a/50)^{4}$ kg \, m$^{-3}$,
$B_{0,{\rm G}} = B_{*} a^2/(2b) = 4.2\times 10^{10} (a/50)^2 (b/3)^{-1}$ T.
In ZEUS-3D (subscript `Z'),
$\rho_{0,{\rm Z}} = c_{\rm s}^{2}/(G h_0^{3}) = 9.6\times 10^{22}{\rm kg \, m}^{-3}$,
$B_{0,{\rm Z}} = (\mu_{0}/G h_0^3)^{1/2}c_{\rm s}^{2} = 3.5\times 10^{14}$ T.
For conversion from the equilibrium code to ZEUS-3D, the factors are
$\rho_{\rm G,Z} =\rho_{0,{\rm G}}/\rho_{0,{\rm Z}} =
G\Ma/c_{\rm s}^{2} = 1.3\times 10^{-7}(m/0.16)(a/50)^{4}$, and
$B_{\rm G,Z} = B_{0,{\rm G}}/B_{0,{\rm Z}} =
B_{*} a^2/(2bc_{\rm s}^{2})(G h_0^3/\mu_{0})^{1/2} = 1.2\times 10^{-4}(a/50)^2 (b/3)^{-1}$.


\subsection{Logarithmic coordinates}
\label{sec:logscale}
PM04 concentrated maximum grid resolution 
near the equator and stellar surface where
$\nabla\rho$ and $\nabla\psi$ are greatest, by employing
logarithmic stretching: 
\begin{equation}
\tilde{x_1} = \log(\tilde{x} + e^{-L_{x}}) + L_{x}\, ,
\end{equation}
\begin{equation}
\tilde{y_1} = -\log[1 - (1 - e^{-L_{y}})\tilde{y}]\, .
\end{equation}
To implement these coordinates in ZEUS-3D, set
{\tt ggen1:x1rat} to
$(X e^{L_x} +1)^{(G_x -1)^{-1}}$,
where 
$0\leq\tilde{x}\leq X$ and
$L_x$ controls the `zoom'.
Radial logarithmic scaling gives less grid resolution near the outer
boundary where the density is least and thus
$\Delta t_{\rm Z}$ is greater and
ZEUS-3D runs for a longer time.

When importing a Grad-Shafranov equilibrium (PM04),
whose grid is linear in $y = \cos\theta$, into
ZEUS-3D, whose grid is (in many cases)
logarithmic in $y$, there
is no trivial multiplicative factor relating
$\tilde{y}_{1}$ and
the required logarithmic $\theta$ scaling through
{\tt ggen1:x2rat}.
The problem can be overcome by rewriting the
Grad-Shafranov code such that its grid is logarithmic in $\theta$.
This problem is an obstacle in certain sorts of numerical experiments
e.g. the bootstrapping method in \S 4.2.

\end{document}